\documentclass[12pt]{iopart}
\usepackage[top=2.5cm,bottom=2.5cm,left=2.5cm,right=2.5cm]{geometry}
\usepackage{graphicx,color}
\usepackage{bm}% bold math
\usepackage{epsfig}
\usepackage{hyperref}
\usepackage{import}

\expandafter\let\csname equation*\endcsname\relax
\expandafter\let\csname endequation*\endcsname\relax
\usepackage{amsfonts} 
\usepackage{amsmath}
\usepackage{iopams}
\usepackage{amssymb}
\usepackage{mathtools}
\usepackage{cancel} 
\usepackage{bm}

\usepackage{caption}

\usepackage{dcolumn}
\usepackage{epic} 
\usepackage{epsfig}
\usepackage{feynmp}
\usepackage{graphicx}
\usepackage{grffile}

\usepackage{mathrsfs}
\usepackage{soul}
\usepackage{wrapfig}
\usepackage{xy} 
\usepackage{xcolor}
\usepackage{amsmath,amssymb,amsfonts,amsthm}
\usepackage[ascii]{inputenc}
\usepackage{makecell}
\allowdisplaybreaks

\newcommand{\be}{\begin{equation}}
	\newcommand{\ee}{\end{equation}}
\newcommand{\ba}{\begin{aligned}}
	\newcommand{\ea}{\end{aligned}}
\newcommand{\bea}{\begin{eqnarray}}
	\newcommand{\eea}{\end{eqnarray}}	
	
\newcommand{\beal}{\begin{align}}
\newcommand{\eal}{\end{align}}

\newcommand{\blue}{\textcolor{black}}

\usepackage{fancyhdr}
\begin{document}

\title[Active particle in a harmonic trap driven by a resetting noise]
{Active particle in a harmonic trap driven by a resetting noise: an approach via Kesten variables}%%%%%%%%%%%%%%%%%%%%%%%%%%%

\author{Mathis Gu\'eneau}
\address{Sorbonne Universit\'e, Laboratoire de Physique Th\'eorique et Hautes Energies, CNRS UMR 7589, 4 Place Jussieu, 75252 Paris Cedex 05, France}
\author{Satya N. Majumdar}
\address{LPTMS, CNRS, Univ.  Paris-Sud,  Universit\'e Paris-Saclay,  91405 Orsay,  France}
\author{Gr\'egory Schehr}
\address{Sorbonne Universit\'e, Laboratoire de Physique Th\'eorique et Hautes Energies, CNRS UMR 7589, 4 Place Jussieu, 75252 Paris Cedex 05, France}
\ead{schehr@lpthe.jussieu.fr}

%%%%%%%%%%%%%%%%%%%%%%%%%%%

%%%%%%%% ABSTRACT %%%%%%%%%

\begin{abstract}
We consider \blue{the statics and dynamics of a single particle trapped in a one-dimensional harmonic potential, and subjected to a driving noise with memory, that is represented by a resetting stochastic process. The finite memory of this driving noise makes the dynamics of this particle ``active''.} At some \blue{chosen times (deterministic or random)}, the noise is reset to an arbitrary position and restarts its motion. We focus on two resetting protocols: periodic resetting, where the period is deterministic, and Poissonian resetting, where times between resets are exponentially distributed with a rate $r$. Between the different resetting epochs, we can express recursively the position of the particle. The random relation obtained takes a simple Kesten form that can be used to derive an integral equation for the stationary distribution of the position. We provide a detailed analysis of the distribution when the noise is a resetting Brownian motion. In this particular instance, we also derive a renewal equation for the full time dependent distribution of the position that we extensively study. These methods are quite general and can be used to study any process harmonically trapped when the noise is reset at random times.

\end{abstract}

\maketitle
\tableofcontents

%\import{Chapters/}{Introduction.tex}

\section{Introduction}

Active particle systems are an exciting and rapidly evolving field of research that offers a unique opportunity to study the emergent collective behavior of out-of-equilibrium systems~\cite{Romanczuk,soft,BechingerRev,Ramaswamy2017,Marchetti2017,Schweitzer}. They encompass a wide variety of physical situations, ranging from micro-organisms like bacteria \cite{Berg2004,Cates2012} or synthetic Janus particles \cite{NS2015} all the way to flock of birds \cite{flocking1, flocking2} and fish-schools \cite{Vicsek,fish}. These systems are composed of self-driven particles that can convert energy into directed motion, leading to a wide range of phenomena, including clustering and jamming \cite{cluster1,cluster2,evans,Locatelli2015}, motility induced phase separation \cite{separation1, separation2, separation3,separation4}, and absence of well defined pressure \cite{Kardar2015}. However, these many-body phenomena are still difficult to describe analytically starting from microscopic models, for which very few exact results have been obtained (see however \cite{evans,leo_active_dbm}).

Hence, several recent works focused on the study of the dynamics of a single or of a few active particles. Indeed, it was realised that active systems exhibit many intriguing features even at the single particle level. In particular, it was shown that, in the presence of external confining potentials, active particles behave very differently from their passive counterparts, exhibiting non-Boltzmann stationary state, clustering near the boundaries of the confining region \cite{Berke2008,Cates2009,Potosky2012,Angelani2013,Solon2015,evans,Angelani2019,ABP2019,DKMSS19,Malakar2019,sebastianActiveNoises,Bressloff,Michel} and unusual dynamical and first-passage properties \cite{Peruani,RTP_free,ABP2018, Singh2019,generalRTP,Debruynes}. Paradigmatic models in this context include for instance the active Brownian motion (ABM), see e.g. \cite{ABP2019}, the run-and-tumble particle (RTP), see e.g. \cite{Tailleur_RTP}, or the active Ornstein-Uhlenbeck (AOU) process, see e.g. \cite{AOUP,Wijland21}. In all these cases, the stochastic dynamics of the active particle is
{\it non-Markovian} and driven by a correlated noise -- at variance with a passive particle which is driven by a white noise. For such non-Markovian dynamics, characterizing the interplay between a confining potential and a colored/correlated noise yields challenging questions such as: what is the nature of the stationary state? How does the system reach the stationary state?

\section{General approach}\label{kesten}

In this paper, we address these questions for the active dynamics of a single particle on a line, whose position is denoted by $x(t)$, in the presence of a harmonic potential $V(x)= \mu \, x^2/2$ and subjected to an active noise $y(t)$ -- which is independent of $x(t)$.  
The overdamped equation of motion thus reads
\begin{equation}
\frac{dx(t)}{dt}=-\mu \, x(t) + y(t)\, ,
\label{lange.1}
\end{equation}
starting from $x(0)=0$ for simplicity. For a passive particle, $y(t)$ is  just a white noise $y(t) = \eta(t)$, of zero mean $\langle \eta(t)\rangle = 0$ and with delta correlation $\langle \eta(t) \eta(t') \rangle \propto \delta(t-t')$: in this case $x(t)$ is just the standard Ornstein-Uhlenbeck process (OUP). On the other hand, if $y(t)$ is itself a OU process, i.e., it evolves via $\dot y= - \gamma\, y+ \zeta(t)$ where $\zeta(t)$ is a Gaussian white noise of zero mean, then Eq. (\ref{lange.1}) represents the so called active Ornstein-Uhlenbeck process (AOUP)~(see e.g. \cite{Wijland21}). Another example is the RTP dynamics where $y(t)= v_0\, \sigma(t)$ where $v_0$ is a constant and $\sigma(t)$ is a telegraphic noise that flips between the two values $\sigma(t) = \pm 1$ at a constant rate $\gamma/2$~\cite{Kac,Weiss,DKMSS19}. In the two latter cases, AOUP and RTP, the correlation of the noise decays exponentially, i.e., $\langle y(t) y(t') \rangle \sim e^{- \gamma|t-t'|}$ for large times $t,t' \gg 1/\gamma$. As $\gamma$ decreases, the dynamics of the particle thus crosses over from a passive behavior, as $\gamma \to \infty$, to a strongly active one as $\gamma \to 0$. In the case $\gamma \to \infty$, the noise $y(t)$ behaves essentially as a white noise with some effective temperature $T^*$ (and consequently $x(t)$ is essentially a passive OUP), while when $\gamma \to 0$, $x(t)$ is ``slaved'' to the noise, i.e., $x(t) \approx y(t)/\mu$. A central observable is the probability density function (PDF) $p(x,t)$ of the position at time $t$. In the presence of a harmonic potential (\ref{lange.1}), it is natural to expect that $p(x,t)$ converges to a stationary distribution $p(x,t \to \infty) = p(x)$. From the remark above, in the limit $\gamma \to \infty$, one thus expects that $p(x)$ is given by the Gibbs-Boltzmann weight associated to the quadratic potential $V(x) = \mu \, x^2/2$, i.e., a Gaussian form $p(x) \propto e^{- \mu x^2/(2 T^*)}$, \blue{where $T^*$ is a model-dependent constant that represents an effective temperature}. On the other hand, in the limit $\gamma \to 0$, assuming that the noise $y(t)$ admits a stationary distribution $p_{\rm noise}(y)$, one anticipates that $p(x) \approx (1/\mu) p_{\rm noise}(x/\mu)$. In the case of the AOUP the stationary distribution of the noise is yet another Gaussian, while for the RTP, $p_{\rm noise}(y)$ is just a sum of two Dirac delta functions $p_{\rm noise}(y) = (1/2)\delta(y-v_0) + (1/2) \delta(y+v_0)$.   

\begin{figure}[t]
\centering
\includegraphics[width=0.7\textwidth]{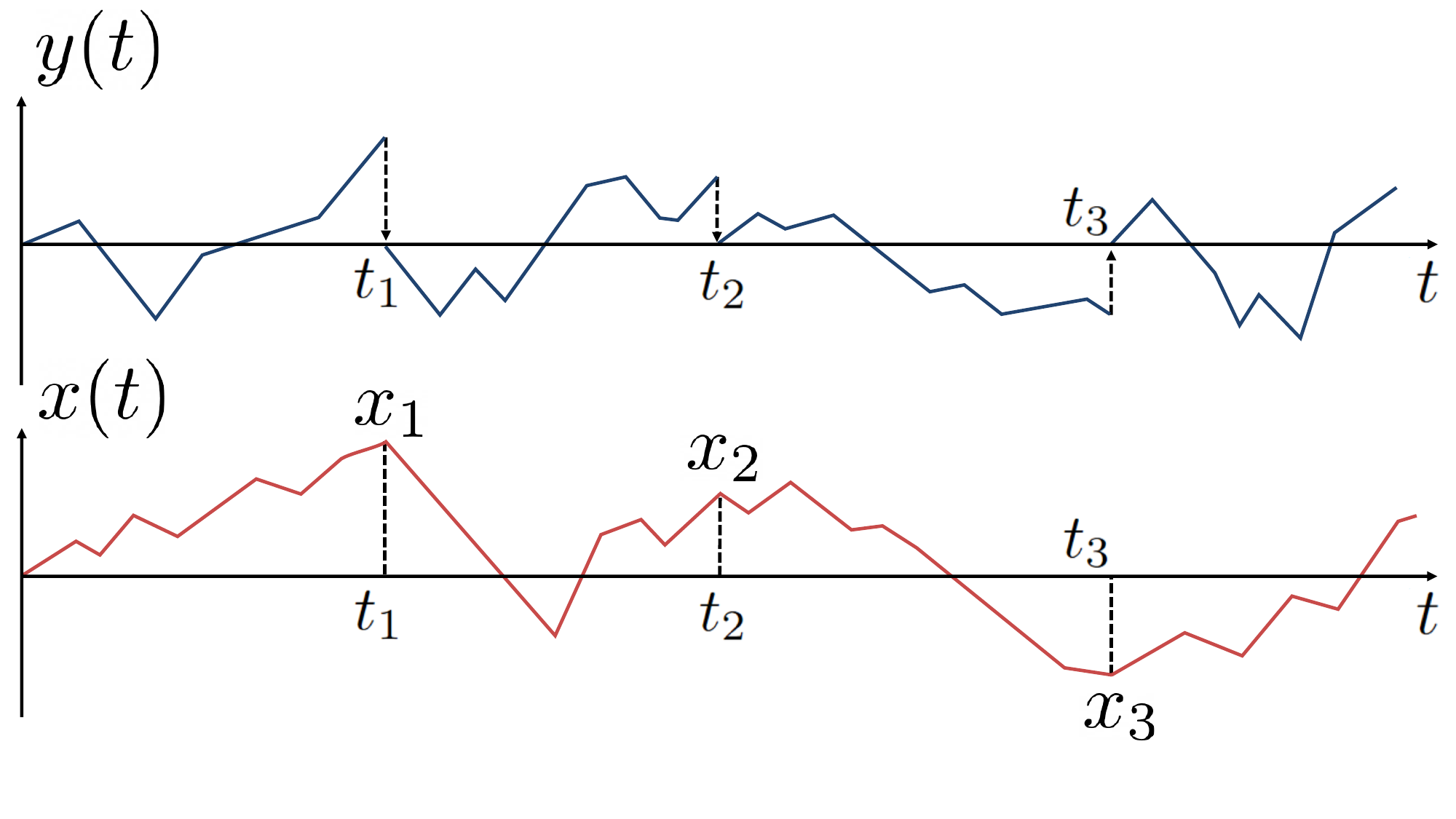}
\caption{A schematic realization of the process $x(t)$ and the noise $y(t)$ in Eq. (\ref{lange.1}). The noise $y(t)$
is reset at epochs $\{t_1,t_2,t_3,\ldots\}$ at which the process $x(t)$ takes values $\{x_1,x_2,x_3,\ldots\}$.
Here, both the process $x(t)$ and the noise $y(t)$ start at the origin.}
\label{fig.process}
\end{figure}

Computing the full crossover between the passive regime $p(x) \propto e^{- \mu x^2/(2 T^*)}$ as $\gamma \to \infty$ and the strongly active one $p(x) \approx (1/\mu) p_{\rm noise}(x/\mu)$ as $\gamma \to 0$ for arbitrary active noise is of course quite challenging. In this paper, we describe a rather general method, based on an approach via Kesten variables, that allows us to describe analytically this crossover of the stationary PDF $p(x)$ for a rather wide class of active noises. We call them "resetting noises", since they bear strong similarities with resetting stochastic processes~\cite{EM2011, Review20, Review22,GuptaReview,Friedman,Besga20,Faisant21}.  
We model the evolution of $y(t)$ as follows. For simplicity it starts from $y(0)=0$. It evolves by its own stochastic dynamics, e.g., just a Brownian or a ballistic motion (with random velocity) or even a (random) constant and then gets reset to $0$ at random epochs $\{t_1, t_2, t_3,\ldots\}$ -- see Fig. \ref{fig.process}. The successive intervals between the resetting events, $\tau_n=t_n-t_{n-1}$ (for $n=1, 2, \ldots$ with $t_0=0$) are statistically independent and each is drawn from a PDF $p_{\rm int}(\tau)$ normalized to unity. Clearly, in the case where the "free" evolution between two successive resettings is a Brownian motion, then $y(t)$ is the well known resetting Brownian motion (rBM), to which the major part of this paper is devoted to. On the other hand, in the case where the "free" evolution is a random constant, say $\pm v_0$ with equal probability, then $y(t)$ corresponds to the telegraphic noise and, consequently, $x(t)$ in (\ref{lange.1}) corresponds to the dynamics of the RTP in the presence of a harmonic potential, which was studied e.g. in \cite{DKMSS19}.  Here we will consider two resetting protocols, namely (i) {\em Poissonian resetting} where $p_{\rm int}(\tau)= r\, e^{-r\, \tau}$~\cite{EM2011} and (ii) {\em periodic resetting} where $p_{\rm int}(\tau)= \delta(\tau-T)$ with $T$ being the period~\cite{PKE16}. In analogy with the AOUP and the RTP discussed above, one thus has $r \sim \gamma$ in the Poissonian case, while $T \sim 1/\gamma$ for periodic resetting. We will see that in the Poissonian case, $x(t)$ approaches a stationary distribution at long times, while in the periodic case $x(t)$ approaches a `time-periodic' stationary state with period $T$. In the context of the rBM, both protocols have been studied theoretically as well as experimentally --and  in fact protocol (ii) turns out to be a bit easier to implement in experiments~\cite{Besga20,Faisant21}. In addition, periodic resetting is also important because it turns out that it is the most efficient resetting protocol in terms of search strategy and it has thus been extensively studied~\cite{BBR16,PR_17,ER20}.

To study the dynamics in Eq. (\ref{lange.1}) driven by an active resetting noise $y(t)$ described above and depicted schematically in Fig. (\ref{fig.process}), it is convenient to introduce $y_n(t)$ which denotes the $y$-process between two successive resetting epochs $t_{n-1}$ and $t_n$. Clearly, $y_n(t)$'s for different $n$'s are statistically independent. 
%For a schematic representation of the two processes $x(t)$ and $y(t)$, see Fig. (\ref{fig.process}).
%
Here we present an approach, based on Kesten variables, to study the stationary distribution of the $x$-process. For a given realization of the resetting epochs $\{t_1,t_2,\ldots\}$, let $x_n$ denote the position of the $x$-process evolving via Eq. (\ref{lange.1}) at the epoch $t_n$, starting from $x_0=0$, see Fig. \ref{fig.process}.  We want to find out the fixed point limiting distribution of $x_n$ as $n\to \infty$. In the Poissonian resetting case, this will coincide with the stationary distribution of the time series $x(t)$. In the periodic case, this will give the limiting distribution of the $x$-process at the end point of a period. Integrating Eq. (\ref{lange.1}) from $t_{n-1}$ to $t_n$ we get a random recursion relation
\begin{equation}
x_n= x_{n-1}\, e^{-\mu \tau_n} + e^{-\mu\, \tau_n}\, \int_0^{\tau_n}d\tau\,  y_n(\tau)\, e^{\mu\, \tau}.
\label{recur.1}
\end{equation}
Note that there are two sources of randomness in this equation. One comes from the realization of the process $y_n(\tau)$ between the two epochs and the second from the randomness of the time interval $\tau_n$ itself. Interestingly, 
this recursion relation is of the generalised Kesten form~\cite{Kesten73,KKS75,DH83,KS84,CLNP85,G91,BDMZ13,GBL21}
\begin{equation}
x_n= U_n\, x_{n-1}+ V_n
\label{kesten.1}
\end{equation}
where $U_n$ and $V_n$ are random variables that may be correlated for a given $n$, but are uncorrelated for different values of $n$. In our case, 
\be \label{def_UV}
U_n= e^{-\mu\, \tau_n} \quad, \quad {\rm and} \quad V_n= e^{-\mu\, \tau_n}\, \int_0^{\tau_n}d\tau\,  y_n(\tau)\, e^{\mu\, \tau}
\ee
 are correlated for a given $n$, since the same random variable $\tau_n$ appears in both $U_n$ and $V_n$, but they are uncorrelated for different $n$'s. In general, finding the stationary distribution of the generalised Kesten recursion (\ref{kesten.1}) is known to be very hard. However, one can make progress in the case where $U_n$ and $V_n$ are jointly distributed according to the joint distribution $P(U,V)$ which is independent of $n$. In this case, using Eq. (\ref{kesten.1}), one can write down a recursion relation
for the position distribution $p(x,n)$ after $n$ ''steps'', namely  
\bea \label{kesten.1bis} 
p(x,n)= \int dU \int dV \int_{-\infty}^{\infty} dx'\,  P(U,V)\, p(x', n-1)\,  \delta(x- U x'-V)\, ,
\eea
where the integration bounds over $U$ and $V$ depend on the joint distribution $P(U,V)$. 
Assuming then that $p(x,n)$ approaches a fixed-point, i.e., $p(x,n) \to p(x)$, as $n \to \infty$, it follows from Eq. (\ref{kesten.1bis}) that the stationary position distribution satisfies the following integral equation
\begin{equation}
p(x)= \int dU \int dV \int_{-\infty}^{\infty} dx'\, P(U,V) \, 
p(x')\,  \delta(x-U\, x'-V)\, .
\label{kesten.2}
\end{equation}
The solution of Eq. (\ref{kesten.2}) is not known for general $P(U,V)$. 
For instance, if we use $y_n(t)=B_n(t)$, i.e., a pure Brownian motion (starting from $0$) between resettings for the noise, and Poissonian resetting $p_{\rm int}(\tau)= r\, e^{-r\, \tau}$, we can easily compute explicitly the joint distribution $P(U,V)$ of $U_n$ and $V_n$. Although it is difficult to solve explicitly the integral equation (\ref{kesten.2}) in this case, we will  see that it is possible to understand the model in details via general methods that can be extended to other resetting noises. Remarkably, we also show that this approach via Kesten variables allows to recover the results for the stationary distribution of the RTP in a harmonic potential (obtained, e.g. in Ref. \cite{DKMSS19} via a completely different method ), and also generalise it to more general velocity distribution (see Appendix \ref{PoissonRTP} for details).

The rest of the paper is organized as follows. In Section \ref{SSperiodic}, we first study the case of periodic resetting and provide explicit solutions of Eq. (\ref{kesten.2}) for various resetting noises. In Section \ref{sectionstatio}, we provide a detailed study of the stationary state in the case where the noise $y(t)$ is the rBM with Poissonian resetting. In Section  \ref{sectionTime}, we analyse the relaxation to the stationary state in this case. Finally we conclude in Section \ref{conclusion}. Technical details have been relegated to Appendices.

\section{Steady state for periodic resetting}\label{SSperiodic}

\subsection{Brownian noise with periodic resetting}
%While Eq. (\ref{kesten.2}) is very complicated in the case of Poissonian resetting, for the periodic resetting protocol, where $p(\tau)= \delta(\tau-T)$ with period $T$ fixed, we can make progress. For the periodic case, Eq. (\ref{recur.1}) reads simply

Here we consider the case where $y_n(\tau) = B_n(\tau)$ is a standard Brownian motion with diffusion coefficient $D$, starting from $0$. In the case of periodic resetting protocol, where $p_{\rm int}(\tau)= \delta(\tau-T)$ with period $T$ fixed, Eq. (\ref{recur.1}) reads 
\begin{equation}
x_n= x_{n-1}\, e^{-\mu\, T} + e^{-\mu\, T}\, \int_0^Td\tau\,  B_n(\tau)\, e^{\mu\, \tau}.
\label{per_Br.1}
\end{equation}
Here $T$ is just a constant. Hence, by iterating this equation, one sees that $x_n$ is just a linear combination of terms each involving a Brownian motion and consequently $x_n$ is a Gaussian variable for any $n$. Its average is clearly zero, $\langle x_n \rangle = 0$ and we only need to compute its variance $\langle x_n^2\rangle $. Here, $\langle \cdots \rangle$ denotes an average over the Brownian noise, which is the only source of randomness here. Squaring Eq. (\ref{per_Br.1}) and taking average gives
\begin{equation}
\langle x_n^2\rangle= \langle x_{n-1}^2\rangle\, e^{-2\,\mu\, T}+ 2D\, e^{-2\, \mu\, T}\, \int_0^Td\tau_1\, \int_0^Td\tau_2\, 
{\rm min}(\tau_1,\tau_2)\, e^{\mu (\tau_1+\tau_2)} \;,
\label{per_Br.2}
\end{equation}
where we have used $\langle B(\tau_1) B(\tau_2)\rangle = 2D\, {\rm min}(\tau_1,\tau_2)$. Note that we have used the fact that $x_{n-1}$ and $B_n(\tau)$ are independent, together with $\langle x_{n-1}\rangle = \langle B_n(\tau)\rangle = 0$. Performing the integral in~(\ref{per_Br.2}) explicitly, we get
\begin{equation}
\langle x_n^2\rangle= \langle x_{n-1}^2\rangle\, e^{-2\,\mu\, T}+
\frac{D}{\mu^3}\, \left[2\,\mu\, T-3 + 4\, e^{-\mu\, T}- e^{-2\, \mu\, T}\right]\, .
\label{per_Br.3}
\end{equation}
\begin{figure}[t]
\centering
\includegraphics[width = 0.5\linewidth]{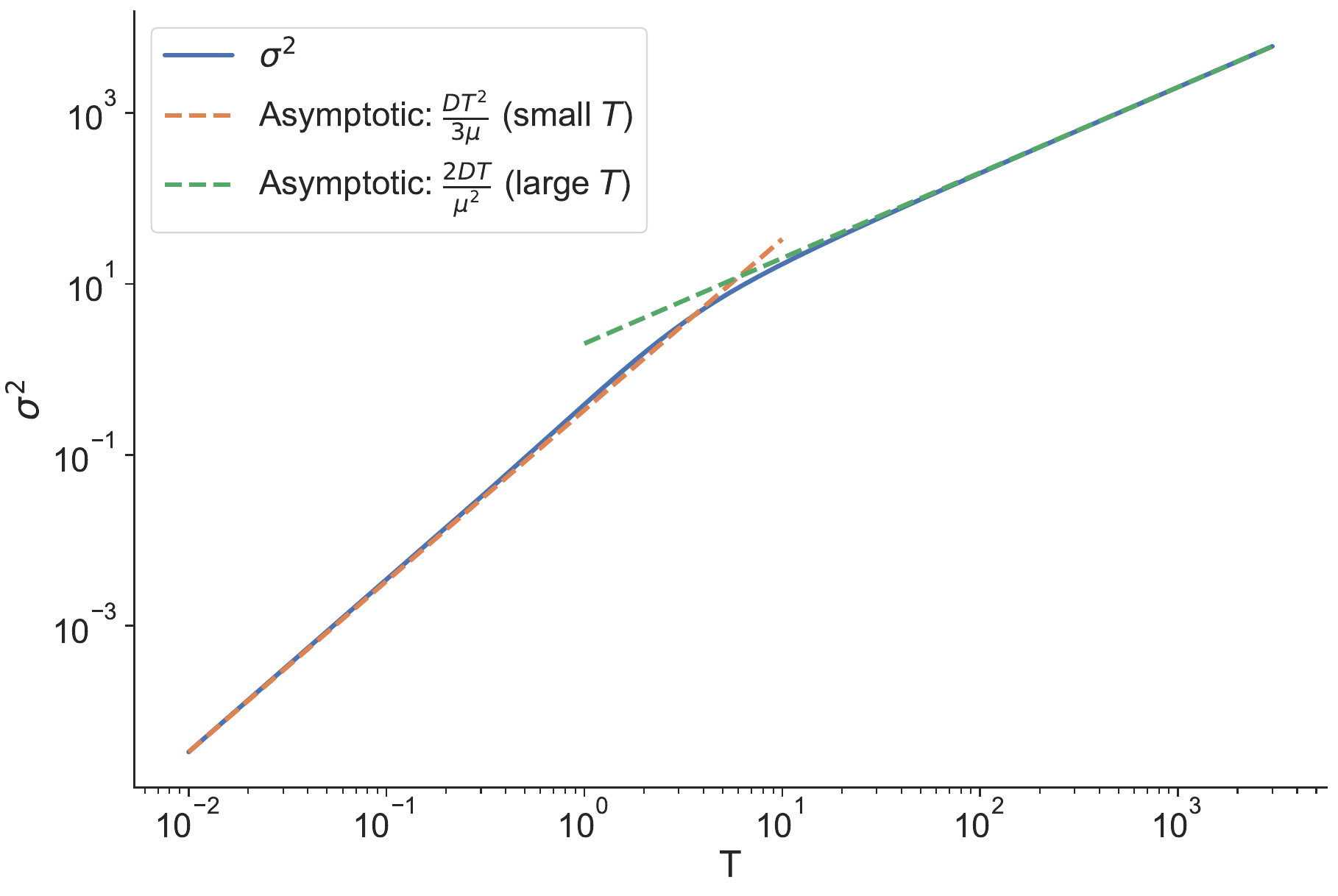}
\caption{The solid blue line shows a plot, on a log-log scale, of $\sigma^2$ given in Eq. (\ref{var_per.1}) vs $T$ for $\mu = 1$ and $D = 1$. The dashed orange and green lines show the asymptotic behaviors of $\sigma^2$ for small and large $T$ respectively, as given in Eq. (\ref{var_asymp}).}\label{plot_sigmaperiodic}
\end{figure}
As $n\to \infty$, the sequence $\{x_n\}$ reaches a stationary Gaussian distribution 
\begin{equation} \label{pst_Gauss}
p(x) = \lim_{n \to \infty} p(x,n) = \frac{1}{\sqrt{2 \pi \sigma^2}} e^{-\frac{x^2}{2\sigma^2}} \;,
\end{equation}
where the variance $\sigma$ is obtained by taking the limit $n\to \infty$ in Eq. (\ref{per_Br.3}). The fixed point variance $\sigma^2 = \lim_{n \to \infty} \langle x_n ^2 \rangle$ is then given explicitly~by
\begin{equation}
\sigma^2= \langle x_{\infty}^2\rangle= \frac{D}{\mu^3}\, \frac{\left[2\,\mu\, T-3 + 4\, e^{-\mu\, T}- e^{-2\, \mu\, T}\right]}{\left[1-e^{-2\, \mu\, T}\right]}\, .
\label{var_per.1}
\end{equation}
In Fig. \ref{plot_sigmaperiodic}, we show a plot of $\sigma^2$ vs $T$ for $\mu=1$ and $D=1$. 
It has the asymptotic behaviors for small and large $T$
\be
\sigma^2 \approx \begin{cases}
\frac{D\, T^2}{3\mu} \quad {\rm as} \quad T\to 0 \\
\\
\frac{2DT}{\mu^2} \quad {\rm as} \quad T\to \infty
\end{cases}
\label{var_asymp}
\ee
Note that the limit $T\to 0$ (rapid resetting) corresponds to the strongly `passive' limit, while $T\to \infty$ (rare resetting) corresponds to the strongly `active' limit. Roughly speaking, these two limits in the periodic resetting correspond respectively to the limits $r\to \infty$ and $r\to 0$ of the Poissonian resetting, since the two protocols are qualitatively similar with the identification $T\sim 1/r$. Thus, as the activity $T$ (period $T$ can be taken as an activity strength) increases, the stationary limit distribution of $x_{\infty}$, while staying Gaussian, has an increasing variance as a function of $T$. Thus the probability mass spreads from the center of the trap outwards as the activity $T$ increases. Therefore activity enhances fluctuations.

Another case with periodic resetting that can be easily solved along the same line is when the particle is driven by an Ornstein-Uhlenbeck noise between resets. As in the case of Brownian motion, the stationary state is a centered Gaussian distribution, with a variance that can also be computed explicitly (see Appendix \ref{ArOUPs}).

\subsection{Ballistic \& Telegraphic noises with periodic resetting}

\noindent{\it Ballistic noise.} Another solvable case corresponds to the ballistic model for the reset process $y(t)$. In this case, after each reset a random velocity $v$ is chosen independently from a symmetric distribution $w(v)$ and the process $y(t)= v\, t$ evolves ballistically with this velocity till it gets reset at the next epoch. Thus in this model the evolution of the noise between the $(n-1)$-th reset and the $n$-th reset is described by $y_n(t)= v_n \, t$ where $v_n$'s are independent and identically distributed (IID) random variables each drawn from $w(v)$. In this case, the recursion relation (\ref{recur.1}) reads
\begin{equation}
x_n =  x_{n-1}\, e^{-\mu\, \tau_n} + v_n\, e^{-\mu \tau_n}\, \int_0^{\tau_n}d\tau \,  \tau \, e^{\mu\, \tau}
=  x_{n-1}\,   e^{-\mu\, \tau_n} + \frac{v_n}{\mu^2}\, \left[ \mu\, \tau_n- 1+ e^{-\mu\, \tau_n}\right]\, .
\label{recur_bal.1}
\end{equation}
Now, this is again of the generalised Kesten form in Eq. (\ref{kesten.1}) with $U_n= e^{-\mu \tau_n}$  and $V_n= \frac{v_n}{\mu^2}\,\left[ \mu\,\tau_n- 1+ e^{-\mu \tau_n}\right]$. 
%As in the Brownian case, here again it is difficult to solve for the limiting distribution when $\tau_n$'s are exponentially distributed, i.e, for the Poissonian resetting case. 
For periodic protocol, $p_{\rm int}(\tau_n)= \delta(\tau_n-T)$, Eq. (\ref{recur_bal.1}) becomes
\begin{equation}
x_n= x_{n-1}\, e^{-\mu\, T}+ \frac{v_n}{\mu^2}\,\left[\mu\, T- 1+ e^{-\mu\, T}\right]\, ,
\label{recur_bal.2}
\end{equation}
where $v_n$ is the only random variable left. This recursion relation is of the form
\begin{equation}
x_n= a\, x_{n-1} + b\, v_n \, ,
\label{AR1.1}
\end{equation}
where $a= e^{-\mu\, T} \leq 1$ and $b= \left(\mu\, T- 1+e^{-\mu\, T}\right)/\mu^2 \geq 0$ are constants, while $v_n$'s are IID random variables, each drawn from a symmetric distribution $w(v)$. Interestingly, Eq. (\ref{AR1.1}) can be thought of as a discrete-time version of an OU process~\cite{Larralde04,MK07,ABRS21}, also known as the AR(1) process (autoregressive process of order $1$) in finance \cite{mckenzie}. In this case, as we will see now, the limiting distribution of $x_\infty$ can be computed explicitly, at least formally, for arbitrary distribution $w(v)$, leading to nontrivial  distributions for the steady state $p(x,n\to \infty)$~\cite{Larralde04,MK07,ABRS21}. 

%Note that in Refs.~\cite{Larralde04,MK07,ABRS21} the authors were mostly interested in the first-passage time distribution of the AR(1) process, rather than the stationary distribution. However, as we will now show, 
%the same techniques used there can be straightforwardly adapted to compute the limiting distribution of Eq. (\ref{AR1.1}) for any distribution $w(v)$ of the $v_n$'s.

\begin{figure}[t]
    \centering
    \includegraphics[width=0.8\textwidth]{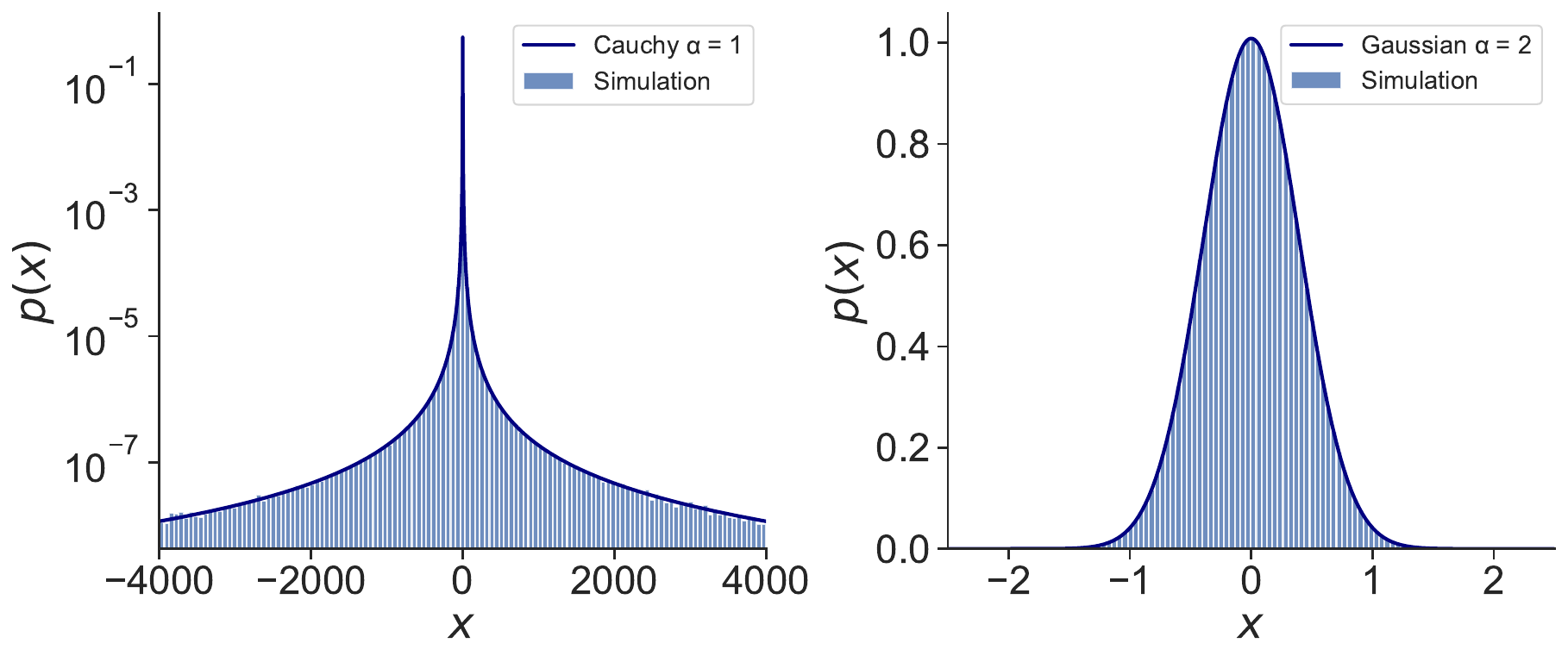}
    \caption{Plot of the stationary distribution for a particle evolving via (\ref{lange.1}) subjected to a resetting ballistic noise $y(t)=v\,t$, where $v$ is distributed according to a Cauchy distribution (left panel) and a Gaussian distribution (right panel). The histograms have been obtained by solving numerically the recursion relations (\ref{recur_bal.2}) after a large number $n$ of iterations, while the solid line corresponds to our exact analytical results in (\ref{distribballistic}).}
    \label{fig:ballisticplot}
\end{figure}

To proceed, it is convenient to define $\tilde v_n = b\, v_n$, whose PDF is simply $\phi(\tilde v_n) =  b^{-1}w(\tilde v_n/b)$, such that the recursion relation becomes \begin{equation}
    x_n= a\, x_{n-1} + \tilde v_n \, .
\end{equation}
The integral equation (\ref{kesten.1bis}) then reads
\begin{equation}
    p(x,n+1) = \int_{-\infty}^{+\infty}dy\, \phi(x-ay)\, p(y,n)\, ,
\label{kestenIEballistic}
\end{equation} where $p(x,n)$ is the distribution of the position after $n$ resets. Thanks to the convolution structure of Eq. (\ref{kestenIEballistic}), we have, in Fourier space,  
\begin{equation}
    \hat{p}(k,n+1) = \hat{\phi}(k)\hat{p}(ak,n)\, ,
\label{recursionBallistic}
\end{equation}
where, for any function $f(x)$, we define its Fourier transform $\hat f(k)$ as
\be \label{def_Fourier}
\hat f(k) = \int_{-\infty}^{+\infty}dx\,  e^{i\,kx} f(x) \;.
\ee
In addition, since $x(0) = 0$ one has $p(x,0) = \delta(x)$ which implies that the initial condition of the recursion relation (\ref{recursionBallistic}) is simply $\hat p(k,0)=1$.
This recursion relation (\ref{recursionBallistic}) has the following solution (see e.g. \cite{Larralde04})
\begin{equation}
    \hat{p}(k,n)=\prod_{m=0}^n\hat{\phi}(a^m k) = \prod_{m=0}^n\hat{w}(a^m \, b\,  k)\,,
\end{equation} 
where we recall that $a= e^{-\mu\, T} \in (0,1)$. Taking the $n\to +\infty$ limit gives 
\begin{equation}
    \hat{p}(k) = \lim_{n \to \infty} \hat{p}(k,n)=\prod_{m=0}^{+\infty}\hat{w}(a^m b\,  k)\, .
\label{fourierperiodicballi}
\end{equation}
Of course, it is not possible to invert explicitly this Fourier transform for arbitrary distribution $w(v)$. However, there is one interesting case for which this inversion can be performed. This is the case of L\'evy distributions of index $\alpha$ and parameter $\lambda$, for which $\hat{w}(k) = e^{-|\lambda k|^\alpha}$ with $0 <\alpha \le 2$. In particular, $\alpha = 1$ (respectively $\alpha =2$) corresponds to the Cauchy distribution (respectively the Gaussian distribution) in real space. For these cases, Eq. (\ref{fourierperiodicballi}) reads
\begin{equation}
    \hat{p}(k) =\prod_{m=0}^{+\infty}\hat{w}(a^m b\,  k) = \prod_{m=0}^{+\infty}\text{exp}\left[- |\lambda \,b \,k|^\alpha (a^\alpha)^m\right] =\text{exp}\left[-|\lambda\, b\, k|^\alpha  \sum_{m=0}^{+\infty}(a^\alpha)^m\right]\, .
\end{equation}
Hence, performing the sum of the geometric series (recalling that $a^\alpha<1$), we have
\begin{equation} \label{phat}
    \hat{p}(k)  = \text{exp}\left[-\frac{|\lambda\, b\,k|^\alpha}{1-a^\alpha} \right]\, .
\end{equation} 
It is again a L\'evy distribution with the same index $\alpha$ as the PDF of the velocity $w(v)$, but with a different parameter $\lambda/(1-a^\alpha)^{1/\alpha}$. Inverting the Fourier transform in (\ref{phat}) one finds
\begin{equation}
    p(x) = \frac{1}{\ell}  \mathcal{L}_\alpha\left( \frac{x}{\ell}\right) \;, \; \ell = \frac{\lambda b}{(1-a^\alpha)^{1/\alpha}} \label{distribballistic}
\end{equation}    
where
\begin{equation}
    \mathcal{L}_\alpha(z)=\int_{-\infty}^{+\infty}\frac{dq}{2\pi}e^{-iqz - |q|^\alpha}\, 
\end{equation} 
is a L\'evy distribution. In Fig. \ref{fig:ballisticplot}, we have compared our analytical prediction (\ref{distribballistic}) with numerical simulation for the cases $\alpha = 1$ and $\alpha = 2$, showing a very good agreement.

Actually, these results can be generalized to a wider class of resetting noises of the form $y(t) = v\, h(t)$, with $h(t)$ an arbitrary function \footnote{Such a noise induces interesting two-time correlations that are computed in Appendix \ref{correlrandomspeed} for Poissonian resetting.}. In this case, the above analysis, starting from
the mapping to the AR(1) process in (\ref{AR1.1}) remains the same with the substitution $b = e^{-\mu T}\, \int_0^{T} h(\tau) \, e^{\mu\, \tau}\, d\tau $, while $a$ remains unchanged.

\noindent{\it Telegraphic noise.} An interesting generalization corresponds to $h(t) = 1$ and while $w(v) = (1/2)\delta(v-v_0) + (1/2) \delta(v+v_0)$ which corresponds to a periodic telegraphic noise, which we now analyse in detail, since it is very similar to a standard RTP model. In the standard RTP model the time $\tau$ between two flips of the noise is a continuous random variable with an exponential distribution. In the present model, this time takes discrete values $\tau = k\,T$ with $k=1, 2, \ldots$ with probability ${\rm Prob.}(\tau = k T) = 2^{-k} = e^{- \gamma_{\rm eff} \, \tau}$, with an effective "tumbling rate" 
\be \label{gamma_eff}
\gamma_{\rm eff} = \frac{\ln 2}{T} \;.
\ee
From the analysis performed above, using $\hat w(k) = \cos(v_0\,k)$
together with $b=  e^{-\mu T}\, \int_0^{T} h(\tau) \, e^{\mu\, \tau}\, d\tau = 1/\mu \left(1-e^{-\mu T}\right)$, 
the Fourier transform of the stationary PDF of the position $\hat p(k)$ is given by
\begin{figure}[]
   % \hspace{-3cm}
  %  \floatbox[{\capbeside\thisfloatsetup{capbesideposition={right,center},capbesidewidth=5.5cm}}]{figure}[\FBwidth]
  \centering
    \includegraphics[width=\textwidth]{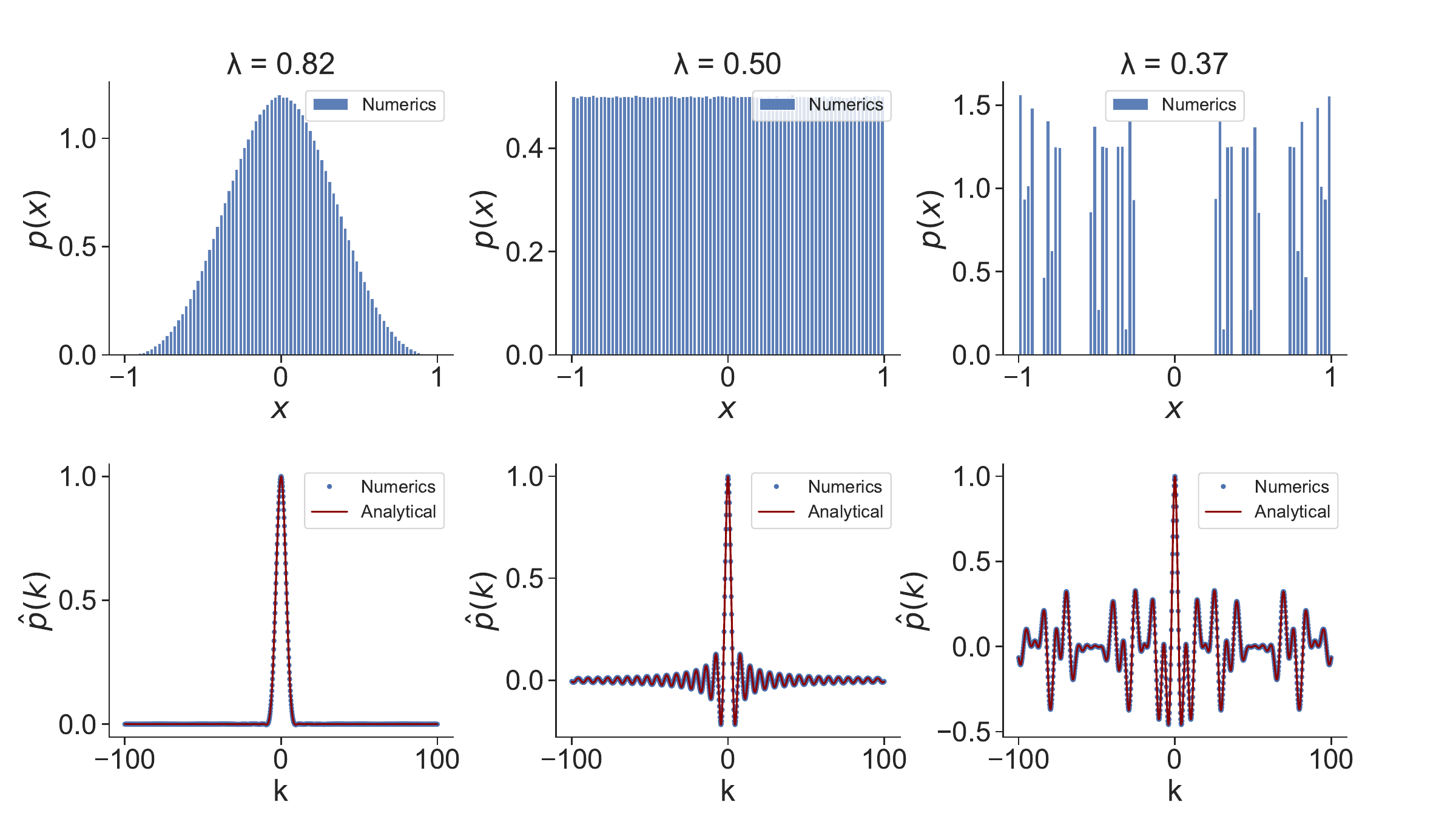}
    {\caption{{\bf Top panel}: Plots of the stationary PDF of the position for periodic run-and-tumble particles in the stationary state governed by the recursion relation (\ref{perRTP1}) for different values of the parameter $\lambda = e^{-\mu T}$. The speed $v_0$ and the strength of the potential $\mu$ are taken to be unity. The support of the distribution is  $[-v_0/\mu, v_0/\mu] = [-1, 1]$. When $\lambda>1/2$ (left panel), the steady state has a bell shape (passive regime), while if $\lambda < 1/2$ (right panel), its support is a Cantor set, and as $\lambda$ decreases, particles are more and more concentrated at the edge of the support, i.e. $x=\pm 1$ (active regime). The transition between the two regimes occurs at $\lambda = 1/2$, when the distribution becomes uniform (middle panel). {\bf Bottom panel:} Plots of the Fourier transform of the stationary PDF of the position for the same values of $\lambda$. The red curves correspond to the analytical product of cosines (\ref{infinitecosperiodicRTP}) -- truncated at a large number of terms. The blue points correspond to a numerical evaluation of the Fourier transform of the density $\left<e^{ikx}\right>$. This has been obtained by averaging over $10^7$ trajectories.}
    \label{fig:periodicRTP}}
 \end{figure}
\begin{equation}
    \hat{p}(k) =\prod_{m=0}^{+\infty}\text{cos}\left\{\frac{v_0}{\mu}\left[\left(e^{-\mu T}\right)^{m}\left(1-e^{-\mu T}\right)\right]k\right\}\, .
\label{infinitecosperiodicRTP}
\end{equation}
As $\left(e^{-\mu T}\right)^{m}<1$ the product converges and is well defined. Interestingly, such distributions~(\ref{infinitecosperiodicRTP}) have appeared in various contexts in the mathematics \cite{PeriodicRTP1,PeriodicRTP3} and physics \cite{PeriodicRTP2,Shrinking} literature and they are known to exhibit a rich and intriguing behavior. To understand it better, it is useful to come back to the Kesten recursion relation in Eq. (\ref{AR1.1}) which reads, in this case
%
%In fact, the problem can be reformulated in a different manner. Between resets, our process exhibits deterministic dynamics starting from $x_0 = 0$ and moving either to the left or right with equal probability. The velocity of the particle fluctuates as a function of its position due to the presence of a harmonic potential, resulting in variations in the size of the deterministic step taken. Specifically, movements that are directed towards the minimum of the potential $V(x)$ are larger than those directed away from it. We have 
%
%
\begin{equation}
x_n= e^{-\mu\, T}\, x_{n-1} \pm \frac{v_0}{\mu} \left(1-e^{-\mu T}\right)\, .
\label{perRTP1}
\end{equation} If we rescale the position such that $\tilde{x}_n = x_n\, \left[\
\frac{v_0}{\mu} \left(1-e^{-\mu T}\right)\right]^{-1}$, the dynamics is as follows
\begin{equation} 
\tilde{x}_n= \lambda\, \tilde{x}_{n-1} + \epsilon_{n} \quad, \quad \epsilon_n = 
\begin{cases}
&+ 1 \quad, \quad {\rm with \; proba.}\; 1/2 \\
& \\
&-1 \quad, \quad {\rm with \; proba.}\; 1/2  
\end{cases}\;,
\label{perRTP2}
\end{equation} 
with $\lambda = e^{-\mu T}$. Hence Eq. (\ref{perRTP2}) is an AR(1) process -- i.e., a discrete version of the OU process -- with Bernoulli $\pm 1$ jumps $\epsilon_n$. This recursion relation (\ref{perRTP2}) can be solved explicitly 
\be \label{sol_rec}
\tilde x_0 = 0 \quad, \quad \tilde x_n = \sum_{m=0}^{n-1} \lambda^m\, \epsilon_{n-m} \;.
\ee
Note that if one denotes $\tilde \epsilon_m = \epsilon_{n-m}$, which are also independent Bernoulli random variables, one can then interpret $\tilde x_n$ as a random walk with shrinking steps, since the size of the $m$-th step is $\lambda^m$ (and we recall that $\lambda < 1$). This is precisely the problem that was studied in Ref.~\cite{PeriodicRTP2}. From Eq. (\ref{sol_rec}), since $\epsilon_{n-m} = \pm 1$, it is clear that the PDF of $\tilde x_n$ has a finite support $[-(1-\lambda^n)/(1-\lambda),+(1-\lambda^n)/(1-\lambda)]$. Indeed, the maximum value of $\tilde x_n$ in (\ref{sol_rec})
is attained for $\epsilon_{n-m} = +1$ for $m=0,1, \cdots, n-1$, leading to $\tilde x_n = \sum_{m=0}^{n-1}\lambda^m = +(1-\lambda^n)/(1-\lambda)$.
Similarly, the minimum corresponds to $\epsilon_{n-m} = -1$ for $m=0,1, \cdots, n-1$, leading to the minimal value $\tilde x_n = -\sum_{m=0}^{n-1}\lambda^m = +(1-\lambda^n)/(1-\lambda)$. In the limit $n \to \infty$, the support of the stationary PDF of $\tilde x_n$ is thus $[-(1-\lambda)^{-1}, (1-\lambda)^{-1}]$. Recalling that $x_n = (v_0/\mu)(1 - \lambda)\, \tilde x_n$, we thus obtain that the stationary PDF $p(x)$ has support over $[-{v_0}/{\mu}, + {v_0}/{\mu}]$. Interestingly, in the case of the standard RTP, the support of the stationary distribution is exactly the same. However, the PDF in the present case turns out to be more exotic. 

To state the main results about $p(x)$, it is useful to rewrite (\ref{infinitecosperiodicRTP}) as 
\begin{equation}
    \hat{p}(k) =\prod_{m=0}^{+\infty}\text{cos}\left(\lambda^m \tilde{k}\right) \;, \; {\rm with} \; \lambda = e^{-\mu T}\quad {\rm and} \quad \tilde{k}=k\,
\frac{v_0}{\mu} \left(1-e^{-\mu T}\right) \;.
\label{bernoulliconvol}
\end{equation} 
In the math literature, such distribution (\ref{bernoulliconvol}) are called ``infinite Bernoulli convolutions'' \cite{PeriodicRTP1,PeriodicRTP3}, since they correspond to the convolution of an infinite number of Bernoulli random variables $\pm \lambda^n$ -- see Eq. (\ref{sol_rec}). For generic values of $\lambda$, this infinite product can not be expressed in a closed form. There are however special values of $\lambda$ for which this infinite product simplifies and consequently $\hat p(k)$ can be inverted. For the simplest case $\lambda=1/2$, thanks to the so called Viete's formula, Eq. (\ref{bernoulliconvol}) reads
\begin{equation} \label{Viete}
    \hat{p}(k) =\prod_{m=0}^{+\infty}\text{cos}\left(\frac{\tilde{k}}{2^m}\right) = \frac{\sin{2\tilde k}}{2\tilde k} \;,
 \end{equation} 
which means that $p(x)$ is simply the uniform distribution over $[-{v_0}/{\mu}, + {v_0}/{\mu}]$ for $\lambda = 1/2$.
In fact, it turns out that $p(x)$ exhibits quite different behaviors as $\lambda$ crosses this special value $\lambda_c = 1/2$. 
For $\lambda > 1/2$, the distribution is regular for almost all $\lambda$ and it has typically a bell-shape, see Fig. \ref{fig:periodicRTP} (in this region, explicit expressions for $p(x)$ can be obtained for $\lambda = 2^{-1/m}$, with $m=1,2, \ldots$~\cite{PeriodicRTP2}): this is a regime that we thus call "passive". In this regime, the density near the boundaries at $\pm v_0/\mu$ vanishes as $p(x) \sim |x\pm v_0/\mu|^{\nu}$ with $\nu = \ln2/|\ln \lambda| - 1$ \cite{PeriodicRTP1}. In terms of the effective tumbling rate (\ref{gamma_eff}) it thus reads $\nu = \gamma_{\rm eff}/\mu - 1$, exactly as in the standard RTP with the same tumbling rate (see e.g. \cite{DKMSS19}). On the other hand, for $\lambda <1/2$ is quite singular. In particular, the support of the distribution is a Cantor set and \blue{is a fractal \cite{PeriodicRTP2}}. In this case, the stationary distribution $p(x)$ exhibits peaks (see Fig. \ref{fig:periodicRTP}) and as $\lambda$ approaches $0$, the support gets restricted to the two points $-v_0/\mu$ and $v_0/\mu$ resulting in particle concentration at the edge of the support. This is a regime that we call ``active''. Recalling that $\lambda = e^{- \mu T}$, we see from Eq. (\ref{gamma_eff}) that 
this transition between the active and the passive regime occurs when the effective tumbling rate $\gamma_{\rm eff}$ crosses the "critical" value $\gamma_{{\rm eff},c} = \mu$, exactly as in the standard RTP model with the same effective tumblings rate (see e.g. \cite{DKMSS19}). It is thus interesting to see that this transition at $\lambda = 1/2$, which is rather well known in the mathematics literature, has a physical interpretation as a transition between a passive phase for $\lambda > 1/2$ and an active one for $\lambda < 1/2$.

%First, the probability distribution of the process (\ref{perRTP1}) has for finite support the interval $[-v_0/\mu, v_0/\mu]$. This can be seen noticing that if the particle only makes positive (respectively negative) jumps, it will end up at the farthest point $x = v_0/\mu$ (respectively $x = -v_0/\mu$). These points are in fact the turning points of the force deriving from the harmonic potential $V(x) = \mu\, x^2/2$. The passive regime corresponds to $\lambda >1/2$. In that case, the distribution has a bell shape and every points in $[-v_0/\mu, v_0/\mu]$ are accessible to the particle. However, when $\lambda <1/2$ the support of the distribution is a Cantor set and is fractal. As $\lambda$ approaches $0$, the support gets restricted to the two points $-v_0/\mu$ and $v_0/\mu$ resulting in particle concentration at the edge of the support. This corresponds to the active regime. The transition between the two regimes occurs when $\lambda = 1/2$ at which point the distribution becomes uniform on the interval $[-v_0/\mu, v_0/\mu]$. These different behaviours are shown in Fig. \ref{fig:periodicRTP}.

%\import{Chapters/}{StationaryState.tex}

\section{Steady state for Poissonian resetting Brownian noise}\label{sectionstatio}

From now on, we focus on the dynamics of a one-dimensional particle in a harmonic potential $V(x) = \mu x^2/2$ that starts its motion at $x(0)=0$ and is subjected to a Poissonian resetting Brownian noise. The position of the particle $x(t)$ thus evolves through the Langevin equation 
\be \label{lange_rbm}
\frac{dx(t)}{dt}=-\mu \, x(t) + \tilde y_r(t) \quad, \quad \tilde y_r(t) = r\, y_r(t) \;,
\ee
with $y_r(t)$ being a resetting Brownian motion (rBM) with resetting rate $r$~\cite{EM2011, Review20}. Note that in Eq. (\ref{lange_rbm}) the factor $r$ in the noise term has been added such that the noise has the dimension of a velocity. More precisely we consider the case where the rBM starts at the origin, i.e., $y_r(0) = 0$, and it is reset at exponential random times also at the origin (as, e.g., in the top panel of Fig.~\ref{fig.process}). During the infinitesimal time interval $[t, t+dt]$, the rBM thus evolves via~\cite{EM2011, Review20} 
\begin{equation} \label{def_rBM}
\displaystyle
y_r(t+dt) = \left\{
    \begin{array}{lll}
        0 \hspace*{3cm}\mbox{ with probability } r\, dt \;,\\
        \\
        y_r(t) + \xi(t)\, dt \hspace*{0.8cm}\mbox{ with probability } (1-r\, dt) \;,
    \end{array}
\right. 
\end{equation}
where $\xi(t)$ is a Gaussian white noise of zero mean $\left<\xi(t)\right> = 0$ and delta-correlations, i.e.,  $\left<\xi(t)\xi(t')\right> = 2D\, \delta(t-t')$. We denote by $\{t_1, t_2, t_3,\ldots\}$ the random times (or epochs) at which the rBM gets reset to $0$. For such a dynamics (\ref{def_rBM}), the time intervals $\tau_n=t_n-t_{n-1}$ (for $n=1, 2, \ldots$ with $t_0=0$) are statistically independent and distributed according an exponential distribution  $p_{\rm int}(\tau_n)= r\, e^{-r\, \tau_n}$: this is called {\it Poissonian resetting}. Therefore we see that the dynamics described by Eqs. (\ref{lange_rbm}) and (\ref{def_rBM}) can be described in the framework relying on Kesten variables as described in Section \ref{kesten}. However, 
at variance with the periodic resetting studied in the previous section, we will see that the integral equation (\ref{kesten.2}) is quite complicated to solve explicitly for Poissonian resetting. Nonetheless, we will see how it can be used to obtain useful detailed information on the stationary state of the Langevin equation~(\ref{lange_rbm}). 

Let us start with a qualitative description of the late time dynamics described by (\ref{lange_rbm}) and (\ref{def_rBM}). At large time, the resetting Brownian noise $\tilde y_r(t)$ converges towards a stationary state, whose limiting PDF is a symmetric exponential distribution, namely \cite{EM2011}
\begin{equation}
P(\tilde y_r) = \frac{1}{2\sqrt{r\,D}} \, e^{- \frac{|\tilde y_r|}{\sqrt{r\,D}}} \, .
\label{statiorBM}
\end{equation}
Besides, its two-time correlation function is given by \cite{correlation resetting}
\begin{equation}\label{correly.1}
  \tilde C_r(t_1,t_2) = \left<\tilde{y}_r(t_1)\tilde{y}_r(t_2)\right> = 2\, D\, r\,  e^{-r\, (t_2-t_1)}\,  (1 - e^{-r\, t_1})  \quad, \quad t_2 > t_1 \;.
\end{equation}
In particular, in the stationary state where $t_1 \to \infty$, keeping $t_2 - t_1 = \tau$ fixed, the correlation function $\tilde C_r(t_1,t_2)$ decays as a pure exponential, i.e.,
\be \label{exp_decay}
\lim_{t_1 \to \infty}  \tilde C_r(t_1,t_1 +\tau) =  2\, D\, r\,  e^{-r\, \tau}\,  \;.
\ee
Hence in the stationary regime, the two-time correlations (\ref{exp_decay}) are very similar to the AOUP or the RTP model, but the one time distribution (\ref{statiorBM}) is different from both models -- since it is Gaussian for the AOUP while it is the sum of two delta-peaks at $\pm v_0$ for the RTP. Because of the similarities with these two models, it is rather natural to expect that PDF of the position $x(t)$ will converge to a stationary distribution $p(x)$ which is the main focus of this section. 

Before we proceed to the detailed analysis of $p(x)$,  
it is useful to note that there are two time scales in the problem: (i) $\tau_\mu = 1/\mu$, which characterises the time scale
of the relaxation within the confining harmonic potential and (ii) $\tau_r = 1/r$ which measures the correlation time of the resetting Brownian noise [see Eq. (\ref{exp_decay})]. Hence it is convenient to use the dimensionless ratio 
\be \label{def_beta}
\beta = \frac{\tau_\mu}{\tau_r} = \frac{r}{\mu} \;. 
\ee
In the limit $\tau_r \ll \tau_\mu$, i.e. $\beta \to \infty$, the system is said to be ``passive'' while for $\tau_r \gg \tau_\mu$, i.e., $\beta \to 0$, the system is strongly active. We will see that the parameter $\beta$ in (\ref{def_beta}) indeed controls the crossover between the passive to active regimes (see Fig. \ref{fig.modelregime}).

\begin{figure}[t]
\centering
\includegraphics[width=0.7\textwidth]{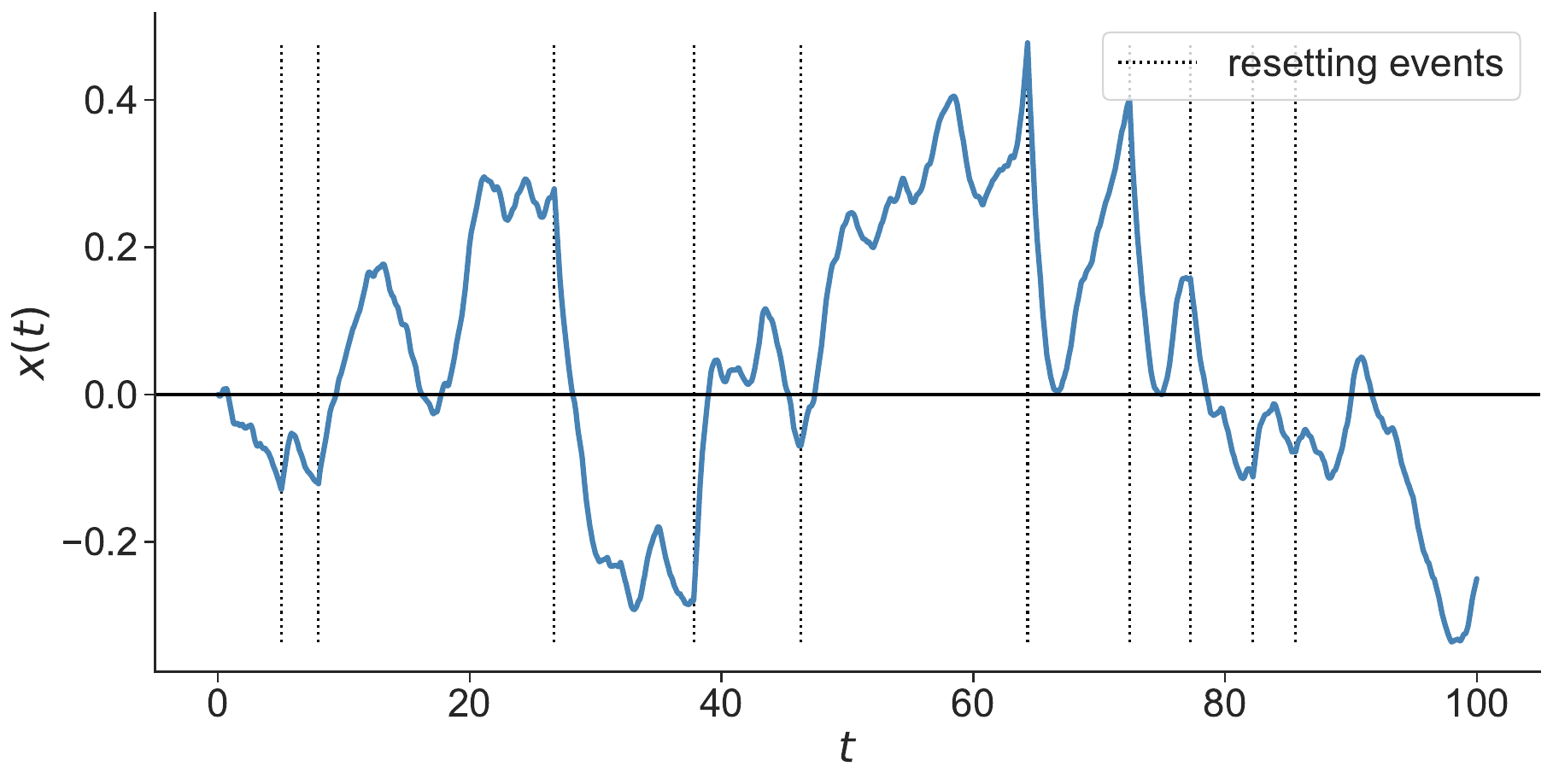}
\caption{A typical realisation of a trajectory for the equation of motion Eq. \eqref{lange_rbm}. Right after a resetting of the noise, one observes a relaxation of the trajectory towards the origin, due to the confining potential. On the other hand it is harder to see the relaxation of the noise $\tilde{y}_r(t)$ itself. While a pure Brownian motion would diverge at large time, one can observe that the resets have a tendency to confine the position of the particle $x(t)$ near the origin. Indeed, when a reset happens, the noise is reset at value $0$ and the speed of the particle is thus $\dot{x} = -\mu\,x$ such that it is directed toward the origin. If the position of the particle is positive (respectively negative) and increasing (respectively decreasing) before the reset, it gets reoriented towards the origin just after the reset. This explains the cusps that we can clearly observe on the trajectory around a resetting event. Here the parameters are $D=1$, $\mu = 1$, $r=0.1$, $x_0=1$. For details on the simulations, see Appendix~\ref{simulationprocedure}.}
\label{fig.traj}
\end{figure}

\begin{figure}[t]
\centering
\includegraphics[width=0.8\textwidth]{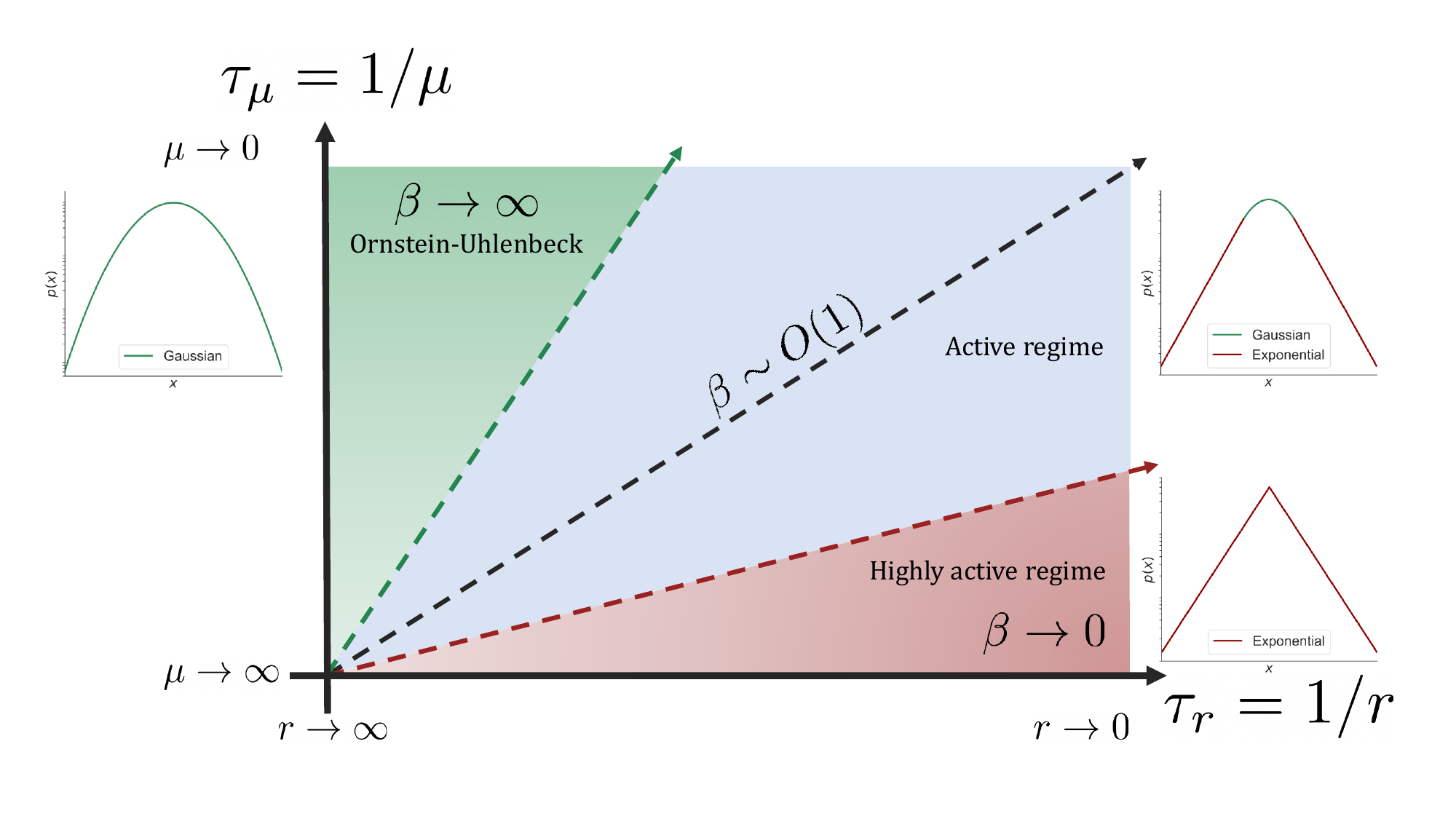}
\caption{Schematic description of the different regimes of the model described by (\ref{lange_rbm}) in the plane $(\tau_r= 1/r, \tau_\mu = 1/\mu)$. When  $\tau_\mu \gg \tau_r$, i.e., in the strongly passive limit $\beta \to \infty$, $x(t)$ is an effective Ornstein-Uhlenbeck process with a diffusion constant $D_{\rm OU}=2D$. In this case the stationary distribution is a Gaussian [see the second line of Eq. (\ref{scalingformlimit})]. On the other hand, when  $\tau_\mu \ll \tau_r$, i.e., in the strongly active limit $\beta \to 0$ the motion is ``slaved'' to the noise $\tilde y_r(t)$ and the distribution is thus a double-exponential [see the first line of Eq. (\ref{scalingformlimit})]. Finally, for intermediate values of $\beta = \tau_\mu/\tau_r = r/\mu$ of order $O(1)$, the tails of distribution are exponential (\ref{largez_F}), while it has a Gaussian shape around the center. For a more comprehensive analysis of these regimes, please refer to Figs. \ref{fig.stationarydistrib}, and \ref{fig.largex}.}
\label{fig.modelregime}
\end{figure}

%After this qualitative description of the model, we will first look at the stationary behaviour of the system. In particular, we will characterise the stationary distribution of the position in the passive and active limit. In section \ref{sectionTime}, we will study the distribution of the model at a finite time $t$, using a renewal argument.

\subsection{Integral equation of the stationary distribution via Kesten variables}

Here, we apply the method of section \ref{kesten} to derive an integral equation for the stationary distribution of the position. We decompose the motion as a sum of sub motions between resetting events that occurred at times $\{t_1,t_2,\ldots\}$. We define $\tau_n = t_n -t_{n-1}$, and integrate Eq. (\ref{lange_rbm}) from $t_{n-1}$ to $t_n$ such that we get a random recursion relation
\begin{equation}
x_n= x_{n-1}\, e^{-\mu \tau_n} + r\, e^{-\mu\, \tau_n}\, \int_0^{\tau_n}d\tau\,  B(\tau)\, e^{\mu\, \tau}\, ,
\label{recur.1bis}
\end{equation}where $B(\tau)$ is a Brownian motion starting from 0, since the noise is reset to zero at each resetting. As noticed in section \ref{kesten}, Eq. (\ref{recur.1bis}) is of the generalised Kesten form \begin{equation}
x_n = U_n\,  x_{n-1} + V_n\, ,
\end{equation} with $U_n= e^{-\mu\, \tau_n}$ and $V_n= r\, e^{-\mu\, \tau_n}\, \int_0^{\tau_n} B(\tau)\, e^{\mu\, \tau}\, d\tau\,$. The stationary state of the position of the particle is given by the following integral equation as in Eq. (\ref{kesten.2})
\begin{equation}
p(x)= \int_{0}^{1} dU \int_{-\infty}^{\infty} dV\,  \int_{-\infty}^{\infty} dx'\, P(U,V) \, 
p(x') \delta(x-U\, x'-V)\, ,
\label{kesten.2rbm}
\end{equation}
and $P(U,V)$ is the joint distribution of $U_n$ and $V_n$. It can be computed using the Bayes's rule
\begin{equation}
    P(U,V) = P(U)P(V|U)\,,
\label{conditionalproba}
\end{equation}
where $P(V|U)$ is the conditional PDF of $V$ given $U$. Using the fact that $p_{\rm int}(\tau)= r\, e^{-r\, \tau}$, we deduce that $U_n = e^{-\mu \tau_n}$ is distributed over the interval $[0,1]$ \blue{according to the PDF}
\begin{equation}
P(U)= \beta\, U^{\beta-1} \quad, \quad 0 \leq U \leq 1 \quad {\rm with}\quad \beta= \frac{r}{\mu}\, .
\label{pdf.urbm}
\end{equation}Because $V_n$ at fixed $\tau_n$ (equivalently at fixed $U_n$) is a linear functional of Brownian motion, it is clear that the distribution of $V_n$, given $U_n$, is a Gaussian random variable. Its mean is clearly zero, while its variance is given by [see Eq. (\ref{per_Br.3})]
\begin{equation} \label{variance_Un}
    \left<V_n^2\right>_{U_n} = e^{-2\mu \tau_n}\frac{D}{r} \beta^3 \left[e^{2\mu \tau_n}(2\mu\tau_n - 3) + 4e^{\mu \tau_n} - 1\right]\, ,
\end{equation}
where $\langle \cdots \rangle_{U_n}$ denotes an average at fixed $U_n$ (equivalently at fixed $\tau_n$). We can then rewrite the variance $ \left<V_n^2\right>_{U_n}$ in (\ref{variance_Un}) as a function of $U_n = e^{-\mu \tau_n}$, leading to
\begin{equation} \label{alpha_U}
\left<V_n^2\right>_{U_n} = \alpha(U_n) = \frac{D}{r} \beta^3 \left[4U_n - U_n^2 -2\,\text{ln}(U_n) - 3\right]\,.
\end{equation}
This shows that
\be \label{Gaussian}
 P(V|U) = \frac{1}{\sqrt{2\pi \alpha(U)}} e^{-\frac{V^2}{2\alpha(U)}} \;.
\ee
Finally, Eq.~(\ref{conditionalproba}) becomes 
\begin{equation}
    P(U,V) = P(U)P(V|U) = \beta\, U^{\beta - 1}\frac{1}{\sqrt{2\pi \alpha(U)}} e^{-\frac{V^2}{2\alpha(U)}}\, ,
\end{equation}
and the integral equation (\ref{kesten.2rbm}) reads explicitly
\begin{equation}
p(x)= \int_{0}^{1} dU \int_{-\infty}^{\infty} dV \int_{-\infty}^{\infty} dx' \, \beta\, U^{\beta - 1}\, \frac{1}{\sqrt{2\pi \alpha(U)}} \, e^{-\frac{V^2}{2\alpha(U)}}\, 
p(x')\,  \delta(x-Ux'-V)\,,
\label{kestenrbm}
\end{equation}
where $\alpha(U)$ is given in (\ref{alpha_U}). Note that we can check that it is normalized by integrating (\ref{kestenrbm}) over $x$. To proceed, 
it is useful to go to Fourier space and introduce the Fourier transform of~$p(x)$ 
\begin{equation}
\hat{p}(k) = \int_{-\infty}^{+\infty}dx\, p(x)\, e^{ikx} \;.
\label{fouriertransform}
\end{equation}
By taking the Fourier transform of Eq. (\ref{kestenrbm}) with respect to $x$, one gets
\begin{equation}
    \hat{p}(k) = \int_{0}^{1} dU \int_{-\infty}^{\infty} dV \int_{-\infty}^{\infty} dx' \,\beta\, U^{\beta - 1}\frac{1}{\sqrt{2\pi \alpha(U)}} e^{-\frac{V^2}{2\alpha(U)}}\, 
p(x') e^{ik(Ux'+V)}\, .
\end{equation}We can perform the integration over $x'$ to simplify further the expression [using Eq. (\ref{fouriertransform})] 
\begin{equation}
    \hat{p}(k) = \int_{0}^{1} dU \int_{-\infty}^{\infty} dV  \,\beta\, U^{\beta - 1}\frac{1}{\sqrt{2\pi \alpha(U)}} e^{-\frac{V^2}{2\alpha(U)}}\, 
\hat{p}(kU) e^{ikV}\, .
\end{equation}
Finally, performing the integral over $V$, one obtains
\begin{equation}\label{fourierstationnaire}
   \hat{p}(k) = \int_{0}^{1} dU \,\beta\, U^{\beta - 1} e^{-\frac{k^2 \alpha(U)}{2}}\, \hat{p}(kU)\, ,
\end{equation}with \begin{equation} \label{def_alpha}
 \alpha(U) = \frac{D}{r} \beta^3 \left[4U - U^2 -2\text{ln}(U) - 3\right]\, .
\end{equation}
Its asymptotic behaviors are given by
\be \label{alpha_asympt}
\alpha(U) =
\begin{cases}
&- \frac{D}{r} \beta^3 \left( 2{\rm ln}(U) + 3  + {\cal O}(U) \right) \quad, \quad \hspace*{0.4cm}U \to 0 \;, \\
& \frac{2D}{3r} \beta^3 (1-U)^3 + {\cal O}((1-U)^4) \quad, \quad U \to 1 \;.
\end{cases}
\ee
Although it seems very hard to solve this integral equation (\ref{fourierstationnaire}), we will see below that many useful information about the stationary distribution $p(x)$ can however be extracted from it. 

We end this section by noting that a similar integral equation (\ref{fourierstationnaire}) can be derived in the case where the noise $\tilde y_r(t)$ in Eq. (\ref{lange_rbm}) is a Gaussian stochastic process (not necessarily Brownian motion) subjected to Poissonian resetting (see e.g. \cite{smith_maj}). In this case the conditional PDF $P(V|U)$ will still be a Gaussian, as in Eq. (\ref{Gaussian}), but with a different function $\alpha(U)$. In Appendix \ref{ArOUPs}, we use this property to treat the case where the noise is a resetting Ornstein-Uhlenbeck process.

\subsection{Moments of the stationary distribution }\label{statiomoments}

We start by deriving a recursion relation that allows to compute the moment of the stationary distribution. This can be done by expanding the left and right hand sides of Eq. (\ref{fourierstationnaire}) in powers of $k$ and identify the corresponding coefficients of $k^m$ on both sides. At this stage, it is useful to recall that both $x(t)$ and $\tilde y_r(t)$ in Eq. (\ref{lange_rbm}) start from the origin, i.e., $x(0)=0$ and $\tilde y_r(0)=0$. Therefore, since the time evolution of both processes are symmetric under $x \to -x$ and $\tilde y_r \to -\tilde y_r$, one expects that the PDF $p(x,t)$ is symmetric, i.e., $p(x,t) = p(-x,t)$ at all times. Therefore, in particular, the stationary PDF is also symmetric, i.e., $p(x) = p(-x)$ which implies that only the even moments $\langle x^{2n}\rangle$ are nonzero. The power expansion in $k$ of the left hand side (LHS) of Eq. (\ref{fourierstationnaire}) thus reads 
\begin{equation}
\hat{p}(k) = \sum_{n=0}^{+\infty} \frac{(ik)^{2n}}{(2n)!}\left<x^{2n}\right> \;.
\end{equation} 
Similarly, by expanding the right hand side of Eq. (\ref{fourierstationnaire}) one gets
%
%Furthermore, 
%\begin{equation}
%e^{\frac{-k^2 \alpha(U)}{2}} = \sum_{m=0}^{+\infty} \frac{(ik)^{2m} \left[\alpha(U)\right]^m}{2^m m!}\, .
%\end{equation}If we inject these expressions in Eq. \ref{fourierstationnaire}, we obtain
\begin{equation}
\sum_{n=0}^{+\infty} \frac{(ik)^{2n}}{(2n)!}\left<x^{2n}\right> = \int_{0}^{1} dU \,\beta U^{\beta - 1} \sum_{m=0}^{+\infty} \frac{(ik)^{2m}\left[\alpha(U)\right]^m}{2^m m!}\, \sum_{p=0}^{+\infty} \frac{(ik)^{2p}U^{2p}}{(2p)!}\left<x^{2p}\right>\, .
\end{equation}One can group the two sums on the r.h.s and lighten the expression,
\begin{equation}
\sum_{n=0}^{+\infty} \frac{(ik)^{2n}}{(2n)!}\left<x^{2n}\right> = \sum_{m=0}^{+\infty}  \sum_{p=0}^{+\infty} \int_{0}^{1} dU \,\beta  U^{\beta - 1 + 2p} \frac{ \left[\alpha(U)\right]^m}{2^m m!(2p)!}\, (ik)^{2(m+p)}\left<x^{2p}\right>\, .
\label{intermediaire-moments}
\end{equation}
%Now we would like to isolate terms of the same power of $k$ on the right and on the left. It is possible by playing with the indices of the sums on the r.h.s. First we make the change of variable $n'=m+p$ such that $\sum_{m,p = 0}^{+\infty} \rightarrow \sum_{p = 0}^{+\infty}  \sum_{n' = p}^{+\infty}\,  $. Then we notice that $ 0 \leq p \leq n' <+\infty$ and re-write the sum as $\sum_{p = 0}^{+\infty}  \sum_{n' = p}^{+\infty} \rightarrow  \sum_{n' = 0}^{+\infty}  \sum_{p = 0}^{n'}\, $. The equation (\ref{intermediaire-moments}) becomes
%\begin{equation}
%\sum_{n=0}^{+\infty} \frac{(ik)^{2n}}{(2n)!}\left<x^{2n}\right> = \sum_{n = 0}^{+\infty}  \sum_{p = 0}^{n} \int_{0}^{1} dU \,\beta  U^{\beta - 1 + 2p} \frac{ \left[\alpha(U)\right]^{n-p}}{2^{n-p} (n-p)!(2p)!}\, (ik)^{2n}\left<x^{2p}\right>\, ,
%\end{equation} where $n' \rightarrow n$, $n'$ being a dummy variable. We select the term of order $k^{2n}$ such that,
By identifying the coefficient of the term $k^{2n}$ on both sides of (\ref{intermediaire-moments}) one obtains, after straightforward manipulations
\begin{equation}
\left<x^{2n}\right> = (2n)! \sum_{p = 0}^{n} \int_{0}^{1} dU \,\beta  U^{\beta - 1 + 2p} \frac{ \left[\alpha(U)\right]^{n-p}}{2^{n-p} (n-p)!(2p)!}\, \left<x^{2p}\right>\, .
\end{equation}
By isolating the terms $\propto \langle x^{2n} \rangle$ to the LHS, we finally obtain the recursion relation
\begin{equation}
\left<x^{2n}\right> = \beta\, (\beta + 2n)\, (2n-1)! \sum_{p = 0}^{n-1}  \frac{\left<x^{2p}\right>}{2^{n-p}\, (n-p)!\, (2p)!}   \int_{0}^{1} dU   \, U^{\beta - 1 + 2p}\, \left[\alpha(U)\right]^{n-p}\, ,
\label{stationary-moments} 
\end{equation}where we recall $\alpha(U) = \frac{D}{r} \beta^3 \left[4U - U^2 -2\text{ln}(U) - 3\right]$. It seems difficult to obtain an explicit expression of $\langle x^{2n} \rangle$ for any arbitrary value of $n$ but the recursion relation in (\ref{stationary-moments}) allows to compute the first few moments -- see Appendix \ref{AppendixMoments}. In fact, we see that the moments take the form 
\be \label{rational}
\left<x^{2n}\right> = L^{2n} R_{n}(\beta) \quad {\rm with} \quad  L = \sqrt{\frac{D}{r}} \frac{\beta}{\sqrt{1+\beta}} \;,
\ee
and $R_{n}$ being a rational function of $\beta$. This form (\ref{rational}) thus suggests to interpret $L$ as the characteristic length scale of the stationary PDF $p(x)$.

\subsection{Stationary distribution of the position}\label{statiowithlength}

%All moments of the stationary distribution can be calculated recursively thanks to Eq. (\ref{stationary-moments}). From the first moments of the stationary distribution (see Appendix \ref{AppendixMoments}), one can see a pattern appearing, and guess that $\left<x^{2n}\right> \propto L^{2n}$, with \begin{equation}
%L = \sqrt{\frac{D}{r}} \frac{\beta}{\sqrt{1+\beta}}\, .
%\end{equation}The length $L$ is the product of two terms. The first one is the typical length a particle travels between two resets under a diffusivity $D$: $\sqrt{D/r}$. The second term is a function of the dimensionless coefficient $\beta = r/\mu$ that depends on the interplay between the two time scales of the process: $\tau_\mu = 1/\mu$  being the mean time needed for the particle to relax in the harmonic potential $V(x) = \mu\, x^2/2$, and $\tau_r = 1/r$ which is the mean time between two resets. To be more specific, $\left<x^{2n}\right> = L^{2n} \, f_n(\beta)$, where $f_n(\beta)$ is a non trivial ratio of polynomials in powers of $\beta$. The exact form of $f_n(\beta)$ is not fully explicit. 

%\blue{GS: use the notation ${\cal F}_s(z;\beta)$, with a subscript $s$ and a ';' between $z$ and $\beta$. Change it consistently everywhere.}

This result (\ref{rational}) suggests that the stationary distribution takes the following scaling form
\begin{equation} \label{scaling_F}
p(x) = \frac{1}{L}\mathcal{F}_s\left(z = \frac{x}{L}; \beta = \frac{r}{\mu}\right)\, ,
\end{equation}
where the subscript `$s$' refers to `stationary'. We thus consider ${\cal F}_s(z;\beta)$ as a function of the variable $z$, depending on the parameter $\beta$. This scaling form can be easily shown directly from Eq. (\ref{fourierstationnaire}). Indeed this scaling form implies that $\hat p(k)$ reads
\be \label{scaling_fourier}
\hat p(k) = \hat{\mathcal{F}}_s\left(q = k\;L, \beta = \frac{r}{\mu}\right) \quad, \quad {\rm where} \quad  \hat{\mathcal{F}}_s\left(q, \beta \right) = \int_{-\infty}^\infty dz\, e^{iqz} {\cal F}_s(z;\beta) \;.
\ee
Inserting this form in (\ref{fourierstationnaire}) one finds that $\hat{\mathcal{F}}_s\left(q; \beta \right)$ satisfies 
\be \label{EqFhat}
\hat{\mathcal{F}}_s\left(q; \beta \right) = \beta \int_0^1 dU \, U^{\beta-1} \, e^{-\frac{1}{2}\beta(1+\beta)q^2\tilde \alpha(U)}\,  \hat{\mathcal{F}}_s\left(q\,U; \beta \right) \quad, \quad \tilde \alpha(U) =  4U - U^2 -2\text{ln}(U) - 3 \;.
\ee
Note that the asymptotic behaviours of $\tilde \alpha(U)$ are simply given by
\be \label{talpha_asympt}
\tilde \alpha(U) =
\begin{cases}
&-  2{\rm ln}(U) - 3  + {\cal O}(U) \quad, \quad \hspace*{0.8cm}U \to 0 \;, \\
& \frac{2}{3} (1-U)^3 + {\cal O}((1-U)^4) \quad, \quad U \to 1 \;.
\end{cases}
\ee
This form (\ref{EqFhat}) will be useful below to discuss the asymptotic behaviors as $\beta \to \infty$ or $\beta \to 0$.

\begin{figure}[t]
\centering
\includegraphics[width=1\textwidth]{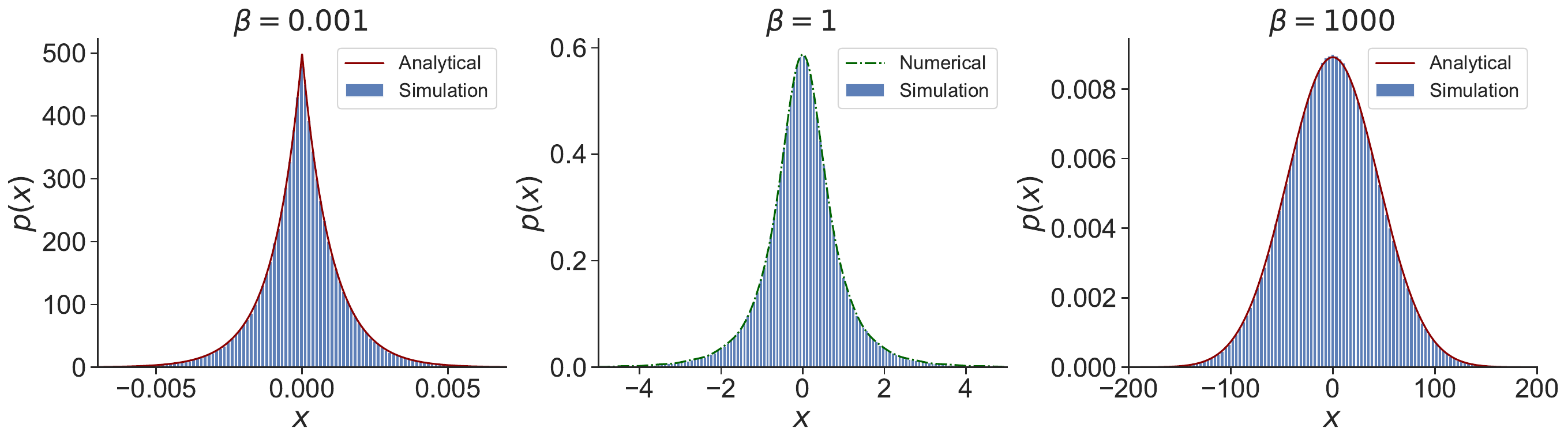}
\caption{These plots show the rich behaviour of the stationary distribution of the position for a particle trapped in a harmonic potential subjected to a colored noise, namely a rBM as in Eq.~(\ref{lange_rbm}). The plot on the left panel corresponds to the strongly active limit $\tau_r \gg \tau_{\mu}$ ($\beta \to 0$) studied in section \ref{expodistribsection}. The distribution is a double-exponential - see Eq. (\ref{expodistri}). It is highly out of equilibrium as in that case, we showed that the effective temperature of the system diverges - see Eq. (\ref{effectivtemp}). The strongly passive limit $\tau_r \ll \tau_{\mu}$ ($\beta \to \infty$) is shown on the right panel. In this limit, the rBM behaves as a white noise, and the underlying distribution is therefore a Gaussian - see Eq. (\ref{gaussiandistrib}). The two red lines on the left and right panels correspond to the theoretical predictions, respectively given in Eq. (\ref{expodistri}) and (\ref{gaussiandistrib}). In the middle panel, for a particular value of $\beta$, we compare our simulations to a numerical solution of the integral equation (\ref{kestenrbm}).  
The agreement is perfect. For details on the numerical procedure, see Appendix \ref{numerics}.}
\label{fig.stationarydistrib}
\end{figure}

Of course, the choice of the couple $L$ and ${\cal F}_s(z;\beta)$ such that $p(x)$ can be written as in Eq.~(\ref{scaling_F}) is not unique but, as we will show below, it ensures that the scaling function ${\cal F}_s(z;\beta)$ has a well defined limit both when $\beta \to 0$ and $\beta \to \infty$. Namely we show below that in these two limit ${\cal F}_s(z;\beta)$ behaves as 
\begin{equation}
\mathcal{F}_s\left(z = \frac{x}{L};\beta\right) \approx \left\{
    \begin{array}{ll}
        \frac{1}{2}\,e^{-|z|} \;,& \quad \mbox{as}  \quad \beta \to 0 \\
        & \\
        \frac{1}{2 \sqrt{\pi}}e^{-\frac{z^2}{4}}\;, & \quad \mbox{as} \quad \beta \to + \infty
    \end{array}
\right. \, .
\label{scalingformlimit}
\end{equation}
For intermediate values of $\beta$, this suggests that the distribution crosses over continuously from an exponential to a Gaussian distribution -- see Figs. \ref{fig.modelregime} and \ref{fig.stationarydistrib}.

\subsubsection{The passive limit  $\beta \to \infty$.}\label{betainf}

In this limit, we anticipate, and check it below, that ${\cal F}_s(z;\beta)$ admits the following expansion
\be  \label{exp_F}
{\cal F}_s(z;\beta) = {\cal F}_{s,0}(z) + \frac{1}{\beta} {\cal F}_{s,1}(z) + \frac{1}{\beta^2} {\cal F}_{s,2}(z) + O(1/\beta^3) \;,
\ee
and similarly for $\hat{\cal F}_s(z;\beta)$, i.e.
\be \label{exp_hatF}
\hat{\cal F}_s(q;\beta) = \hat{\cal F}_{s,0}(q) + \frac{1}{\beta} \hat{\cal F}_{s,1}(q) + \frac{1}{\beta^2} \hat{\cal F}_{s,2}(q) + O(1/\beta^3) \;.
\ee
We next insert this expansion (\ref{exp_hatF}) in Eq. (\ref{EqFhat}). It turns out that for $\beta \to \infty$, the integral over $U$ in (\ref{EqFhat}) is dominated by $U \approx 1$. By performing the change of variable $U = 1 - u/\beta$, and injecting the expansion in Eq. (\ref{EqFhat}), 
it is then rather straightforward to obtain a set of differential equations that can be solved recursively. 
The first two of this hierarchy of equations read
\bea
&&\hat{\cal F}_{s,0}'(q) + 2 q \hat{\cal F}_{s,0}(q) =0 \label{eqF0}\\
&& \hat{\cal F}_{s,1}'(q) + 2 q \hat{\cal F}_{s,1}(q) = 2 q \left(20 q^2+2\right) \hat{\cal F}_{s,0}(q)+\left(8 q^2+1\right) \hat{\cal F}_{s,0}'(q)+q \hat{\cal F}_{s,0}''(q) \;, \label{eqF1}
\eea
supplemented by the boundary conditions, obtained from $\hat{\cal F}_s(q=0;\beta) =1$ (since the PDF ${\cal F}_s(z;\beta)$ is normalized to unity)
\be \label{CI}
\hat{\cal F}_{s,0}(0) = 1 \quad, \quad \hat{\cal F}_{s,1}(0) = 0 \;.
\ee 
The solution to these equations (\ref{eqF0})-(\ref{CI}) reads
\be
\hat{\cal F}_{s,0}(q) = e^{-q^2} \quad, \quad \hat{\cal F}_{s,1}(q) = 7q^4\,e^{-q^2} \;. 
\ee
Taking the inverse Fourier transform, one finds
\be \label{Frealspace}
{\cal F}_{s,0}(z)  = \frac{1}{2 \sqrt{\pi}} e^{-\frac{z^2}{4}} \quad, \quad {\cal F}_{s,1}(z) = \frac{7}{32 \sqrt{\pi }} \left(z^4-12 z^2+12\right)\,e^{-\frac{z^2}{4}} \;.
\ee
The leading term ${\cal F}_{s,0}(z)$ thus gives the result announced in the second line of Eq. (\ref{scalingformlimit}), while ${\cal F}_{s,1}(z)$ provides the first correction to this limiting behavior. 

This Gaussian limiting behavior of ${\cal F}_s(z;\beta)$ in the limit $\beta = r/\mu \to \infty$ can be easily understood by considering the limit $r \to \infty$ at fixed $\mu$. Indeed, in this limit, the noise $\tilde y_r(t)$ in Eq. (\ref{lange_rbm}) converges to a white noise in the sense that
\begin{equation}
\left<\tilde{y}_r(t)\tilde{y}_r(t + |t_2-t_1|)\right>  \underset{t \to \infty}{\sim}  4D\, \frac{r}{2}\, e^{-r|t_2-t_1|}\xrightarrow[r \to \infty]{}4D\, \delta(t_2-t_1)\, .
\label{whitnoiserBM}
\end{equation} 
In this limit, one thus expects that, at large time, Eq. (\ref{lange_rbm}) behaves similarly to a (passive) Ornstein-Uhlenbeck process with a diffusion constant $D_{\rm OU} = 2D$. It is then natural to expect that 
the system will eventually converge to a Boltzmann equilibrium described by the stationary PDF 
\begin{equation}\label{gaussiandistrib}
p(x) = \sqrt{\frac{\mu}{2\pi D_{\rm OU}}}e^{-\frac{\mu}{2 D_{\rm OU}}x^2} = \sqrt{\frac{\mu}{4\pi D_{}}}e^{-\frac{\mu}{4 D_{}}x^2}\, .
\end{equation}
In the large $\beta$ limit, one has $L \sim \sqrt{D/\mu}$ (see Eq. (\ref{rational})), hence $p(x)$ in Eq. (\ref{gaussiandistrib}) can be re-written as
$p(x) \sim e^{-z^2}/(L \sqrt{4 \pi})$ with $z = x/L = \sqrt{\mu/D}\; x$, which yields the second line of Eq. (\ref{scalingformlimit}). Note however that even in this limit $r \to \infty$, the noise term $\tilde y_r(t)$ remains non-Gaussian (see Eq. (\ref{statiorBM})), but, as shown by our explicit computation [see Eq. (\ref{Frealspace})], this does not modify the nature of the stationary state.

\subsubsection{The active limit $\beta \to 0$.}\label{expodistribsection}

To analyse this limit, it is convenient to perform the change of variable $z = U^\beta$ in Eq. (\ref{EqFhat}), which yields
\be \label{F_beta_0}
\hat{\cal F}_s(q;\beta) = \int_0^1 dz\, z^{q^2 + \beta q^2} e^{-\frac{1}{2} \beta(1+\beta) q^2 (4 z^{1/\beta} - z^{2/\beta} - 3)} \hat{\cal F}_s(q\, z^{1/\beta}; \beta) \;.
\ee
Since $z \in [0,1]$, $z^{1/\beta} \ll 1$ in the limit $\beta \to 0$ and one thus expects from this equation (\ref{F_beta_0}) 
that $\hat{\cal F}_s(q,\beta)$ admits an expansion in powers of $\beta$ as $\beta \to 0$. This expansion is however a bit cumbersome and we restrict our analysis here to the leading term $\hat{\cal F}_s(q;\beta=0)$. Indeed using $\hat{\cal F}_s(q\, z^{1/\beta}; \beta) \to  \hat{\cal F}(0, 0) = 1$ as $\beta \to 0$, one easily obtains from Eq. (\ref{F_beta_0}) that $\hat{\cal F}(q,0)$ is simply given by
\be \label{expr_F_beta_0}
\hat{\cal F}_s(q;0) = \int_0^1 dz\, z^{q^2} = \frac{1}{1+q^2} \;.
\ee
By inverting this Fourier transform, one immediately obtains the result announced in the first line of Eq. (\ref{scalingformlimit}). 

This limiting behavior in the limit $\beta = r/\mu \to 0$ can be rationalised by considering the limit $r \to 0$ for fixed $\mu$. In this limit, the typical time between two successive resettings of the noise $\tilde y_r(t)$ is $\tau_r \approx 1/r \gg 1$. Therefore, after such a long time, the amplitude of the resetting Brownian noise is $\tilde y_r(\tau_r) \sim r \, \sqrt{\tau_r} =   \sqrt{r}$. The equation of motion (\ref{lange_rbm}) thus suggests that the process $x(t)$ is ``slaved'' to the noise, i.e., $x(\tau_r) \sim \tilde y_r(\tau_r) /\mu \sim \sqrt{r}$, such that the term in the left hand side of Eq. (\ref{lange_rbm}), i.e., $\dot {x}(\tau_r) \sim r/\sqrt{\tau_r} \sim r^{3/2}$, is indeed subleading in the limit $r \to 0$. Using the fact that the stationary PDF is a symmetric exponential given in Eq. (\ref{statiorBM}) it follows that $x(t) \sim \tilde y_r(t)/\mu$ is given by
\begin{equation}\label{expodistri}
p(x) \sim \frac{1}{2}{\frac{\mu}{\sqrt{D\, r}}}e^{-{\frac{\mu}{\sqrt{D\,r}}}|x|}\, .
\end{equation}
In the limit $\beta \to 0$, the length is $L \sim \sqrt{D\, r} / \mu$ (see Eq. (\ref{rational})) and therefore the PDF in (\ref{expodistri}) can be re-written as $p(x) = 1/(2 L)\, e^{-|z|}$, as given in first line of Eq. (\ref{scalingformlimit}).

\subsection{\blue{Asymptotic behaviours} of the stationary distribution $p(x)$} \label{largexsection}

\begin{figure}[t]
\centering
\includegraphics[width=0.7\textwidth]{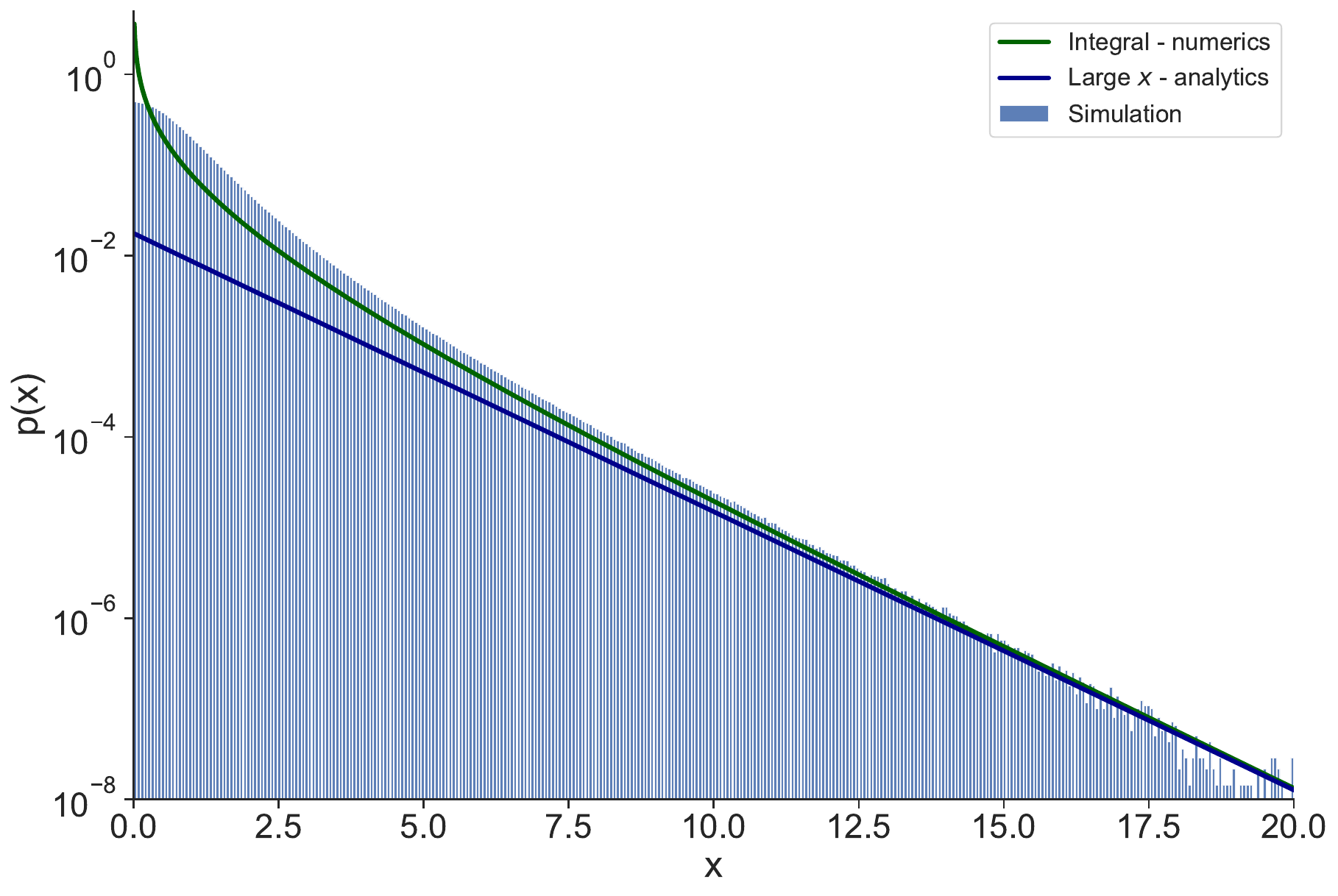}
\caption{Plot of the stationary distribution $p(x)$ vs $x$ for $\beta = 2$. The histogram shows the result of numerical simulations of Eq. (\ref{lange_rbm}). The green solid line corresponds to the function $h(x)$ in Eq.~(\ref{largex_2}), which is supposed to be a reasonably good approximation of $p(x)$ for $x$ large enough, while the blue straight line corresponds to our analytical prediction of the exact large $x$ behavior (\ref{largex_3}). There are no fitting parameters. As expected, both solid lines agree very well with the numerical results for large $x$ (here for $x \geq 12.5$), while the green one gives a very good estimate of $p(x)$ already for $x \simeq 7.5$. The parameter used for this plot are $\mu =1$, $r=2$, $D=1$, and $x_0=0$.} \label{Fig_comp_largex}
\end{figure}

To extract the large $x$ behavior of $p(x)$, it turns out to be more convenient to start from the integral equation in real space, i.e., Eq. (\ref{kestenrbm}). Performing the integral over $V'$ in (\ref{kestenrbm}), one obtains
\be \label{largex_1}
p(x) = \beta \int_0^1 \frac{dU}{\sqrt{2 \pi \alpha(U)}} U^{\beta-1} \int_{-\infty}^{\infty} dx'\, e^{-\frac{(x-Ux')^2}{2 \alpha(U)}} p(x') \;.
\ee
\blue{The function $p(x)$ is symmetric in $x$, which can be checked self-consistently from Eq. \eqref{largex_1}. Hence below we will focus on $x \geq 0$ only.} In the limit of large $x$, one can simply replace $e^{-\frac{(x-Ux')^2}{2 \alpha(U)}}$ by $ e^{-\frac{x^2}{2 \alpha(U)}}$, to leading order for large~$x$. This is because, in that limit, the integral in Eq. (\ref{largex_1}) is dominated by $U = O(e^{-x}) \ll 1$. Therefore, one can thus expand, for small $U$,
\bea \label{expansion}
\int_{-\infty}^{\infty} dx'\, e^{-\frac{(x-Ux')^2}{2 \alpha(U)}} p(x') = \int_{-\infty}^{\infty} dx'\ p(x') e^{-\frac{x^2}{2 \alpha(U)}} \left( 1 + \frac{U}{\alpha(U)}xx' + \cdots\right) \;.
\eea
Keeping only the first term in this expansion and performing then the integral over $x'$, using $\int_{-\infty}^{\infty} dx' p(x') = 1$, one obtains
\be \label{largex_2}
p(x) \approx h(x) \quad {\rm with} \quad h(x) = \beta \int_0^1 \frac{dU}{\sqrt{2 \pi \alpha(U)}} U^{\beta-1} e^{-\frac{x^2}{2 \alpha(U)}} \;.
\ee
The large $x$ behavior can finally be obtained via a saddle-point analysis [see Appendix \ref{saddleana} for details], leading to
\be\label{largex_3}
p(x) \approx \frac{e^{-3\beta/2}}{2 \beta} \sqrt{r/D} \, e^{- \frac{x}{\beta}\sqrt{r/D}} \quad, \quad x \to \infty \;.
\ee 
From this, we can read off the large $z$ behavior of the scaling function ${\cal F}_s(z;\beta)$ defined in Eq. \eqref{scaling_F}, namely
\be \label{largez_F}
{\cal F}_s(z;\beta) \approx \frac{e^{-3\frac{\beta}{2}}}{2\sqrt{1+\beta}}\, e^{- \frac{z}{\sqrt{1+\beta}}} \quad, \quad z \to \infty \;.
\ee

\begin{figure}[t]
\centering
\includegraphics[width=1\textwidth]{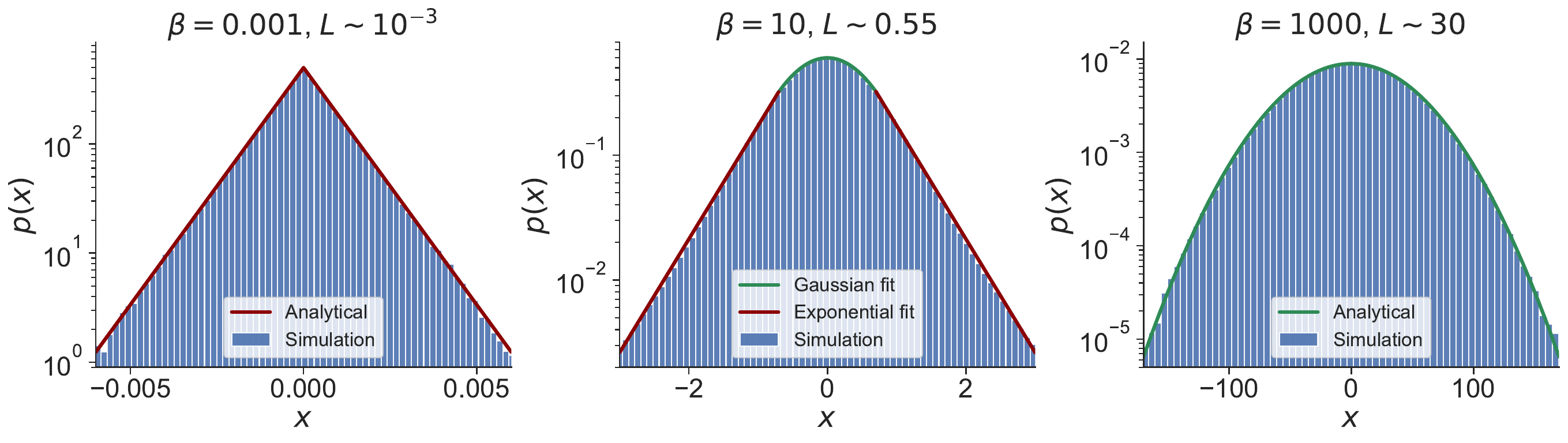}
\caption{When $\beta$ approaches the value $0$, we showed that the distribution becomes a symmetric double exponential, while when $\beta \to + \infty$, $p(x)$ converges to a Gaussian - see the plots on the left and right panel (the plots shown on these two panels are the same as the one shown in Fig. \ref{fig.stationarydistrib} but there are shown here in a log-linear scale). In blue, we show the histograms of simulations of many trajectories -- the solid red and green lines correspond respectively to the double exponential form (\ref{expodistri}) and the Gaussian form (\ref{gaussiandistrib}). The middle panel shows a plot of $p(x)$ for an intermediate value of $\beta = 10$. The solid green line in the center corresponds to a Gaussian fit, while the red solid line corresponds to an exponential fit. It shows a nice crossover from a Gaussian to an exponential behavior.}
\label{fig.largex}
\end{figure}

\blue{One can also ask about the small $x$ behavior of $p(x)$. In fact one can easily see, by expanding the right hand side of Eq. \eqref{kestenrbm} around $x=0$ that $p(x)$ behaves for small $x$ as $p(x) \approx p(0) - a_2\, x^2$. Hence, one can summarize the asymptotic behavior of $p(x)$ in the limits of small and large $x$ as
\bea \label{asympt_summary_intro}
p(x) \approx
\begin{cases}
& p(0) - a_2\, x^2 \quad, \quad \hspace*{1.7cm} x \to 0 \;, \\
& \\
& \frac{e^{-3\beta/2}}{2 \beta} \sqrt{r/D} \, e^{- \frac{x}{\beta}\sqrt{r/D}} \quad, \quad x \to \infty \;.
\end{cases}
\eea
We recall that $p(x)$ is symmetric around $x=0$. 
}
In Fig. \ref{Fig_comp_largex}, we show a comparison between our numerical simulations and our predictions of the large $x$ behaviour of $p(x)$ in Eqs. (\ref{largex_2}) and (\ref{largex_3}), showing a very nice agreement. In the middle panel of Fig. \ref{fig.largex}, we show a plot of $p(x)$ as a function of $x$ for a moderate value of $\beta = 10$ in log linear scale. \blue{These figures, in agreement with Eq. \eqref{asympt_summary_intro} demonstrate that the stationary PDF crosses over from a Gaussian behavior in the center (i.e., for moderate value of $x$) to an exponential tail behavior (\ref{largex_3}) for large $x$}. In the large $\beta$ limit, by matching the Gaussian behavior in (\ref{scalingformlimit}) with the exponential tail in (\ref{largez_F}), one finds that this crossover occurs for $z \sim \sqrt{\beta}$, i.e., for $x \sim \beta$.

%\blue{GS: here there are two aspects that still need to be discussed
%\begin{enumerate}
%\item{The non-commutation of the limits $z \to \infty$ and $\beta \to \infty$ -- compare Eqs. (\ref{scalingformlimit}) and (\ref{largez_F}).}
%\end{enumerate}}

Let us comment on the fact that the two limits $z \to \infty$ and $\beta \to \infty$ do not commute. This can be easily seen by comparing Eqs. (\ref{largez_F}) and the second line of Eq.~(\ref{scalingformlimit}). This is certainly an interesting point that may deserve further investigations and that probably requires a finer investigation of the integral equation (\ref{fourierstationnaire}).  

\blue{Let us conclude this subsection by remarking that the crossover from a Gaussian form at small $x$ to an exponential decay at large $x$ has been seen in a variety of theoretical models, for example in the time-dependent position distribution in models of diffusing diffusivity, \cite{CSMS2017,LG2018,BB2020} and also in certain experimental systems \cite{WABG2009,WGLFGH2019}.}

%In section \ref{statiowithlength}, we have introduced a characteristic length of the system $L$ of system that we identified noticing  $\left<x^{2n}\right> \propto L^{2n}$. Its expression is \begin{equation}
%L(\beta) = \sqrt{\frac{D}{r}} \frac{\beta}{\sqrt{1+\beta}}\, .
%\end{equation}
%In the stationary state, we saw that the system presents two extreme behaviours. First, when $\beta \to +\infty$ the stationary distribution of the position is Gaussian - see Eq. (\ref{gaussiandistrib}), \begin{equation}
%    \text{log}\left[p(x)\right] \propto \left(\frac{x}{L}\right)^2\, ,
%\end{equation}
%and when $\beta \to 0$ the stationary distribution of the position is exponential - see Eq. (\ref{expodistri}), \begin{equation}
%    \text{log}\left[p(x)\right] \propto \left|\frac{x}{L}\right|
%\end{equation}
%There is this transition from an exponential distribution to a Gaussian as $\beta$ grows, and when $\beta$ grows, $L(\beta)$ grows. As numerical simulations suggest, for an arbitrary value of $\beta$, we believe that there is a core region around $x=0$ where the distribution can be approximate by a Gaussian, while the tails are exponential - see Fig \ref{fig.largex}.  
%
%In other words, when $L \ll 1$, the Gaussian region is small compared to the typical fluctuation of the position, and if $L \gg 1$, the Gaussian region is very large, much larger than the typical position fluctuation. For an arbitrary value of $L$, the Gaussian region is of size of order $O(L)$. Outside this Gaussian region, the large $x$ region is exponential.

\subsection{Violation of the fluctuation-dissipation theorem in the steady-state}\label{FDT}

We end our study of this non-equilibrium steady state (NESS) by studying the violation of the fluctuation-dissipation theorem (FDT) -- {note that it has been recently studied for a pure Brownian motion under resetting in \cite{SokolovTemp}}. For that purpose, it is useful to compute the stationary correlation $C_{\rm st}(\tau)$ and response $R_{\rm st}(\tau)$ functions defined respectively as
\be \label{def_CR}
C_{\rm st}(\tau) = \lim_{t \to \infty} \langle x(t) x(t+\tau)\rangle \quad, \quad R_{\rm st}(\tau) = \lim_{t \to \infty} \left.\left<\frac{\delta \tilde{x}(t+\tau)  }{\delta h(t)}\right>\right|_{h=0}\, ,
\ee
where $\tilde{x}(t)$ evolves as in (\ref{lange_rbm}) in the presence of an additional external force field $h(t)$, i.e., 
$\dot{\tilde{x}}(t)=-\mu\, \tilde{x}(t) +r\, y_r(t) + h(t)\,$.  

For the present model (\ref{lange_rbm}), $C_{\rm st}(\tau)$ and $R_{\rm st}(\tau)$ can be computed straightforwardly since the equation of motion is linear, leading to (see Appendix \ref{appendixcorrelations} for details)
\begin{equation} \label{exp_RC}
    C_{\rm st}(\tau)= \frac{2\, D\, r\, (\mu\, e^{-r\tau} - re^{-\mu\tau})}{\mu (\mu^2 - r^2)}\, \quad, \quad R_{\rm st}(\tau) = e^{-\mu \tau} \;.
\end{equation}
If the system were at equilibrium in contact with a bath at temperature $T^*$, $C_{\rm st}(\tau)$ and $R_{\rm st}(\tau)$ would be related via the relation 
$R(\tau) = -\frac{1}{T^*} \frac{dC(\tau)}{d\tau}$, which corresponds to the FDT. To quantify the violation of FDT, it is convenient to introduce an effective temperature via the relation~\cite{cugliandolo_kurchan}
\be \label{Teff}
T_{\rm eff}(\tau) = - \frac{1}{R_{\rm st}(\tau)}  \frac{dC(\tau)}{d\tau} \;.
\ee
Inserting the explicit expressions from (\ref{exp_RC}) in Eq. (\ref{Teff}), one finds
\begin{equation}
T_{\text{eff}}(\tau) = \frac{2D\, r^2}{r^2-\mu^2}\left[1-e^{-(r-\mu)\tau}\right]\, .
\label{effectivtemp}
\end{equation}
When $r < \mu$, the effective temperature grows exponentially with $\tau$, which indicates a highly out-of-equilibrium situation. On the other hand, when $r> \mu$, the effective temperature $T_{\rm eff}(\tau)$ converges to a finite value $T_{\rm eff}(\tau) \to T_{\rm eff} = {2D\, r^2}/{(r^2-\mu^2)}$. In particular, in the passive limit $r \to \infty$, one finds $T_{\rm eff} \to 2D$ which, given the relation in Eq. (\ref{whitnoiserBM}), can be considered as the equilibrium temperature. 
Finally, when $r = \mu$, $T_{\rm eff}(\tau) = D\, r\, \tau$, which also grows unboundedly with $\tau$. 

In conclusion, this computation of $T_{\rm eff}(\tau)$ indicates that this simple active dynamics is always out-of-equilibrium (see Eq. (\ref{effectivtemp})). In addition, this effective temperature is a growing function of the activity in the system (quantified here by $\mu/r = 1/\beta$). In fact, the equilibrium situation is recovered only in the passive limit $r \to \infty$.

\vspace*{0.5cm}
\noindent {\bf Remark:} One can easily show that the relation obtained here for the dynamics described by Eq. (\ref{lange_rbm}) also holds for the RTP dynamics with a telegraphic noise with rate $\gamma$, as studied e.g. in \cite{DKMSS19}. In fact the results for the RTP are given by Eqs. (\ref{exp_RC}) and (\ref{effectivtemp}) with the substitution $\sqrt{2\, D\, r} \to v_0$, and $r \to 2\gamma$.

%The behaviour found in Eq. (\ref{relaxation}) is quite close to the one already observed for run-and-tumble particles in \cite{DKMSS19}. For a RTP, the decay of $d(t)$ is also exponential, but the relaxation $\lambda_0$ is different as the noises of the two models are not exactly the same, but quite similar with the identification $v_0 \to \sqrt{2\, D\, r}$, and $2\gamma \to r$, as mentioned in section \ref{model}. Here we consider that the motion starts at $x_0 = 0$. The variance of a run-and-tumble particle has been computed in \cite{DKMSS19} and it gives \begin{equation}
%\text{Var}_{RTP}\left(x(t)\right) = v_0^2\, \left[\frac{1}{2\gamma \mu + \mu^2} + \frac{2e^{-t(\mu + 2\gamma)}}{4\gamma^2 -\mu^2}+\frac{e^{-2\mu t}}{\mu(\mu - 2\gamma)}   \right]\, .
%\label{RTPvariance}
%\end{equation}If we perform the change of variables $v_0 \to \sqrt{2\, D\, r}$, and $2\gamma \to r$, some terms are similar and the difference of the two variances (\ref{RTPvariance}), and (\ref{variance_time}) yields \begin{equation}
%\text{V}(t)- \text{Var}_{RTP}\left(x(t)\right) \sim e^{-\lambda_0\,  t}\, ,
%\end{equation}with $\lambda_0$ defined in Eq. (\ref{lambdarelax}). Therefore, at large time, the variances of the two models are approximately equal. To probe more precise differences between the two models, it is probably necessary to study the correlation functions at more than two points.

\section{Time-dependent properties for Poissonian resetting Brownian noise}\label{sectionTime}

In the previous section, we focused on the stationary state of a particle evolving via Eq. (\ref{lange.1}) in the presence of a Poissonian resetting Brownian noise. In this section, we focus on the time dependent properties, before the steady state is reached. 
%{\color{blue}At finite time, PDF of processes with active noises have been shown to deviate from the Gaussian distribution \cite{sebastianActiveNoises}}. 
As we will see, our analytical approach in this case is totally different from the one based on Kesten variables studied before but 
relies instead to a large extent on a renewal approach, well known in the context of resetting processes (see e.g.~\cite{Review20}).

%In the last sections we studied the stationary state of a particle in a harmonic trap subjected to a noise that is reset at random times. For this purpose, we divided the motion of the particle in sub-motions between resets that we know how to describe. In particular, in section \ref{sectionstatio} we focused on the case where the reset noise is a resetting Brownian motion. Here we still consider the same process, although the methods can be generalized to any reset processes, and we aim at studying the full time distribution $p(x,t)$ of the process. In order to accomplish this, one can derive a renewal integral equation for $p(x,t)$. We are going to derive an analytical expression of $p(x,t)$ in the small $\mu$ and small $r$ regimes, and even though the method is completely different from the Kesten approach, we are going to obtain results that are consistent with the one derived in the last section.
%

\subsection{Relaxation to the steady state}\label{relaxationtoSS}

\begin{figure}[t]
\centering
\includegraphics[width=0.9\textwidth]{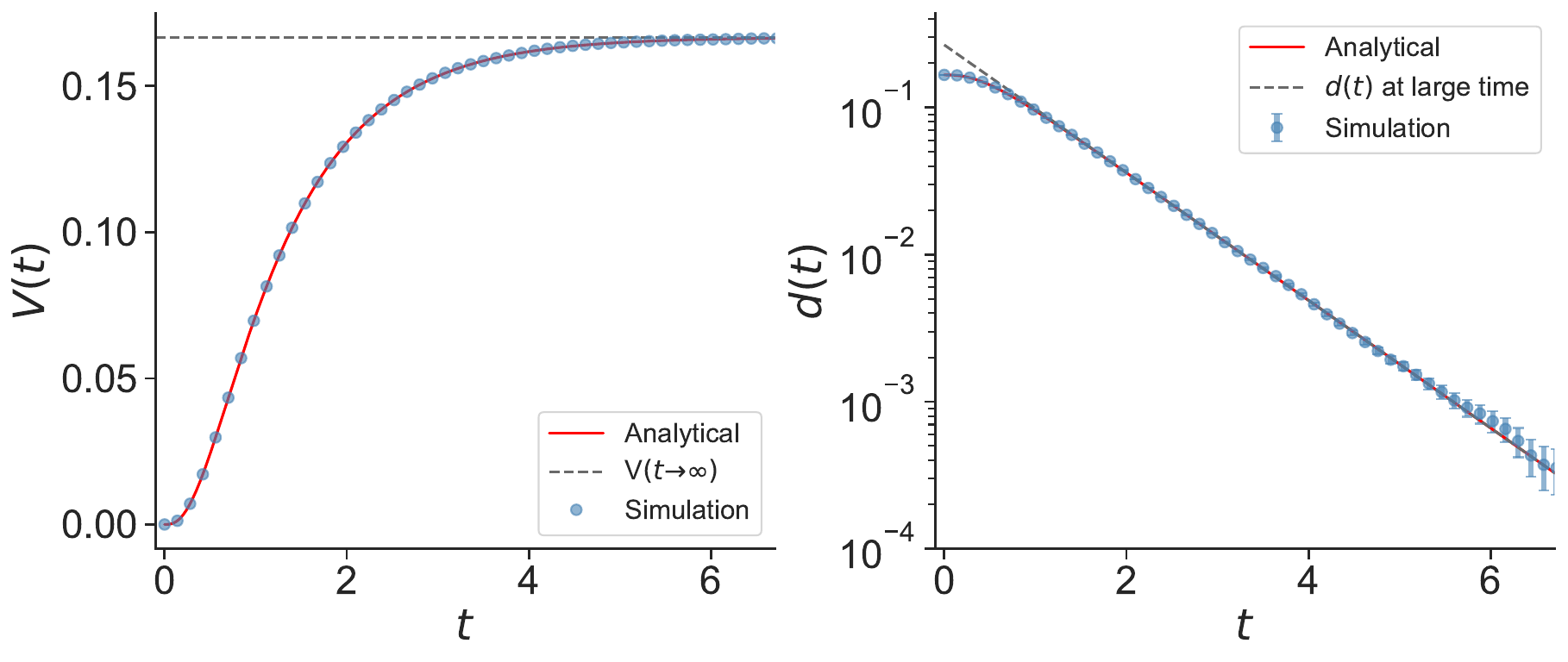}
\caption{On the left panel, we show a plot of ${\rm V}(t)$ vs. $t$ for $x_0 = 0$. The blue dots correspond to numerical simulations while the solid line is our exact analytical result (\ref{variance_time}) with $x_0=0$. The agreement is perfect and in particular, we see that as  
$t\to\infty$, ${\rm V}(t)$ converges to the asymptotic value ${\rm V}(t \to \infty)$ given in Eq. (\ref{Vinf}) and indicated as a dotted line on the figure. On the right panel, we show a plot of $d(t)$ on a linear-log plot, showing clearly the exponential decay predicted in Eq.~(\ref{relaxation}). The parameter used for these plot are $\mu =3$, $r=1$, $D=1$, and $x_0=0$.}
\label{fig.variance}
\end{figure}

Let us first study some simple observables that allow us to characterise the relaxation of the system to the steady state. For this purpose, it is convenient to rewrite the equation of motion (\ref{lange.1}) as
\begin{equation}
\frac{dx(t)}{dt}=-\mu \, x(t) + r\, y_r(t)=-\frac{1}{\tau_\mu} \, x(t) + \frac{1}{\tau_r}\, y_r(t)\, ,
\label{lange.2_bis}
\end{equation}
in terms of the two time scales $\tau_\mu = 1/\mu$ and $\tau_r = 1/r$. 
The explicit solution of the equation of motion (\ref{lange.2_bis}) is
\begin{equation} \label{exact_sol}
x(t) = x_0\, e^{-\mu t} +  r\, e^{- \mu t}\int_0^t dt'\, e^{\mu t'}\, y_r(t')\, .
\end{equation} 
The RBM $y_r(t)$ is reset at position $0$ and is therefore symmetric, i.e., $\langle y_r(t')\rangle = 0$ for all time $t'$. Hence, by taking the average in Eq. (\ref{exact_sol}) one finds simply
\begin{equation} \label{x0_av}
\left<x(t)\right> = x_0 \, e^{-\mu t}\, .
\end{equation}
The computation of the variance of $x(t)$, namely $\text{Var}\left(x(t)\right) = \text{V}(t)= \left<x(t)^2\right> - \left<x(t)\right>^2$,  is a bit more complicated, since it involves the two-time correlation function $\langle y_r(t') y_r(t'') \rangle$ at two different times $t'$ and $t''$. Fortunately, this correlation function can be computed \cite{correlation resetting} and we get eventually an explicit expression for ${\rm V}(t)$ (see Appendix \ref{appendixcorrelations} for details) 
\begin{equation}
\text{V}(t)  = \frac{2\, D\, r}{\mu}\left[\frac{1}{\mu + r} +\frac{2\, r \, e^{-(\mu + r)t}}{\mu^2-r^2} - \frac{r\, e^{-2\mu t}}{(\mu-r)(2\mu-r)}-\frac{2\, e^{-rt}}{2\mu -r}\right] - x_0^2 \, e^{-2\mu t}\, \;,
\label{variance_time}
\end{equation}
which is valid for $r \neq \mu$ and $r \neq 2 \mu$. For these two specific values, one finds instead
\begin{equation} \label{rmu}
\text{V}(t) = \frac{D}{\mu}\left(e^{-2\mu t}(3 + 2\mu t) - 4 e^{-\mu t} +1\right) - x_0^2 \, e^{-2\mu t}\, \quad , \quad {\rm for} \quad r = \mu 
\end{equation}
while  
\begin{equation} \label{r2mu}
\text{V}(t)= \frac{4D}{3\mu}\left(e^{-2\mu t}(3 - 6\mu t) - 4 e^{-3\mu t} +1\right)- x_0^2 \, e^{-2\mu t}\,  \quad , \quad {\rm for} \quad r =2 \mu \;.
\end{equation}
%Note that in the large time limit, the process has no memory of the initial conditions as we have\begin{equation}
%    \lim_{t \to \infty} \left<x(0)x(t)\right> - \left<x(0)\right>\left<x(t)\right> = 0\, .
%\end{equation}
In particular, in the large time limit, one finds from Eqs. (\ref{variance_time})-(\ref{r2mu})
\begin{equation} \label{Vinf}
\text{V}(t \to \infty) = \frac{2\, D\, r}{\mu\, (\mu + r)}\, \quad, \quad {\rm for \; all} \quad (r, \mu) \;.
\end{equation}
One can check that this limiting value coincides with the one obtained in the stationary state and given in Eq. (\ref{x2_app}) with $\beta = r/\mu$, as it should. From Eqs.(\ref{variance_time})-(\ref{r2mu}), one also gets a measure of the approach to this steady state value, namely
\begin{equation}
d(t)= \text{V}(t \to \infty) -\text{V}(t) \sim \exp{[-\min(2\mu,r)\,t]} \quad, \quad t \to \infty \;.
\label{relaxation}
\end{equation} 
In Fig. \ref{fig.variance} we show a comparison of our analytical expressions (\ref{variance_time}) and (\ref{relaxation}), showing a very good agreement.

From Eqs. (\ref{x0_av}) and (\ref{relaxation}), one can get an estimate of the largest relaxation time $1/\lambda_0$. For $x_0 = 0$, we see that $\langle x(t)\rangle = 0$ independently of $t$ and the relaxation is completely controlled by $d(t)$, and hence $\lambda_0 = \text{min}(2\mu,r)$. However, for a generic $x_0 \neq 0$, the decay of $\langle x(t) \rangle \sim e^{- \mu\,t}$ introduces a new time scale, namely $\mu^{-1}$. Hence in this case, $\lambda_0 = \min(\mu,r)$. Therefore to summarise
\begin{equation}
\displaystyle
\lambda_0= \left\{
    \begin{array}{ll}
    	\text{min}(2\mu,r)  \quad \mbox{ if } \; x_0 = 0 \\
        \text{min}(\mu,r)   \quad \;\, \mbox{ if } \; x_0 \neq 0 
    \end{array}
\right.
\;,
\label{lambda-relax}
\end{equation} 
such that $1/\lambda_0$ gives an estimate of the relaxation time to the steady state. Thus, both for $x_0 = 0$ and $x_0 \neq 0$, the slowest relaxation mode $\lambda_0$ exhibits a transition as a function of $r$ respectively at $r =2\mu$ and $r= \mu$. This is in contrast with the standard RTP, where a transition occurs only for $x_0 = 0$ (see Ref. \cite{DKMSS19}).

\subsection{A renewal equation for the Fourier transform of the distribution of the positions}

We consider the dynamics defined by Eqs. (\ref{lange_rbm}) and (\ref{def_rBM}) with $x(0) = 0$ and $y_r(0) = 0$. 
The explicit solution of Eq. \eqref{lange_rbm} thus reads
\begin{equation}
x(t) =\int_{0}^{t} dt'e^{-\mu(t-t')} \tilde{y}_{r}(t') =r\, e^{-\mu t}\int_{0}^{t} dt'e^{\mu t'} y_{r}(t')\, .
\label{explicitsol}
\end{equation}
This section aims at studying the time dependent distribution of the position $p(x,t)$. For this purpose, we are going to use a renewal argument.

\subsubsection{A renewed quantity between resets.}

Given a realization of the resetting epochs $\{t_1,t_2,\ldots\}$, the weighted area\begin{equation}
a_{n-1} = \int_{t_{n-1}}^{t_n} dt'e^{\mu t'} y_{r}(t')\, , \label{a1}
\end{equation} is clearly independent of \begin{equation}
a_n = \int_{t_{n}}^{t_n+1} dt'e^{\mu t'} y_{r}(t')\,, \label{a2}
\end{equation}
since $y_r(t')$ before the resetting at $t_n$ is independent of $y_r(t')$ after this resetting event. 
%This useful property will allow us to derive a renewal equation. 
Let us now define the weighted area $A_r(t)$ over the whole interval $[0,t]$, i.e., 
\begin{equation}
A_r(t) = \int_{0}^{t} dt'e^{\mu t'} y_{r}(t')\, .
\label{areadefintion1}
\end{equation}%Notice that the function $A_r(t)$ is not linear in its argument. We have indeed
For later purpose, it is useful to decompose $A_r(t_1 + t_2)$ as
\begin{equation}
A_r(t_1+t_2) = \int_0^{t_1+t_2}dt'e^{\mu t'}y_r(t') = \int_0^{t_1}dt'e^{\mu t'}y_r(t') + \int_{t_1}^{t_1+t_2}dt'e^{\mu t'}y_r(t')\, .
\end{equation}
Let us choose $t_1$ such that a reset occurs at $t=t_1$. Performing the change of variable $t''=t'-t_1$ one gets
\begin{equation}
\int_{t_1}^{t_1+t_2}dt'e^{\mu t'}y_r(t') = \int_{0}^{t_2}dt''e^{\mu (t''+t_1)}y_r(t''+t_1) = e^{\mu t_1}\int_{0}^{t_2}dt''e^{\mu t''}y_r(t'')\,,
\end{equation}
where the second equality holds in distribution. Indeed, since a resetting happens at $t_1$, one has $y_r(t''+t_1) = y_r(t'')$ in distribution. 
 Therefore, the "addition" law for $A_r(t)$ reads
\begin{equation}\label{area}
A_r(t_1+t_2) =A_r(t_1) +  e^{\mu t_1}A_r(t_2)\, ,
\end{equation} where $A_r(t_1)$ is independent of the second term $e^{\mu t_1}A_r(t_2)$.

We also note the area $A_0(t)$ corresponding to the case where there is no reset between initial time and $t$,
\begin{equation}
A_0(t) = \int_{0}^{t} dt'e^{\mu t'} B(t')\, ,
\end{equation}with $B(t)$ a Brownian motion. 
We define $\mathcal{P}_r(A,t)$ and $\mathcal{P}_0(A,t)$ the probability density functions of respectively the random variables $A_r(t)$, and $A_0(t)$. It is easy to see that the PDF of $x(t)$, denoted as $p(x,t)$, is related to $\mathcal{P}_r(A,t)$. Indeed, from Eqs. \eqref{explicitsol} and (\ref{areadefintion1}), one has simply
\begin{equation}
    x(t) = r\, e^{-\mu t}\, A_r(t) \;.
\end{equation}
Since $p(x,t)\, dx = \mathcal{P}_r(A,t)\, dA$, we obtain
\begin{equation}\label{density rel1}
   p(x,t) = \frac{e^{\mu t}}{r}\mathcal{P}_r\left(\frac{x}{r} e^{\mu t},t\right) \, .
\end{equation}We will now derive a renewal equation for $\mathcal{P}_r(A,t)$, which will give access to $p(x,t)$ via Eq.~(\ref{density rel1}). 
%the random variable $A_r(t)$. Thanks to Eq. (\ref{density rel1}), we will rewrite it as an equation for $\hat{p}(k,t)$, the Fourier transform of $p(x,t)$.

\subsubsection{Renewal approach.}\label{renewal area} 
\begin{figure}[t]
\centering
\includegraphics[width=0.7\textwidth]{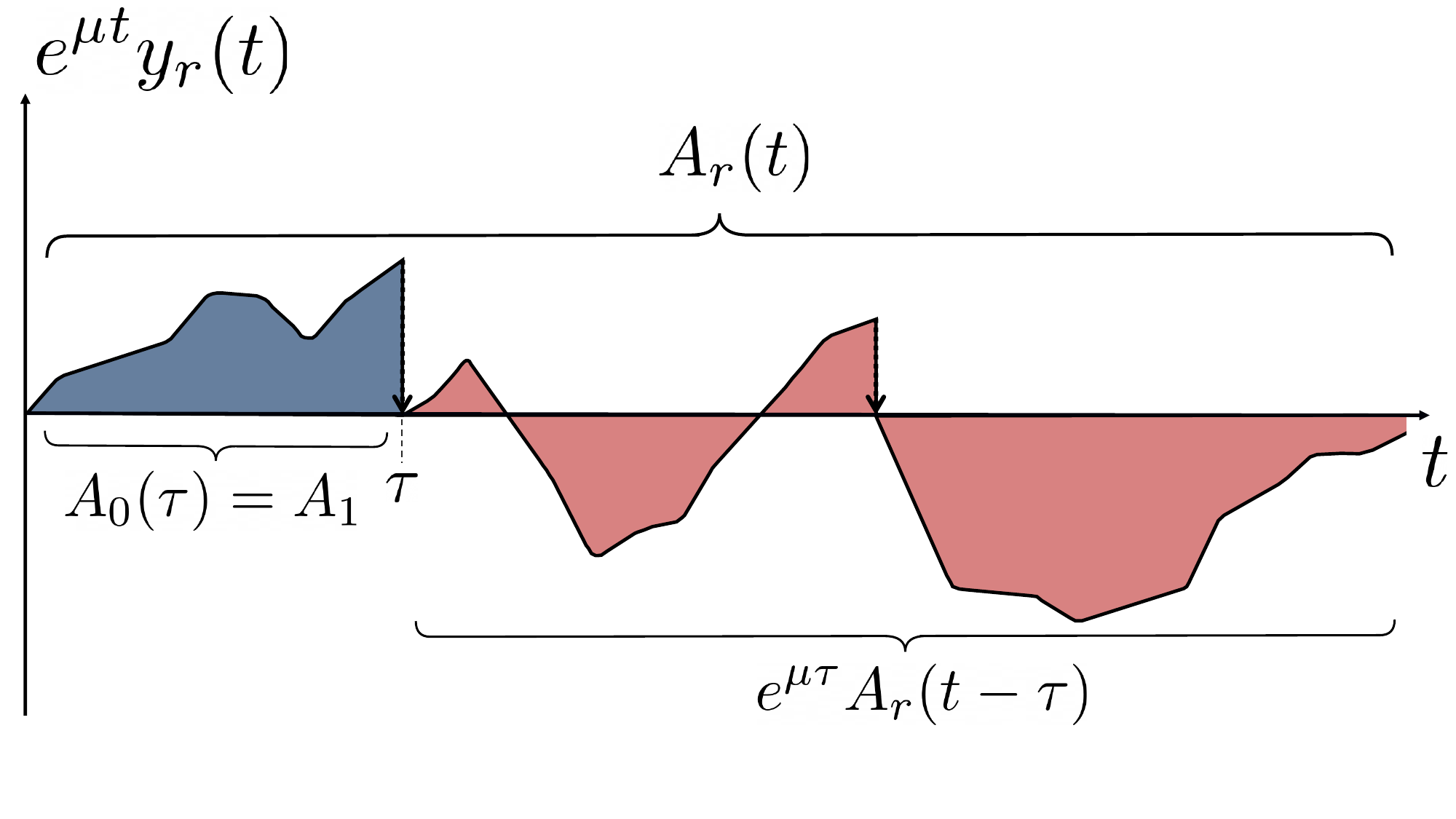}
\caption{Illustration of a resetting process, here $y_r(t)$. The first reset happens at time $\tau$. In our case, using the addition law \eqref{area}, the area under the curve up to time $t$ is $ A_r(t) = A_0(\tau) + e^{\mu \tau} A_r(t-\tau)$. The two areas $A_0(\tau)$ and $e^{\mu \tau} A_r(t-\tau)$ are independent. Hence, it is possible to use the renewal argument \eqref{areaintroductionA1} to compute the distribution $\mathcal{P}_r(A,t)$. In Eq. (\ref{areaintroductionA1}) we integrate over all possible areas $A_1=A_0(\tau)$ without resetting.}
\label{fig.areaexplanation}
\end{figure}

To write a renewal equation for $\mathcal{P}_r(A,t)$ we consider all the trajectories on the time interval $[0,t]$ and divide them into two groups: (i) the trajectories that experienced no resetting and (ii) the trajectories that experienced at least one resetting. In that case, we denote by $\tau$ the time of the first resetting (see Fig. \ref{fig.areaexplanation}). With this decomposition, we can write the following renewal equation (see also Ref. \cite{denHoll} for a closely related approach in the context of linear functionals of the rBM)
\begin{equation}
    \mathcal{P}_r(A,t) = e^{-rt}\, \mathcal{P}_0(A,t) +r\,  \int_{-\infty}^{+\infty}dA_1\int_{0}^{t}d\tau \, e^{-(r+\mu)\tau}\mathcal{P}_0(A_1,\tau)\, \mathcal{P}_r(e^{-\mu \tau}(A-A_1),t-\tau)\, .
\label{areaintroductionA1}
\end{equation}
The first term in the right hand side (RHS) comes from the trajectories (i) that experienced no resetting, since the probability that there is no resetting up to time $t$ is just $e^{-rt}$. To understand the contribution of the trajectories of the second group (ii), namely the second term in the RHS of (\ref{areaintroductionA1}), it is useful to write $A_r(t) = A_r(\tau + (t-\tau)) = A_0(\tau) + e^{\mu \tau} A_r(t-\tau)$, where we have used the addition law in Eq. (\ref{area}). Here, as we have seen above [see Eqs. (\ref{a1}) and (\ref{a2})], $A_0(\tau)$ and $e^{\mu \tau} A_r(t-\tau)$ are independent random variables drawn from the PDF $\mathcal{P}_0(A, \tau)$ and $e^{-\mu \tau}\mathcal{P}_r(e^{-\mu \tau} A,t-\tau)$ respectively. Hence the distribution of $A_r(t)$ is a convolution of these two distributions, as can be seen in Eq. (\ref{areaintroductionA1}). Finally, we need to average over all the possible values of $\tau$, with a PDF given by $r \,e^{-r \tau}$, which explains the integral over $\tau$ in Eq. (\ref{areaintroductionA1}).

Because of the convolution structure of Eq. (\ref{areaintroductionA1}), it is natural to take the Fourier transform of this equation with respect to $A$. 
For this purpose, we first perform the change of variable $A_1'=A-A_1$ in (\ref{areaintroductionA1}) to obtain
\begin{equation}\label{renewalArea}
    \mathcal{P}_r(A,t) = e^{-rt}\, \mathcal{P}_0(A,t) +r\,  \int_{-\infty}^{+\infty}dA_1'\int_{0}^{t}d\tau \, e^{-(r+\mu)\tau}\mathcal{P}_0(A- A_1',\tau)\, \mathcal{P}_r(e^{-\mu \tau}A_1',t-\tau)\, .
\end{equation}
Taking the Fourier transform with respect to $A$ on both sides of this equation one gets
\begin{equation}
    \widehat{\mathcal{P}}_r(k,t) = e^{-rt}\, \widehat{\mathcal{P}}_0(k,t) + r\, \int_{0}^{t}d\tau \, e^{-(r+\mu)\tau}  \int_{-\infty}^{+\infty}dA \int_{-\infty}^{+\infty}dA_1' \, e^{ikA}\, \mathcal{P}_0(A- A_1',\tau)\, \mathcal{P}_r(e^{-\mu \tau}A_1',t-\tau)\, .
\end{equation}Noticing that $e^{ikA}=e^{ikA_1'}e^{ik(A-A_1')}$ we get
\begin{equation}
    \widehat{\mathcal{P}}_r(k,t) = e^{-rt}\, \widehat{\mathcal{P}}_0(k,t) + r\, \int_{0}^{t}d\tau \, e^{-(r+\mu)\tau}  \int_{-\infty}^{+\infty}dA_1' \, e^{ikA_1'}\, \widehat{\mathcal{P}}_0(k,\tau)\, \mathcal{P}_r(e^{-\mu \tau}A_1',t-\tau)\, .
\end{equation}Hence, with another change of variable $A_1''=e^{-\mu \tau}A_1'$ we finally obtain
\begin{equation}\label{afterfourier}
    \widehat{\mathcal{P}}_r(k,t) = e^{-rt}\, \widehat{\mathcal{P}}_0(k,t) + r\, \int_{0}^{t}d\tau \,  e^{-r\tau}  \widehat{\mathcal{P}}_0(k,\tau)\, \widehat{\mathcal{P}}_r(ke^{\mu \tau},t-\tau)\, .
\end{equation}
To specify fully the renewal equation \eqref{afterfourier} we need to compute the Fourier transform of $\mathcal{P}_0(A,t)$. We recall that the area $A_0(t)$ is defined as
\begin{equation} \label{A0}
A_0(t) = \int_{0}^{t} dt'e^{\mu t'} B(t')=\int_{0}^{t}dt'e^{\mu t'}\int_{0}^{t'}ds\, \eta(s)\, \;.
\end{equation}
Here $B(t)$ is a standard Brownian motion, i.e., $\dot{B}(t) = \eta(t)$ where $\eta(t)$ is a Gaussian white noise of zero mean and with delta correlations $\left<\eta(t)\eta(t')\right> = 2D\delta(t-t')$. Since $B(t)$ is a Gaussian random variable, $A_0(t)$ is also  a Gaussian random variable, being the sum (or integral) of Gaussian random variables. Hence, $\mathcal{P}_0(A,t)$ is fully specified by the mean and variance of $A_0(t)$.  
They can be computed straightforwardly form (\ref{A0}), leading to 
\begin{equation}
\left<A_0(t)\right> = 0 \quad, \quad \left<A_0(t)^2\right> = \frac{D}{\mu^3}\, \left[e^{2\mu t}(2\mu t -3) + 4e^{\mu t} -1\right] \;.
\end{equation} 
Hence, the Fourier transform of the probability density function $\mathcal{P}_0(A,t)$ is
\begin{equation}\label{FourierP0A}
{\widehat{\mathcal{P}}_0(k,t) = e^{-\frac{k^2D}{2\, \mu^3} \,  \, [e^{2\mu t}(2\mu t -3) + 4e^{\mu t} -1]}\, .}
\end{equation}

Let us finally rewrite Eq. \eqref{afterfourier} in terms of $\hat p(k,t)$, namely the Fourier transform of $p(x,t)$, using the relation in (\ref{density rel1}). Indeed, from (\ref{density rel1}), one has  
\begin{align} \label{pP}
\hat{p}(k,t) = \widehat{\mathcal{P}}_r(k\, r\, e^{-\mu t},t)\, \;.
\end{align}
Finally, combining Eqs. (\ref{pP}) together with (\ref{afterfourier}) one finally obtains a closed integral equation for $\hat p(k,t)$, namely
\begin{equation}\label{Fourierp(x,t)}
\hat{p}(k,t) = e^{-rt}\, \widehat{\mathcal{P}}_0(k\, r\, e^{-\mu t},t) + r\, \int_{0}^{t}d\tau\,  e^{-r\tau}\,  \widehat{\mathcal{P}}_0(k\, r\, e^{-\mu t},\tau)\, \hat{p}(k,t-\tau)\, ,
\end{equation}with $\widehat{\mathcal{P}}_0(k,t)$ given in Eq. \eqref{FourierP0A}. Note that $\widehat{\mathcal{P}}_0(k\, r\, e^{-\mu t},\tau)$ can be conveniently written as
\be \label{conv1}
 \widehat{\mathcal{P}}_0(k\, r\, e^{-\mu t},\tau) = e^{-\frac{k^2}{2} e^{2 \mu(\tau -t)}\alpha(e^{-\mu \tau})} \;,
\ee
in terms of the function $\alpha(U)$ defined in Eq. (\ref{def_alpha}).

Solving explicitly this integral equation (\ref{Fourierp(x,t)}) seems of course very difficult. In fact, it seems already very difficult to show that this equation (\ref{Fourierp(x,t)}) reduces to the integral equation for the stationary distribution in Eq. (\ref{fourierstationnaire}) that we have derived using Kesten variables. It is however possible to compute recursively the time-dependent moments $\langle x^{2n}(t)\rangle$ from Eq.~(\ref{Fourierp(x,t)}) -- see Appendix \ref{appendixmomentstimedep} for details. It is then possible to check explicitly that, in the large time limit $t \to \infty$, these moments coincide with the one computed in the stationary state from Eq.~(\ref{fourierstationnaire}). 

At finite time $t$, it is natural to expect that $p(x,t)$ takes the scaling form 
\be \label{scaling_t}
p(x,t) = \frac{1}{L} {\cal F}\left(z = \frac{x}{L}, {\tilde t}= r\,t; \beta = \frac{r}{\mu} \right) \:,
\ee
which generalises the scaling form found in the steady state (\ref{scaling_F}). Equivalently, in Fourier space, one has
\be \label{scaing_tF}
\hat p(k,t) = {\cal F}(q = k L, \tilde t=r\,t; \beta) \quad {\rm where} \quad  {\cal F}(q, \tilde t; \beta) = \int_{-\infty}^\infty dz\,  {\cal F}(z,\tilde t; \beta) e^{i q z}  \;.
\ee
To derive the equation satisfied by ${\cal F}(q, \tilde t; \beta)$, we start from Eq. (\ref{Fourierp(x,t)}) and perform the change of variable $\tau' = r \tau$. 
After some simple algebra, using in particular Eq. (\ref{conv1}), one finds 
\begin{equation} \label{scaled_F}
\hat{\cal F}(q,\tilde t;\beta) = e^{-\tilde t-\frac{q^2}{2}\beta(1+\beta) \tilde \alpha(e^{-\tilde t/\beta})} + \int_{0}^{\tilde t}d\tau'\, e^{-\tau'}\,  e^{-\frac{q^2}{2}\beta(1+\beta) \tilde\alpha(e^{-\tau'/\beta})e^{2(\tau'-\tilde t)/\beta}}\hat{\cal F}(q,\tilde t-\tau';\beta) \;,
\end{equation} 
where we recall that $\tilde \alpha(U)$ is defined in Eq. (\ref{EqFhat}). Below, we analyse this integral equation in the two limits $\beta \to 0$ and $\beta \to \infty$ separately.

\subsection{The strongly active limit $\beta \to 0$}\label{smallrtimesection}

\begin{figure}[t]
\centering
\includegraphics[width=0.9\textwidth]{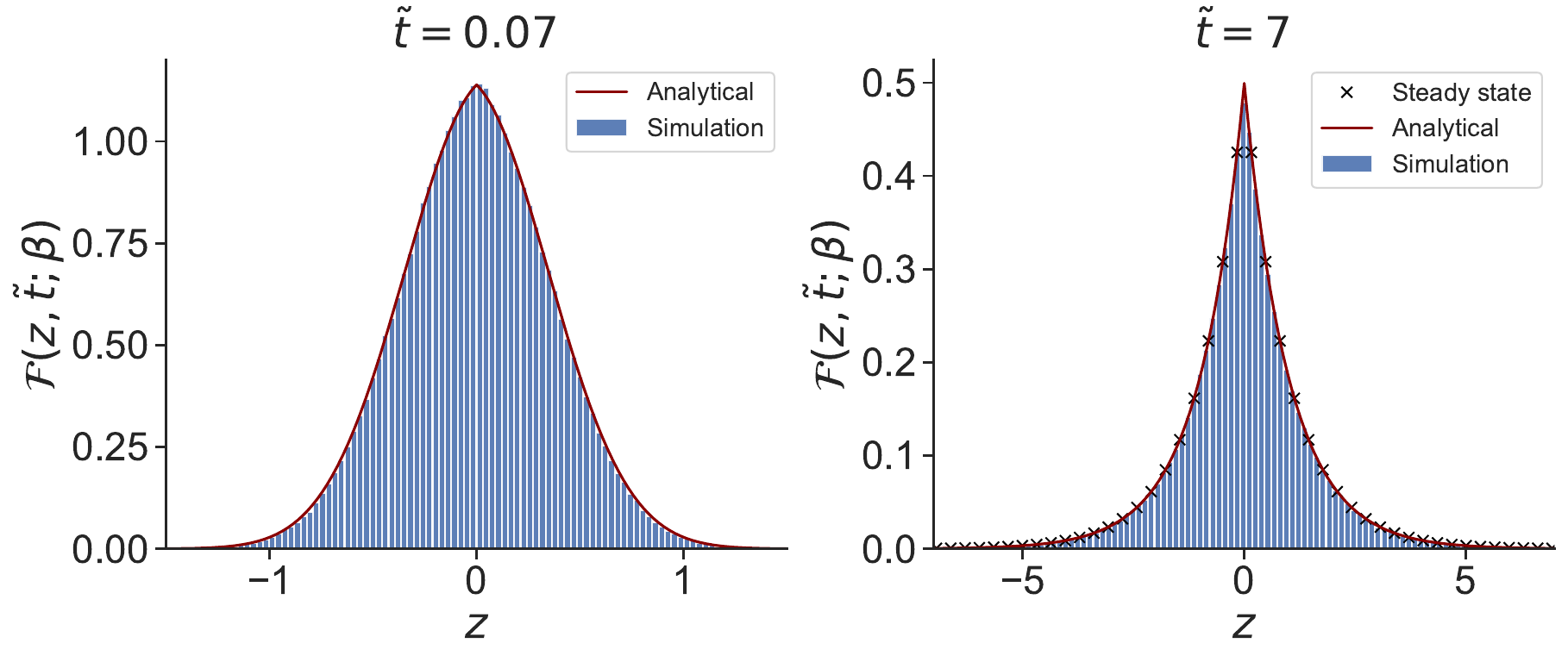}
\caption{In the left panel, we show a plot of the scaled PDF ${\cal F}(z,\tilde t;\beta)$ vs. $z$ for a small value of (rescaled) time $\tilde t =0.07$ and $\beta = 0.001$ (corresponding to $t=70$). The histograms in blue indicate the results of numerical simulations of Eq. (\ref{lange.2_bis}) while the solid red line corresponds to our exact analytical result in Eq. (\ref{inv_Fourier}). The agreement is very good. In the right panel, we show a plot of ${\cal F}(z,\tilde t;\beta)$ vs. $z$ with $\tilde t=7$ and $\beta = 0.001$ (corresponding to $t=7000$). Here again, the solid red line corresponds to our exact analytical result in Eq. (\ref{inv_Fourier}) while the black crosses correspond to the stationary state (\ref{expodistri}). Note that here the stationary state is expected to be reached at a time of order  $t \sim 1/\lambda_0 = 1/r=1000$ - see section \ref{relaxationtoSS} as well as Eq. (\ref{lambda-relax}), which explains why the black crosses fit perfectly with the data. In both panels, the parameters used for the simulations are $r = 0.001$, $\mu = 1$, and $D=1$.}
\label{fig.smallrdistrib}
\end{figure}

%In Appendix \ref{appendixmomentstimedep}, we provide a recursive derivation of the time dependent moments of the distribution $\left<x^{2n}(t)\right>$ starting from Eq. (\ref{Fourierp(x,t)}).

\blue{As discussed in the introduction, in this strongly active limit $\beta \to 0$, we expect the $x$-process to be completely slaved to the resetting noise, i.e., $x(t) \approx r\, y_r(t)/\mu$, where $y_r(t)$ represents the resetting Brownian motion. We have already demonstrated this property in the $t \to \infty$ limit, i.e., in the stationary state in Section \ref{expodistribsection}. In this section, we show that this property actually holds at all time $t$, and not just at large times.} In the limit $\beta \to 0$, using the asymptotic behavior of $\tilde \alpha(U)$ in Eq. (\ref{talpha_asympt}), one easily obtains the asymptotic behaviors of the different factors in Eq. (\ref{scaled_F}). In particular one finds
\be \label{asympt_term1}
\lim_{\beta \to 0} e^{-\tilde t-\frac{q^2}{2}\beta(1+\beta) \tilde \alpha(e^{-\tilde t/\beta})} = e^{-\tilde t(1+q^2)} \;,
\ee
while
\be \label{asympt_term2}
\lim_{\beta \to 0} e^{-\frac{q^2}{2}\beta(1+\beta) \tilde\alpha(e^{-\tau'/\beta})e^{2(\tau'-\tilde t)/\beta}} = 1 \quad {\rm for} \quad 0 \leq \tau' \leq \tilde t \;.
\ee
Using these asymptotic behaviors (\ref{asympt_term1}) and (\ref{asympt_term2}) in Eq. (\ref{scaled_F}), one finds that $\tilde{\cal F}(q,\tilde t;\beta = 0)$ satisfies the following integral equation
\be \label{int_beta0}
\hat{\cal F}(q,\tilde t;0) = e^{-\tilde t(1+q^2)} + \int_0^{\tilde t} d\tau'\, e^{-\tau'} \hat{\cal F}(q,\tilde t-\tau';0) \;.
\ee
This integral equation has a nice convolution structure and can thus be solved by taking the Laplace transform on both sides of Eq. (\ref{int_beta0}), which leads straightforwardly to
\be \label{inv_Laplace}
\hat{\cal F}(q,\tilde t;0)  = {\cal L}^{-1}_{s \to \tilde t} \left( \frac{1+s}{s(1+q^2+s)}\right) = \frac{1}{1+q^2} + q^2\frac{e^{-(1+q^2)\tilde t}}{1+q^2} \;.
\ee
\begin{figure}[t] 
    \centering
    \includegraphics[width=0.75\linewidth]{{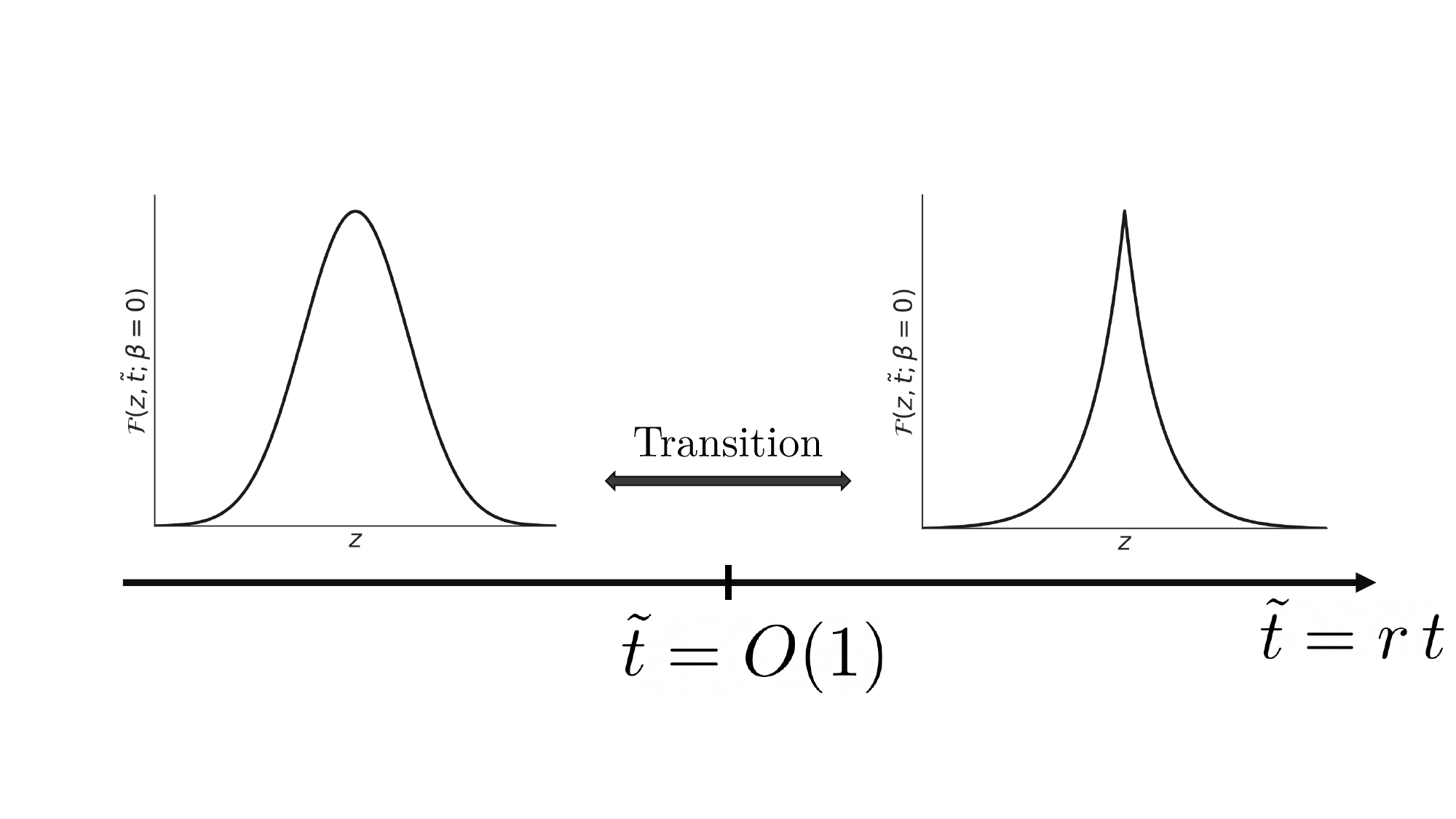}} 
    \caption{A sketch of the scaling function ${\cal F}(z,\tilde t;\beta=0)$ vs. $z$ as $\tilde t = r t$ is varied for the model described by Eq. (\ref{lange.2_bis}). It
   shows a transition from a Gaussian to a double-exponential distribution at large time, and small resetting rate $r$, i.e., $\beta \to 0$ [see Eq.~(\ref{gaussian_exp})].} 
    \label{fig7:a} 
\end{figure} 
By expanding this Fourier transform in powers of $q$, one can compute the moments of the distribution ${\cal F}(z,\tilde t;\beta =0)$ and, via Eq. (\ref{scaling_t}), the moments of $x(t)$. Skipping the details, one finds
 \begin{equation}
\left<x^{2n}\right> =  \gamma(n,\tilde t) \frac{(2n)!}{(n-1)!} L^{2n}\, ,
\end{equation}
where we recall that $L$ is given in Eq. (\ref{rational}) while $\gamma(n,\tilde t)$ is the lower incomplete gamma function of order $n$, 
\begin{equation}
\gamma(n,\tilde t) = \int_0^{\tilde t} du\,  u^{n-1} e^{-u} \, .
\end{equation} 
In fact the Fourier transform in (\ref{inv_Laplace}) can eventually be inverted to give
\bea \label{inv_Fourier}
\hspace*{-2cm}{\cal F}(z,\tilde t;\beta = 0) &=& \frac{1}{\sqrt{4 \pi\,\tilde t}} e^{-\frac{z^2}{4\tilde t} - \tilde t} \nonumber \\
&+& \frac{1}{4} \left[e^{|z|}\left({\rm erf}\left( \frac{|z|}{2 \sqrt{\tilde t}} + \sqrt{\tilde t}\right)-1 \right) -  e^{-|z|}\left( {\rm erf}\left( \frac{|z|}{2 \sqrt{\tilde t}} - \sqrt{\tilde t}\right)-1 \right) \right] \;,
\eea
where ${\rm erf}(y) = (2/\sqrt{\pi}) \int_0^y e^{-t^2}\,dt$ is the error function. \blue{In fact, one immediately recognizes that this is indeed the time-dependent position distribution of a one-dimensional resetting Brownian motion (rBM) with a resetting rate rescaled to unity \cite{Review20,NESS rBM} [see in particular Eq. (14) of the Supplementary Material of \cite{BMS22}].} In Fig. \ref{fig.smallrdistrib} we show a comparison of our analytical result (\ref{inv_Fourier}) with numerical simulations for small but nonzero value of $\beta$, showing a very good agreement. In fact, this distribution (\ref{inv_Fourier}) interpolates between a Gaussian behavior at short times $\tilde t \to 0$ and an exponential distribution as $\tilde t \to \infty$, which corresponds to the steady state [see the first line of Eq. (\ref{scalingformlimit})]. This can be easily seen from the explicit expression (\ref{inv_Fourier}) by using the asymptotic behaviors of the error function which we recall are given by  
\begin{equation} \label{asympt_erf}
\text{erf}(x) \underset{x \to \pm \infty}{\sim} \pm\left(1 - \frac{e^{-x^2}}{\sqrt{\pi} x}\right) \;.
\end{equation} 
Using these asymptotic behaviors (\ref{asympt_erf}) in (\ref{inv_Fourier}), one indeed finds (see Fig. \ref{fig7:a})
\be \label{gaussian_exp}
{\cal F}(z,\tilde t;\beta = 0) \approx
\begin{cases}
&\frac{1}{\sqrt{4 \pi \tilde t}} e^{-\frac{z^2}{4 \tilde t}} \;, \; \quad \tilde t \to 0 \\
& \\
&\frac{1}{2} e^{-|z|} \;, \; \quad\quad \;\; \tilde t \to \infty \;.
\end{cases}
\ee 
\begin{figure}[t]
\centering
\includegraphics[width=0.7\textwidth]{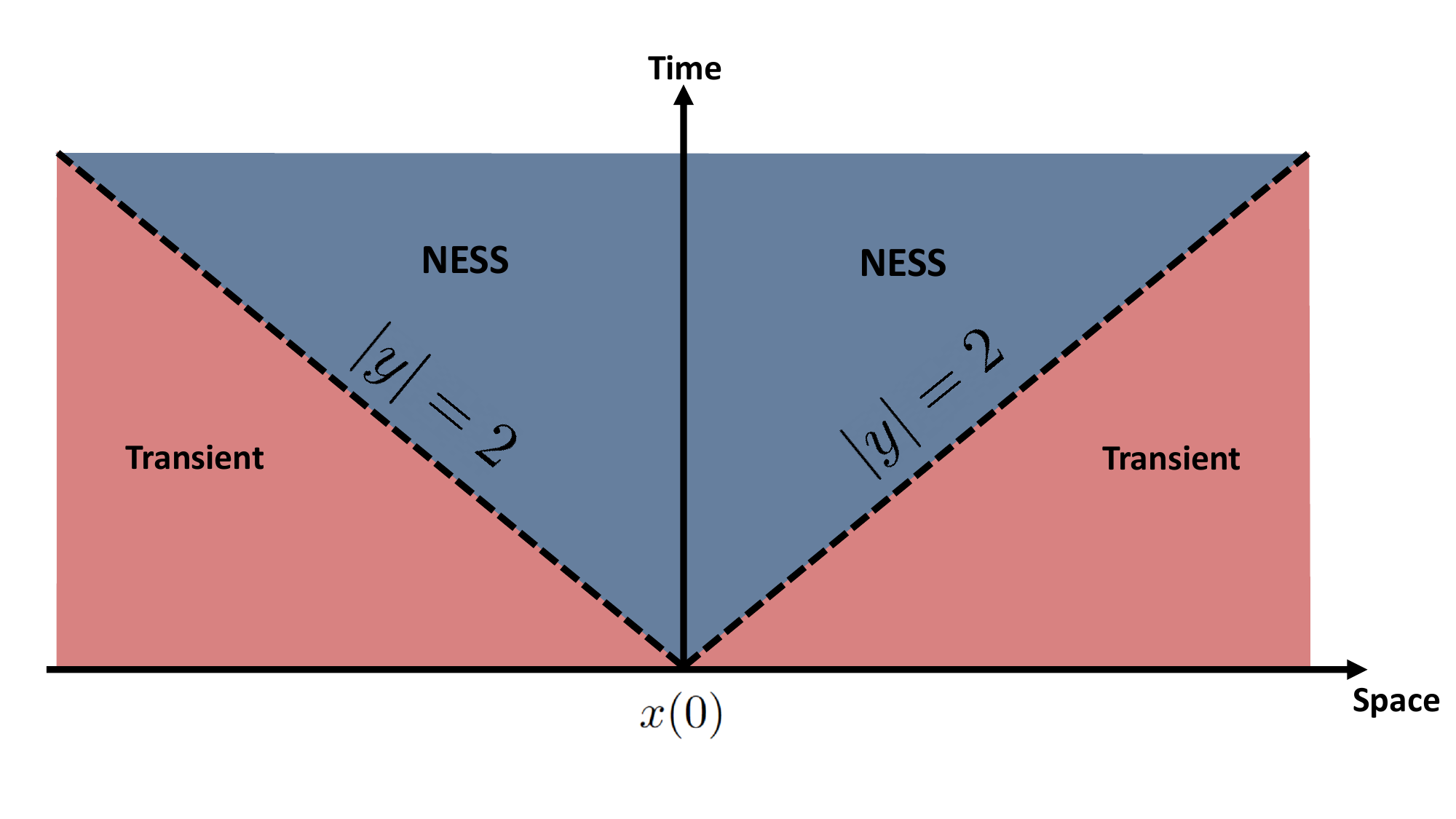}
\caption{Space-time phase diagram of the model (\ref{lange.2_bis}) in the strongly active limit $\beta \to 0$. In this limit, the scaled PDF ${\cal F}(z,\tilde t;0)$ takes a large deviation form (\ref{LDF}), characterised by a large deviation function $I(y=z/\tilde t)$ given in Eq. (\ref{rate_LDF}). This function exhibits a second order phase transition at $|y|=2$, indicated by the dashed lines on the figure. They separate a region (at small distance) where the system has reached a non-equilibrium steady state (NESS) from a region at large distance which is still time-dependent. A similar transition was also found for the rBM itself in Ref. \cite{NESS rBM}.}
\label{fig.NESS}
\end{figure}
In fact, one finds from Eq. (\ref{inv_Fourier}) that there is an interesting scaling regime where both $z$ and $\tilde t$ are large, but the ratio $y = z/\tilde t$ is kept fixed. Indeed, in this regime one finds that ${\cal F}(z,\tilde t;\beta = 0)$ takes a large deviation form
\begin{equation} \label{LDF}
{\cal F}(z,\tilde t;\beta=0) \sim \text{exp}\left[-\tilde t\, I\left(y=\frac{z}{\tilde t}\right)\right]\,,
\end{equation}
where the large deviation function $I(y)$ reads
\begin{equation} \label{rate_LDF}
I(y) = \left\{
    \begin{array}{ll}
    1 + \dfrac{y^2}{4} \quad \quad \mbox{for} & \; |y| > 2 \\
    	& \\
|y| \quad \quad \quad \quad \mbox{for} & \; |y| < 2 \\
                \end{array}
\right. \, .
\end{equation}
These two behaviors smoothly match the short and large time limiting form found above in Eq. (\ref{gaussian_exp}). Interestingly, although $I(y)$ as well as its first derivative are continuous at $y=\pm 2$, its second derivative is discontinuous since $I''(y=2^-) = 0$ while $I''(y=2^+) = 1/2$ (and similarly at $y=-2$). This thus corresponds to a second order dynamical phase transition. It separates the trajectories that have experienced only a small number of resettings (corresponding to large values of $|y| >2$) from trajectories who has been reset many times are already essentially in the NESS (corresponding to $|y|<2$). The frontiers separating them grows linearly with $\tilde t$. This transition is illustrated in Fig. \ref{fig.NESS}. \blue{In fact, this second order dynamical phase transition in the rate function $I(y)$ in Eq. \eqref{rate_LDF} was first obtained for the rBM in Ref. \cite{NESS rBM} using a saddle-point method.}

\subsection{The strongly passive limit $\beta \to \infty$}
\begin{figure}[t]
\centering
\includegraphics[width=0.9\textwidth]{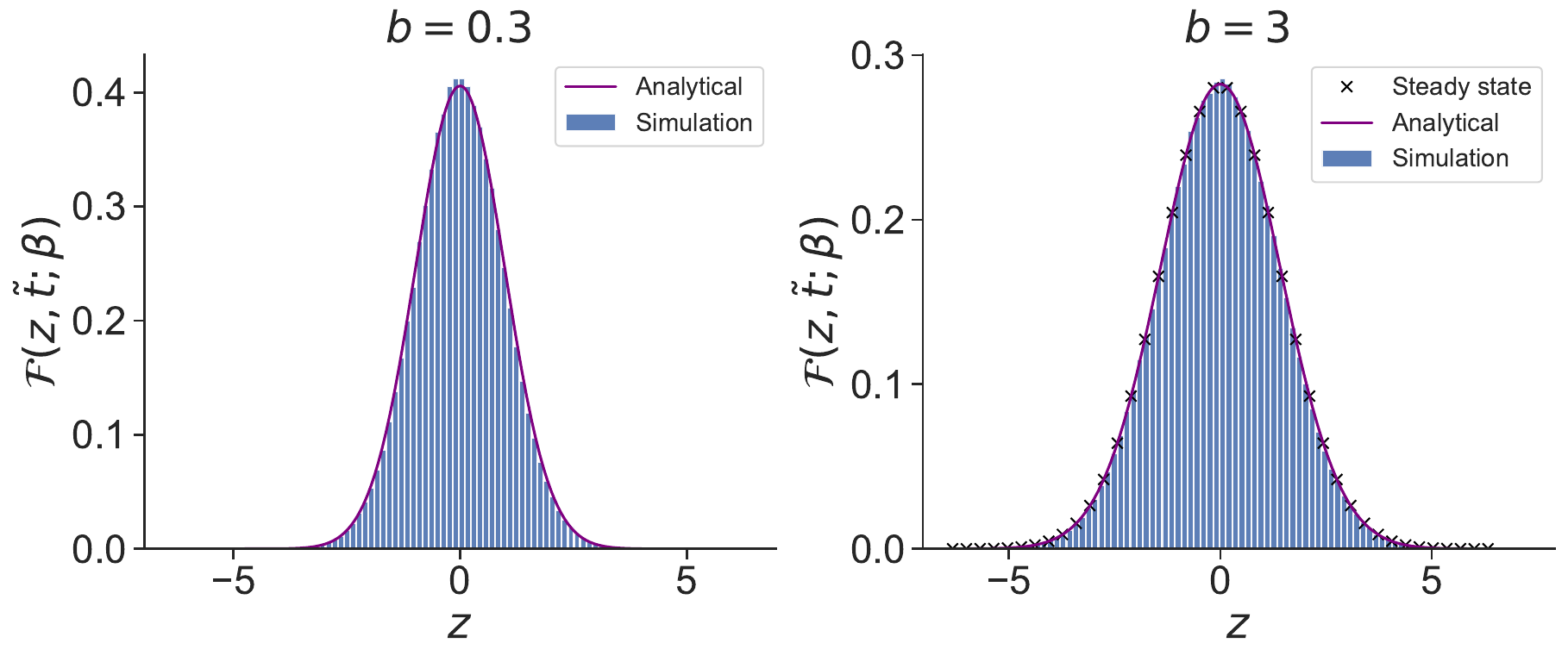}
\caption{For two different times, we plot the distribution $p(x,t)$ (purple), in the approximation of small $\mu$ for times of order $O(1/\mu)$ and compare it to simulations (histograms in blue). We showed that in this particular limit, the distribution  Eq. (\ref{smallmutimedistrib}) is the one of a Ornstein-Uhlenbeck process. The parameters of the simulations are $\mu = 0.001$, $r=1$, and $D=1$. Hence, it is supposed to be a good approximation for time of order $1/\mu = 1000$ . In addition, the system should have reached the stationary state near $t \sim 1/\lambda_0 = 500$ - see section \ref{relaxationtoSS}. The stationary state should therefore be reached on the right plot. To confirm this statement, we have plotted at some points (black crosses) the stationary state for small values of $\mu$ derived in Eq. (\ref{gaussiandistrib}). For small $z$, we see that at short time (left plot), the approximation is less accurate. This is probably due to the fact that the scaling form in Eq. (\ref{final_F0_real}) holds for $z \sim O(1)$, i.e, not too small.}
\label{fig.smallmudistrib}
\end{figure}

In this strongly passive limit $\beta \to \infty$, we will see that we also need to take the limit $\tilde t \to \infty$ but keeping the ratio $b=\tilde t/\beta$ fixed such that $\hat{{\cal F}}(q,\tilde t;\beta)$ takes the scaling form
\be \label{scaling_passive}
\hat{{\cal F}}(q,\tilde t;\beta) = \hat{{\cal F}}_0(q,b=\tilde t/\beta) + {\cal O}(1/\beta)\quad, \quad \beta \to \infty \;,
\ee
which is thus the generalisation of the form found in the stationary state in Eq. (\ref{exp_hatF}). In particular, one expects that this form (\ref{scaling_passive}) yields back the stationary state in the limit $b \to \infty$, i.e.
\be \label{match_stat}
\lim_{b \to \infty} \hat{{\cal F}}_0(q,b=\tilde t/\beta) = e^{-q^2} \;.
\ee
To proceed and compute the scaling function $\hat{{\cal F}}_0(q,b)$, we inject this scaling form (\ref{scaling_passive}) in the integral equation (\ref{scaled_F}) and
then performing the expansion for large $\tilde t$ and large $\beta$, keeping $b= \tilde t/\beta$ fixed. The computations are then very similar to the one performed in the stationary state (see Section \ref{betainf}). Skipping the details, we finally obtain a differential equation for $\hat{{\cal F}}_0(q,b)$ which reads
\be \label{ODE}
\partial_b  \hat{{\cal F}}_0(q,b) + 2 q^2\, e^{-2b}  \hat{{\cal F}}_0(q,b) = 0 \;.
\ee
It can straightforwardly be solved as $\hat{{\cal F}}_0(q,b) = A(q)e^{q^2 e^{-2b}}$. To determine the constant $A(q)$, we simply match this form with the result found in the stationary state (\ref{match_stat}), which simply gives $A(q) = e^{-q^2}$. Therefore $\hat{{\cal F}}_0(q,b)$ is given by
\be \label{final_F0}
 \hat {\cal F}_0(q,b) = e^{-q^2(1- e^{-2b})} \;.
\ee
By performing the inverse Fourier transform, one finds, in real space
\be \label{final_F0_real}
{\cal F}(z,b=\tilde t/\mu) \sim \frac{1}{\sqrt{4 \pi(1-e^{-2b})}} e^{-\frac{z^2}{4(1-e^{-2 b})}} \;.
\ee
It is instructive to write the corresponding result for the unscaled PDF $p(x,t)$, using Eqs.~(\ref{scaling_t}) and (\ref{final_F0_real}), namely
\begin{equation}
p(x, t) {\sim}  \sqrt{\frac{\mu }{4 \pi D(1-e^{-2 \mu\,t})}} e^{-\frac{\mu}{4 D(1-e^{-2\mu t})}x^2}\, .
\label{smallmutimedistrib}
\end{equation}
Hence in this strongly passive limit, the PDF $p(x,t)$ is exactly the one of a Ornstein-Uhlenbeck process with a diffusion constant $D_{\rm OU} = 2D$. As already mentioned, this can be easily understood by considering that the limit $\beta \to \infty$ can be realised at fixed $\mu$ and large $r$ where the noise becomes a white noise with a diffusion constant $2D$ [see Eq. (\ref{gaussiandistrib})].  In Figure \ref{fig.smallmudistrib}, the result of Eq. (\ref{smallmutimedistrib}) is compared to numerical simulations, showing a very good agreement.

\section{Conclusion}\label{conclusion}

In this paper, we have introduced and studied a class of models of an active particle where the dynamics is driven by a resetting noise in the presence of an external quadratic potential. We first focused on the stationary state distribution of the position of the particle $p(x)$ and showed that it can be studied within the framework of {\it Kesten variables}. This allowed us to derive an integral equation satisfied by $p(x)$, which can be explicitly written for different types of 
resetting noises. This includes in particular the telegraphic noise, in which case the model studied here coincides with the well known run-and-tumble particle. In this case, the integral equation satisfied by $p(x)$ can be solved explicitly and allows us to recover, by a completely different method, the well known result for the corresponding stationary distribution $p(x)$. We also showed that this integral equation can be solved explicitly for a wide class of resetting noises when the resetting protocol is periodic -- sometimes called "sharp resetting" or "stroboscopic resetting". For other resetting protocols, in particular in the case where the noise is a Brownian motion in the presence of Poissonian resetting -- i.e., the standard resetting Brownian motion -- it is very hard to solve exactly this integral equation. Nevertheless, we showed that it can be used to compute the moments of $p(x)$, as well as its asymptotic behavior for large $x$. Interestingly, we showed that $p(x)$ exhibits an exponential tail, which is markedly different form the Gaussian tail obtained in the case of a white noise -- instead of the resetting noise considered here. For Poissonian resetting noise, we also analyzed the full time-dependent distribution of the position $p(x,t)$ through the use of a renewal equation that can be generalized to other types of resetting noises. This model exhibits a rather rich dynamical behavior. In particular, in the strongly passive limit, $p(x,t)$ exhibits a second-order dynamical transition, which is akin to a similar transition found for the resetting Brownian motion itself (which is the noise term in this case).

The approach developed here based on Kesten variables is quite appealing and it would be nice to extend it in various directions. A first class of models where this approach could be useful are the "potential resetting" problems, studied for instance in Refs. \cite{MBMS20,MBM22,Santra}. In these models, a single Brownian particle is subjected to an external confining potential that is switched on and off stochastically, as in the recent experiments on resetting using optical tweezers \cite{Besga20,Faisant21}. One may also wonder whether this approach can be extended to two or several particles. A possible starting point could be the two RTPs model studied in \cite{bound}. It would 
also be interesting to use this framework to study the recently introduced active Dyson Brownian motion \cite{leo_active_dbm}.  
Similarly, it is natural to ask whether this approach can be extended to study models of active particles in two and higher dimensions, for which there exists very few analytical results. One could also study different resetting protocols acting on the noise term. Here we mainly focused on the periodic/sharp and Poissonian protocols but we could consider protocols which are more realistic from the experimental point of view \cite{Besga20}. In such protocols, the resetting is not instantaneous and it will be interesting to see how this feature would modify the results presented here. Finally, one can wonder whether this Kesten variables approach can be used to obtain information about the large deviation form of the stationary distribution, a question that has recently attracted some attention~\cite{Gradenigo,Mori_LD,MGM21,NaftaliActive}.   

\vspace*{0.5cm}
\noindent{\bf Acknowledgements:} We thank Uwe Ta\"uber for his many contributions to nonequilibrium statistical physics.

%\import{Chapters/}{Supplementary.tex}

%\newpage

\appendix

\section*{Appendices}

\renewcommand\thesection{\Alph{section}}

\section{Active resetting Ornstein-Uhlenbeck particle}\label{ArOUPs}
\begin{figure}[t]
\centering
\includegraphics[width = 0.5\linewidth]{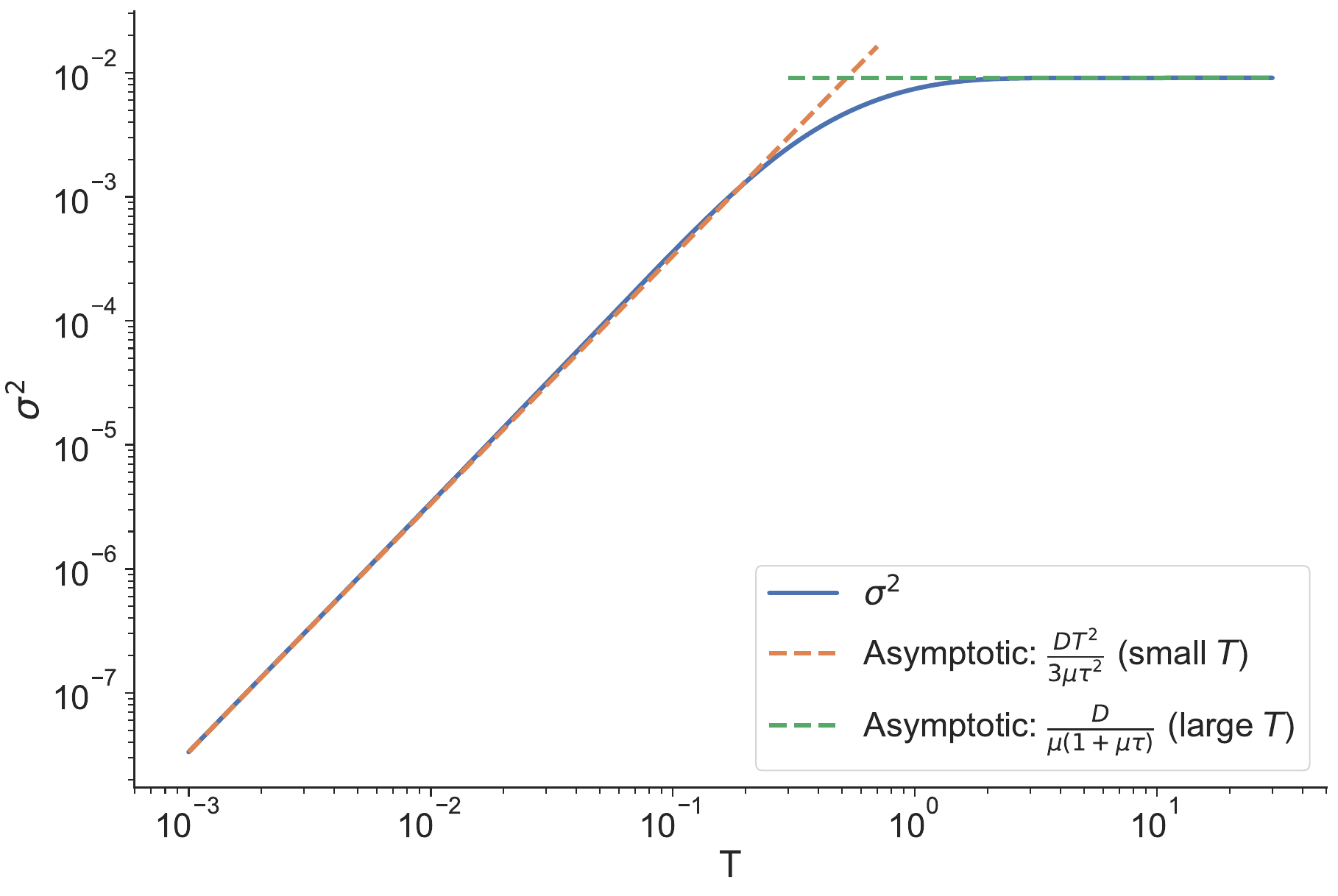}
\caption{The solid blue line shows a plot, on a log-log scale, of $\sigma^2$ given in Eq. (\ref{var_per.ArOUP}) vs $T$ for $\mu = 10$, $D = 1$ and $\tau=1$. The dashed orange and green lines show the asymptotic behaviors of $\sigma^2$ for small and large $T$ respectively, as given in Eq. (\ref{var_asymp_app}).}\label{plot_sigmaperiodicOUP}
\end{figure}

In this appendix, we generalise the approach presented in section \ref{kesten} to study the case of a resetting Ornstein-Uhlenbeck (rOU) noise (instead of a resetting Brownian noise discussed in details in the main text). We thus consider the dynamics of an active particle in the presence of a harmonic potential $V(x)= \mu x^2/2$ in one dimension. The position of the particle at time~$t$ is denoted $x(t)$, while the noise $y(t) = v_r(t)$ is a resetting Ornstein-Uhlenbeck (rOU) noise. The equation of motion reads
\begin{equation}
\frac{dx(t)}{dt}=-\mu \, x(t) + v_r(t)\, \;.
\label{langerOU.1}
\end{equation}
For simplicity, we choose as initial conditions $x(0)=0$ and $v_r(0)=0$. The dynamics of the process $v_r(t)$ is 
\begin{equation}
\begin{aligned}
v_r(t+dt) = \begin{cases}
0 &\text{, with probability } r\, dt \\
v_r(t) + \frac{dt}{\tau}\left(-v_r(t) + \sqrt{2D}\, \eta(t)\right) &\text{, with probability } (1-r\, dt)
\end{cases} \, .
\end{aligned}
\label{langerOU.2}
\end{equation}
where $\tau$ is the persistence time, and $\eta(t)$ a standard white noise with correlations $\left<\eta(t_1)\eta(t_2)\right> = \delta(t_2-t_1)$. 
Integrating Eq. (\ref{langerOU.1}) between two resetting epochs $t_{n-1}$ to $t_n$ we get 
\begin{equation}
x_n= x_{n-1}\, e^{-\mu \tau_n} + e^{-\mu\, \tau_n}\, \int_0^{\tau_n}d\tau\,  v(\tau)\, e^{\mu\, \tau}\, ,
\label{recur.1aroups}
\end{equation}
where $v(\tau)$ is a pure Ornstein-Uhlenbeck process (without resetting) whose dynamics is described by
\begin{equation}
    \tau\, \frac{dv(t)}{dt}=-v(t) + \sqrt{2D}\, \eta(t)\, .
\end{equation}
One can compute the two-time correlation of a OU process and obtain
 \begin{equation}
\left<v(t)v(t')\right> = \frac{D}{\tau}\left[ e^{-\frac{|t-t'|}{\tau}} - e^{-\frac{(t+t')}{\tau} }   \right] \;.
\label{correlOUv}
\end{equation}
Let us assume a periodic resetting with period $T$. The Kesten relation is given by
\begin{equation}
x_n= x_{n-1}\, e^{-\mu T} + e^{-\mu\, T}\, \int_0^{T}d\tau\,  v(\tau)\, e^{\mu\, \tau}\, ,
\label{recurOU.1aroups}
\end{equation}
where now, the stochasticity only comes from the OU process $v(\tau)$ whose distribution is a Gaussian. As a linear combination of Gaussians, the limiting distribution of $x$ will be Gaussian too. Hence, we just need to compute the second moment $\left<x_n^2\right>$. Squaring Eq. (\ref{recurOU.1aroups}) and taking average gives
\begin{equation}
\langle x_n^2\rangle= \langle x_{n-1}^2\rangle\, e^{-2\,\mu\, T}+ \, e^{-2\, \mu\, T}\, \int_0^Td\tau_1\, \int_0^Td\tau_2\, 
\left<v(\tau_1)v(\tau_2)\right> \, e^{\mu (\tau_1+\tau_2)}\, .
\label{per_Br.2_app}
\end{equation}
Using the two time correlations of $v(t)$ given by Eq.(\ref{correlOUv}), we find \begin{equation}
\langle x_n^2\rangle= \langle x_{n-1}^2\rangle\, e^{-2\,\mu\, T}+ \frac{D}{\tau}\, e^{-2\, \mu\, T}\, \int_0^Td\tau_1\, \int_0^Td\tau_2\, 
 \left[ e^{-\frac{|\tau_1-\tau_2|}{\tau}} - e^{-\frac{(\tau_1+\tau_2)}{\tau} }   \right]\, e^{\mu (\tau_1+\tau_2)}\, .
\end{equation}
Performing explicitly this integral, we get
\begin{equation}
\begin{split}
\langle x_n^2\rangle &= \langle x_{n-1}^2\rangle\, e^{-2\,\mu\, T} +\frac{D}{\tau}\,  e^{-2\,\mu\, T}\, \left\{\frac{\tau}{\mu^2 \tau^2 -1} \left[\frac{1}{\mu}\left(1 + (\mu \tau - 1)e^{2\mu T}\right) + \tau\left(1 - 2\tau\, e^{(\mu - \frac{1}{\tau})T} \right) \right]  \right. \\
&\left. - \frac{\tau^2}{(\mu \tau -1)^2} \left[e^{(\mu -\frac{1}{\tau})T}-1\right]^2  \right\}\, .
\end{split}
\label{per_Br.3aroups}
\end{equation}
As $n\to \infty$, the sequence $\{x_n(t)\}$ reaches a stationary Gaussian distribution. Taking $n\to \infty$
limit in Eq. (\ref{per_Br.3aroups}), the fixed point variance is then given explicitly by
\begin{equation}
\begin{split}
   \sigma^2 = \langle x_{\infty}^2\rangle &= \frac{D}{\left[e^{2\, \mu\, T}-1\right]}\, \left\{\frac{1}{\mu^2 \tau^2 -1} \left[\frac{1}{\mu}\left(1 + (\mu \tau - 1)e^{2\mu T}\right) + \tau\left(1 - 2\tau\, e^{(\mu - \frac{1}{\tau})T} \right) \right]  \right. \\
    &\left. -\frac{\tau}{(\mu \tau -1)^2} \left[e^{(\mu -\frac{1}{\tau})T}-1\right]^2\right\}\, .
\end{split}
\label{var_per.ArOUP}
\end{equation}
A plot of $\sigma^2$ vs $T$ is shown in Fig. \ref{plot_sigmaperiodicOUP} for $\mu=10$, $D=1$ and $\tau=1$. It has the following asymptotic behaviors for small and large $T$
\be
\sigma^2 \approx \begin{cases}
\frac{D\, T^2}{3\mu \tau^2} \quad \quad {\rm as} \quad T\to 0 \\
\\
\frac{D}{\mu(1+\mu \tau)} \quad {\rm as} \quad T\to \infty
\end{cases}\, .
\label{var_asymp_app}
\ee
The rare resetting limit $T\to \infty$ corresponds to the stationary distribution of an active OU particle - See Eq. (10) of \cite{AOUP}.
In the diffusive limit, i.e. $\tau \to \infty$ while $\sqrt{D}/\tau = \sqrt{\tilde{D}}$, $v_r(t)$ becomes a resetting Brownian motion and we retrieve the expected variance (\ref{var_per.1})
\begin{equation}
\sigma^2=  \frac{\tilde{D}}{\mu^3}\, \frac{\left[2\,\mu\, T-3 + 4\, e^{-\mu\, T}- e^{-2\, \mu\, T}\right]}{\left[1-e^{-2\, \mu\, T}\right]}\, .
\end{equation}
In the case of a Poissonian resetting protocol, results of section \ref{sectionstatio} can be used replacing the variance from Eq. (\ref{variance_Un}) by
\begin{equation}
\begin{split}
\left<V_n^2\right> &= \frac{D}{\tau}\,  e^{-2\,\mu\, \tau_n}\, \left\{\frac{\tau}{\mu^2 \tau^2 -1} \left[\frac{1}{\mu}\left(1 + (\mu \tau - 1)e^{2\mu \tau_n}\right) + \tau\left(1 - 2\tau\, e^{(\mu - \frac{1}{\tau})\tau_n} \right) \right]  \right. \\
&\left.- \frac{\tau^2}{(\mu \tau -1)^2} \left[e^{(\mu -\frac{1}{\tau})\tau_n}-1\right]^2  \right\}\, ,
\end{split}
\end{equation}
with $V_n = e^{-\mu\, \tau_n}\, \int_0^{\tau_n} v_n(\tau)\, e^{\mu\, \tau}\, d\tau\,$. Concerning section \ref{sectionTime}, it is possible to apply the same methods as when the noise is a rBM by changing the definition (\ref{areadefintion1}) of the area $A_r(t)$ to \begin{equation}
A_r(t) = \int_{t_n}^{t_{n-1}}dt'\, e^{\mu t'}v_r(t)\, ,
\end{equation}
where $v_r(t)$ is a resetting Ornstein-Uhlenbeck process with a resetting rate $r$.

\section{Run-and-tumble models}\label{PoissonRTP}

\subsection{Telegraphic noise : the run-and-tumble particle}
Consider the telegraphic noise $y(t)= v_0\, \sigma(t)$, where $v_0$ is a constant and $\sigma(t)=\pm 1$ with flip rate $r$. Note that in this case $y(t)$ alternates between the values $+v_0$ and $-v_0$ between successive resettings. Applying the Kesten approach, the limiting distribution of $x_n$ in Eq.~(\ref{kesten.1}) will depend on whether the $n$-th run of $x$ is for a positive $y=+v_0$ or negative $y=-v_0$. In this case,  we need to use $y_n(\tau)= v_0 \sigma_n$ in our general recursion relation (\ref{recur.1}), where $\sigma_n=1$ if the $n$-th run is with a positive velocity  and $\sigma_n=-1$ if the $n$-th run is with a negative velocity. In this case, Eq. (\ref{recur.1}) reduces to
\begin{equation}
x_n= x_{n-1}\, e^{-\mu\, \tau_n}+ \sigma_n\, \frac{v_0}{\mu}\,\left(1- e^{-\mu \tau_n}\right) \;.
 \label{rtp_recur.1}
\end{equation}
It is convenient to re-write it as
\begin{equation}
x_n= x_{n-1}\, U_n + \sigma_n\, \frac{v_0}{\mu}\,(1-U_n)\, \quad {\rm where}\quad U_n= e^{-\mu \tau_n}\, .
\label{rtp_recur.2}
\end{equation}
Given that the distribution of $\tau$ is $p_{\rm int}(\tau)=r\, e^{-r\, \tau}$, it follows that the variables $U_n$ are distributed over $U\in [0,1]$ with the following PDF
\begin{equation}
P(U)= \beta\, U^{\beta-1}\, \quad {\rm with}\quad \beta= \frac{r}{\mu}\, .
\label{pdf.u}
\end{equation}
Let $p_+(X)$ and $p_-(X)$ denote the limiting distributions of $x_n$ (respectively at the end of a positive and a negative run). Noting that since the intervals alternate, if $x_n$ denotes the end position of a positive run, then $x_{n-1}$  in Eq. (\ref{rtp_recur.2}) necessarily represents the end position of a negative run and $\sigma_n=1$ (and vice versa), we can easily write down the pair of Kesten integral equations for the two limiting distributions, namely
\begin{eqnarray}
p_{+}(x) & = & \int_{-\infty}^{\infty} dx' \int_0^{1} dU\, P(U)\, p_{-}(x')\, 
\delta\left(x- U\, x'- \frac{v_0}{\mu}\, (1-U)\right) \;,
\label{plus.1} \\ 
p_{-}(x) & = & \int_{-\infty}^{\infty} dx' \int_0^{1} dU\, P(U)\, p_{+}(x')\, 
\delta\left(x- U\, x'+ \frac{v_0}{\mu}\, (1-U)\right) \;,
\label{minus.1}
\end{eqnarray}
where $P(U)$ is given in Eq. (\ref{pdf.u}). To simplify the algebra, we rescale $x= (v_0/\mu)\, Z$ and denote the limiting PDF's $P_{\pm}(v_0 Z/\mu)= Q_{\pm}(Z)$. Then using $P(U)$ from Eq. (\ref{pdf.u}), and carrying out the integration over the $x'$ variables, the pair of integral equations (\ref{plus.1}) and (\ref{minus.1}) in the $Z$ variables reduce to
\begin{eqnarray}
Q_{+}(Z) & = & \beta\, \int_0^{1} dU\, U^{\beta-2}\, Q_{-}\left(1- \frac{(1-Z)}{U}\right)\,,\, \label{plus.2} \\
Q_{-}(Z) & = &  \beta\, \int_0^{1} dU\, U^{\beta-2}\, Q_{+}\left(\frac{(1+Z)}{U}-1\right)\,, \, \label{minus.2}
\end{eqnarray}
where we recall that $\beta=\gamma/\mu$. In fact by taking derivatives of these two equations (\ref{plus.2}) and (\ref{minus.2}) with respect to $Z$ and performing integration by parts in the integrals over $U$, one can transform these two coupled integral equations into two coupled linear differential equations, namely
\bea 
(1-Z)Q_+'(Z) = \beta Q_-(Z) + (1-\beta) Q_+(Z) \label{plus.2_ed1} \\
(1+Z)Q_-'(Z) = -\beta Q_+(Z) - (1-\beta) Q_-(Z)  \label{minus.2_ed1} \;.
\eea  
It is convenient to rewrite these equations as
\bea
&& 0 = -\frac{\partial}{\partial Z}[(-Z+1) Q_+(Z)] + \beta Q_-(Z) - \beta Q_+(Z) \label{plus.2_ed2} \\
&&0 = -\frac{\partial}{\partial Z}[(-Z-1) Q_-(Z)]  +\beta Q_+(Z) - \beta Q_-(Z)  \label{minus.2_ed2} \;.
\eea
Under this form, one can check that these equations are exactly identical to the ones found for the standard RTP model (see e.g. Eqs. (5) and (6) in Ref. \cite{DKMSS19}) using a completely different method. In particular, one can show that $Q_{\pm}(Z)$ have a finite support over $Z\in [-1,1]$ where they take the form
\begin{eqnarray}
Q_{+}(Z) &=& A\, (1+Z)^{\beta}\, (1-Z)^{\beta-1}\, , \quad\,\, -1\le Z\le 1 \label{Q_plus.1} \\
Q_{-}(Z) &=& A\, (1+Z)^{\beta-1}\, (1-Z)^{\beta}\, , \quad\,\, -1\le Z\le 1 \label{Q_minus.1} \;,
\end{eqnarray}
where $A = 1/(4^\beta B(\beta, \beta))$ where $B(x,y)$ is the beta function. This demonstrates that this Kesten approach allows to recover this well known result (\ref{Q_plus.1}))-(\ref{Q_minus.1}) using a method which is quite different from the usual one, relying on a more standard Fokker-Planck approach.

\subsection{The generalised run-and-tumble particle with an arbitrary speed distribution}

Let us consider a generalized model of a run-and-tumble particle evolving in a harmonic trap with initial position $x(0)=0$. After each tumble, the speed $v$ is now drawn from an arbitrary distribution $w(v)$, see e.g. \cite{generalRTP}, and the equation of motion is then $\dot{x}=-\mu\, x + v$. Another way to describe the motion is to consider the deterministic motion between the different resetting epochs $\{t_1,t_2,...t_n\}$ such that if we integrate between $t_n$ and $t_{n-1}$, and denote $\tau_n = t_n -t_{n-1}$, we have 
\be
    x_n = x_{n-1}e^{-\mu \tau_n} + \frac{v_i}{\mu}\left[e^{-\mu\tau_n} - 1\right]\, ,
\ee with $p_{\rm int}(\tau_n)=r\, e^{-r \tau_n}$. This recursion relation is of the generalised Kesten form $x_n = U_n x_{n-1} + V_n$, with $U_n = e^{-\mu \tau_n}$ and $V_n = \frac{v_i}{\mu}\left[e^{-\mu\tau_n} - 1\right] = \frac{v_i}{\mu}\left[U_n - 1\right]$. The joint law of $U$ and $V$ can be written as follows 
\be
    P(U,V)=P(U)P(V|U) =\beta\, U^{\beta-1}\, \frac{\mu}{U-1}\, w\left(\frac{\mu V}{U-1}\right)\, .
\ee
From Eq. (\ref{kesten.2}) in the main text, we deduce that the stationary distribution verifies the integral equation \begin{equation}
p(x)= \int_{0}^{1} dU \int_{-\infty}^{\infty} dV\int_{-\infty}^{\infty} dx'\, \beta\, U^{\beta-1}\, \frac{\mu}{U-1}\, w\left(\frac{\mu V}{U-1}\right)\, 
p(x')\,  \delta(x-U\, x'-V)\, .
\label{kesten.RTPG1}
\end{equation}
Performing a Fourier transform with respect to the variable $x$, one gets \begin{equation}
\hat{p}(k)= \int_{0}^{1} dU \int_{-\infty}^{\infty} dV\int_{-\infty}^{\infty} dx'\, \beta\, U^{\beta-1}\, \frac{\mu}{U-1}\, w\left(\frac{\mu V}{U-1}\right)\, 
p(x')\,  e^{ik(Ux'+V)}\, .
\label{kesten.RTPG2}
\end{equation} Next, we do the integration on the variable $x'$ and obtain
\begin{equation}
\hat{p}(k)= \int_{0}^{1} dU \int_{-\infty}^{\infty} dV\, \beta\, U^{\beta-1}\, \frac{\mu}{U-1}\, w\left(\frac{\mu V}{U-1}\right)\, 
\hat{p}(kU)\,  e^{ikV}\, .
\label{kesten.RTPG3}
\end{equation}
Finally, we apply the change of variables $V' = \frac{\mu V}{U-1}$ and we integrate over the new variable $V'$. It leads to 
\begin{equation}
\hat{p}(k)= \int_{0}^{1} dU\, \beta\, U^{\beta-1}\, \hat{w}\left(\frac{U-1}{\mu}\, k\right)\, 
\hat{p}(kU)\, .
\label{kesten.RTPG4}
\end{equation} 
%Integral equation (\ref{kesten.RTPG4}) may take a simpler form in some limits. For instance, consider $\mu \to +\infty$ while the ratio $\beta = r/\mu$ fixed, the integral does not depend on the distribution $w$ anymore. 
It is surely very interesting to study the integral equation (\ref{kesten.RTPG4}), but for a general $w(v)$ it is a very hard problem.

\begin{figure}[t]
\centering
\includegraphics[width=0.6\textwidth]{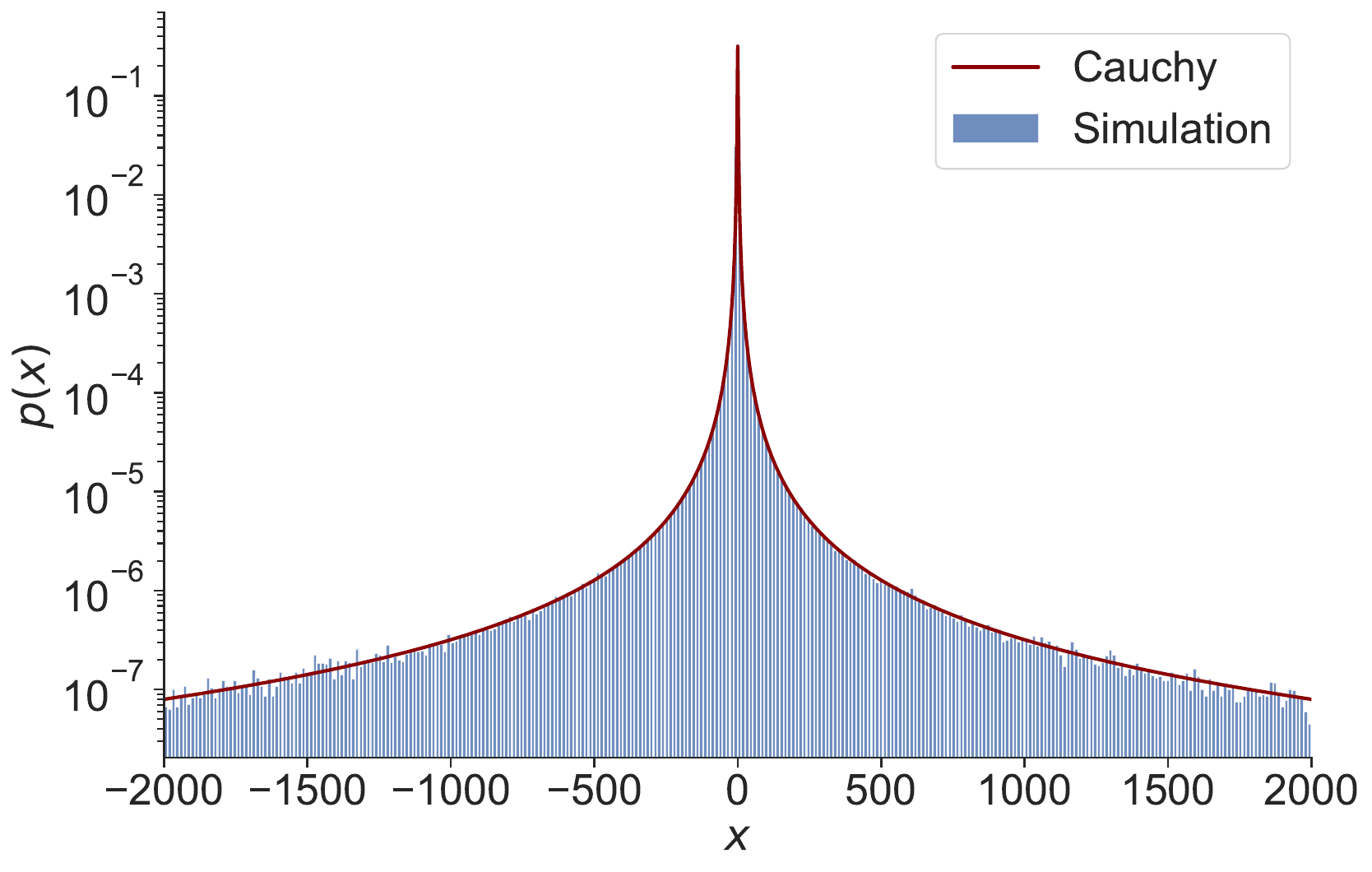}
\caption{Plot of the stationary distribution of the position $p(x)$ for a generalised run-and-tumble particle where the velocity has 
a Cauchy distribution. The blue histogram is the result of numerical simulations while the solid line indicates our analytical prediction in (\ref{Cauchyresultappendix}), showing an excellent agreement. Here $\lambda = \mu = 1$.}
\label{fig.appendixcauchy}
\end{figure}

There is however one case which can be solved exactly, namely the case where $w(v)$ is a Cauchy distribution with parameter $\lambda >0$, i.e. $\hat{w}(k) = e^{-\lambda|k|}$. In this case Eq. (\ref{kesten.RTPG4}) reads
\begin{equation}
\hat{p}(k)= \int_{0}^{1} dU\, \beta\, U^{\beta-1}\, e^{-\lambda|\frac{U-1}{\mu}\, k|} 
\hat{p}(kU) \;.
\end{equation}
Noticing that $|U-1| = 1-U$, we can rewrite it as
\begin{equation}
e^{\frac{\lambda\, |k|}{\mu}}\, \hat{p}(k)= \int_{0}^{1} dU\, \beta\, U^{\beta-1}\, e^{\lambda\, \frac{U}{\mu}\, |k|}\, 
\hat{p}(kU)\, .
\end{equation}
If we introduce $q(k) = e^{\frac{\lambda\, |k|}{\mu}}\, \hat{p}(k)$, the equation takes a simpler form
\begin{equation}
q(k)= \int_{0}^{1} dU\, \beta\, U^{\beta-1}\, q(kU)\, .
\end{equation}
One can indeed perform the change of variable $v=k\, U$, and it leads to
\begin{equation}
k^\beta\, q(k)= \beta\, \int_{0}^{k} dv\, v^{\beta-1}\, q(v)\, .
\end{equation}
Taking the derivative of this equation with respect to $k$ gives directly
\begin{equation}
    k^\beta\, \partial_kq(k) = 0\, .
\end{equation}
As a consequence, for all values of $k$ we have $q(k)=c$, with $c$ a constant. However, the distribution $p$ is normalised so that $\hat{p}(0) = q(0) = 1$. Hence $c=1$ and $\hat{p}(k)=e^{-\frac{\lambda\, |k|}{\mu}}$. The stationary distribution is therefore a Cauchy distribution with parameter $\lambda/\mu$, namely 
\begin{equation}
    p(x) = \frac{\lambda\, \mu}{\pi(\lambda^2+\mu^2\, x^2)}\,.
\label{Cauchyresultappendix}
\end{equation}
In Fig. \ref{fig.appendixcauchy} we compare this analytical prediction (\ref{Cauchyresultappendix}) to numerical simulations, showing a very good agreement.

\section{Correlation of the reset random time-dependent speed noise}\label{correlrandomspeed}

We consider the process $v(t) = v\, h(t)$ that is reset at exponentially distributed time $t_i$ such that $p(t_i)=re^{-rt_i}$. The constant speed $v$ is changed at every reset and the $v_i$'s are drawn from a distribution $w(v)$. For Poissonian resetting, the two-time correlation function is \cite{correlation resetting}
\begin{equation}
    C_r(t_1,t_2) = e^{-r(t_2-t_1)}\left[\int_0^{t_1} d\tau\, r\, e^{-r\tau}\, C_0(\tau,t_2-t_1+\tau)+e^{-rt_1}C_0(t_1,t_2)\right]\, ,
\end{equation}where $C_0$ is the correlation function of the same process without reset. Hence here we have, 
\begin{eqnarray}
    C_0(t_1,t_2) = \left<v^2\, h(t_1)\, h(t_2) \right>_{w(v)} = \left<v^2\right>_{w(v)}\, h(t_1)\, h(t_2)\, .
\end{eqnarray}
We deduce directly
\begin{equation}
    C_r(t_1,t_2) =\left<v^2\right>_{w(v)}\,  e^{-r(t_2-t_1)}\left[\int_0^{t_1} d\tau\, r\, e^{-r\tau}\, h(\tau)\, h(t_2-t_1+\tau)+e^{-rt_1}\, h(t_1)\, h(t_2)\right]\, .
\end{equation}
If $h(t)=1$, we recover the noise of the generalised run-and-tumble particle,
\begin{equation}
    C_r(t_1,t_2) =\left<v^2\right>_{w(v)}\,  e^{-r(t_2-t_1)}\left[\int_0^{t_1} d\tau\, r\, e^{-r\tau}+e^{-rt_1}\right] =\left<v^2\right>_{w(v)}\,  e^{-r(t_2-t_1)}\, .
\end{equation}
If $h(t)=t$, the reset process is the noise of the ballistic model with random speed $v$. The correlation function is in this case
\begin{equation}
    C_r(t_1,t_2) =\frac{\left<v^2\right>_{w(v)}}{r^2}\,  \left[e^{-r(t_2-t_1)}\left(2+r(t_2-t_1)\right) -e^{-rt_2}\left(2+r(t_1+t_2)\right)   \right]\, .
\end{equation}
Notice that when $h(t)=1$, $C_r(t,t+\tau)$ depends only on $\tau$. This property is not verified anymore when the motion becomes ballistic, i.e. $h(t) = t$.

\section{Numerical results}\label{numerics}

\subsection{Simulation of trajectories}\label{simulationprocedure}

The numerical data shown in the different plots of the main text have been obtained by simulating trajectories of the motion evolving according to 
\begin{equation}
\frac{dx(t)}{dt}=-\mu \, x(t) + r\, y_r(t)\, ,
\end{equation}where  \begin{equation}
\displaystyle
y_r(t+dt) = \left\{
    \begin{array}{ll}
        0 \hspace*{3cm}\mbox{ with proba.} \; rdt \\
        y_r(t) + \xi(t)\, dt \hspace*{0.8cm}\mbox{ with proba.} \; (1-rdt)
    \end{array}
\right.
,
\end{equation} and $\xi(t)$ is a white noise with a diffusivity $D$. 
Trajectories evolve therefore under the following rule
\begin{equation}
\begin{split}
  x(t+dt) &= x(t)\,  +\, dt\, \left(-\mu\, x(t) + r\, y_r(t)\right) \\
  &=\left\{
    \begin{array}{ll}
         x(t) - dt\, \mu\, x(t)\mbox{ with probability } rdt \\
         x(t) + dt\, \left[-\mu\, x(t) + r\, \left(y_r(t-dt) + \sqrt{2\, D\, dt} \, {\rm N}(0,1)\right)\right]\mbox{ with proba. } (1-rdt)
    \end{array}
\right.
\end{split}
\end{equation} and ${\rm N}(0,1)$ is a random number drawn a centered Gaussian distribution of variance unity. 

\subsection{Numerical resolution of the integral equation}\label{numericprocedure}
In the middle panel of Fig. \ref{fig.stationarydistrib}, we show the results which have been obtained by solving numerically the integral equation (\ref{fourierstationnaire}) to compute eventually $p(x)$. We recall that we want to solve 
\begin{equation} \label{int_eq_app}
   \hat{p}(k) = \int_{0}^{1} dU \,\beta\, U^{\beta - 1} e^{-\frac{k^2 \alpha(U)}{2}}\, \hat{p}(kU)\, .
   \end{equation}We are going to approximate the value of $\hat{p}(k)$ by discretizing the interval $[0,k]$ in $n$ small sub-intervals of size $\delta k$. We then consider that in the interval $[m\, \delta k, (m+1)\, \delta k]$ the value $\hat{p}(kU)$ is constant and, consequently, the discretized version of (\ref{int_eq_app}) reads
  \begin{equation}\label{fourierdiscret}
   \hat{p}(n \delta k) \sim \sum_{m=0}^{n-1}\hat{p}(m \delta k) \int_{\frac{m}{n}}^{\frac{(m+1)}{n}} dU \beta\, U^{\beta -1} e^{-\frac{(n \delta k)^2}{2} \alpha(U)} \, .
   \end{equation}
In view of computing the inverse Fourier transform to obtain a numerical estimate of $p(x)$, we can interpolate, using a calculation software, the list computed from Eq. \ref{fourierdiscret} such that we have a smooth function $\hat p(k)$ that has values for every $k$ on an interval $[-N \delta k, N \delta k]$, with $N$ being a large integer number (and using $\hat p(-k) = \hat p(k)$). We can then finally compute $p(x)$ from    
    \begin{equation}
   p(x)  \sim \int_{-N\delta k}^{+N\delta k}dk\,  \frac{e^{-i k x}}{2 \pi} \hat{p}(k) \;. 
   \end{equation}
In the middle panel of Fig. \ref{fig.stationarydistrib}, we show a plot of $p(x)$ which has been evaluated numerically using this procedure and compare it to a direct simulation of Eq. (\ref{lange_rbm}) for $\beta = 1$. We see that the agreement is excellent.   
   
\section{First moments of the stationary distribution $p(x)$}\label{AppendixMoments}
The equation (\ref{stationary-moments}) in section \ref{statiomoments} provides a recursion relation that allows us to compute the moments in the stationary state. We recall that this recursion relation reads (for $n \geq 1$)
\begin{equation}
\left<x^{2n}\right> = \beta\, (\beta + 2n)\, (2n-1)! \sum_{p = 0}^{n-1}  \frac{\left<x^{2p}\right>}{2^{n-p}\, (n-p)!\, (2p)!}   \int_{0}^{1} dU   \, U^{\beta - 1 + 2p}\, \left[\alpha(U)\right]^{n-p}\,,
\label{stationary-moments-appendix} 
\end{equation}
with $\alpha(U) = \frac{D}{r} \beta^3 \left[4U - U^2 -2\text{ln}(U) - 3\right]$. It is easy to get the first few moments (e.g., using Mathematica) from (\ref{stationary-moments-appendix}). One gets for instance
\begin{equation}
\left<x^2\right> = \frac{2D\beta^2}{r(1+\beta)}\, , \label{x2_app}
\end{equation}
\begin{equation} \label{x4_app}
\left<x^4\right> = \frac{12D^2\beta^4(12 +19\beta + \beta^2)}{r^2(1+\beta)^2(2+\beta)(3+\beta)}\, ,
\end{equation}\begin{equation}
\left<x^6\right> = \frac{120D^3 \beta^6 (4320+16344 \beta+22614 \beta^2+12443 \beta^3+2393 \beta^4+61 \beta^5+\beta^6)}{r^3(1+\beta)^3 (2+\beta)^2 (3+\beta)^2 (4+\beta) (5+\beta)}\, , \label{x6_app}
\end{equation}
which can indeed be written under the scaling form (\ref{rational}) given in the text.
%\begin{equation}
%\begin{split}
%&\left<x^8\right> = \frac{1680  D^4 \beta^8}{ r^4(1+\beta)^4 (2+\beta)^3 (3+\beta)^3 (4+\beta)^2 (5+\beta)^2 (6+\beta) (7+\beta)}\\
%&(87091200+551681280 \beta+1514343168 \beta^2+2345559408 \beta^3+2223640020 \beta^4+1328039864 \beta^5\\
%&+499588025 \beta^6 +114847458 \beta^7+14893875 \beta^8+891252 \beta^9+11583 \beta^{10}+130 \beta^{11}+\beta^{12})\, .
%\end{split}
%\end{equation}

\section{Saddle-point analysis for the large $x$ behaviour of the stationary distribution}\label{saddleana}

The purpose of this appendix is to derive the large $x$ behaviour of the stationary distribution in the case of a Poissonian resetting Brownian noise given in the text in (\ref{largex_1}). In particular, we will derive the asymptotic behaviour given in Eq.~(\ref{largex_3}). In the large $x$ limit, we showed that,

\be \label{largex_2_appendix}
p(x) \approx h(x) \;, \; h(x) = \beta \int_0^1 \frac{dU}{\sqrt{2 \pi \alpha(U)}} U^{\beta-1} e^{-\frac{x^2}{2 \alpha(U)}} \;.
\ee
First, we apply the change of variable $Y=\alpha(U)$ to obtain
\be \label{largex_3_appendix}
h(x) = \beta \int_0^{+\infty} \frac{dY}{\sqrt{2 \pi Y}} \frac{1}{|\alpha'\left(\alpha^{-1}(Y)\right)|}\left[\alpha^{-1}(Y)\right]^{\beta-1} e^{-\frac{x^2}{2 Y}} \;.
\ee
Then, another change of variable $Z=Y/x$ leads to
\be \label{largex_4_appendix}
h(x) = \beta \int_0^{+\infty} \frac{dZ}{\sqrt{2 \pi x\, Z}} \frac{x}{|\alpha'\left(\alpha^{-1}(x\, Z)\right)|}\left[\alpha^{-1}(x\, Z)\right]^{\beta-1} e^{-\frac{x}{2 Z}} \;.
\ee
The large $x$ behaviour is dominated by small $U$ values, thus large $Y$ values. We therefore consider $Y = Z\, x$ to be large. Now, notice that $\alpha\left(\alpha^{-1}(Y)\right) = Y \to +\infty$ when $ Y \to \infty$. From the behaviour of $\alpha(Y)$ we deduce that $\alpha^{-1}(Y)\to 0$ when $Y \to \infty$. Using the first asymptotic behaviour of $\alpha(U)$ in Eq.~(\ref{alpha_asympt}) one can show
\be\label{alphaeq1}
\alpha^{-1}(Y) = e^{-\frac{3}{2}-\frac{r}{2\, \beta^3\, D} Y}\left(1+O(e^{-Y})\right) \, ,
\ee
and the function $\alpha'\left(\alpha^{-1}(x\, Z)\right)$ is dominated by
\be \label{alphaeq2}
\alpha'\left(\alpha^{-1}(x\, Z)\right) \approx \frac{2D\beta^3}{r}e^{\frac{3}{2}+\frac{3r}{2\beta^3}x\, Z}\, .
\ee
Injecting Eq.~(\ref{alphaeq1}) and (\ref{alphaeq2}) in Eq.~(\ref{largex_4_appendix}) gives after simplifications
\be\label{saddlepointintegral}
h(x) \approx \sqrt{\frac{x}{2 \pi}} \frac{r}{2D\beta^2} \int_0^{+\infty} \frac{dZ}{\sqrt{Z}} e^{-x\left[\frac{r}{2\beta^2 D} Z + \frac{1}{2 Z}\right]}\, .
\ee
A saddle-point analysis on Eq.~(\ref{saddlepointintegral}) gives the expected large $x$ behaviour of the function $h(x)$ (\ref{largex_3}),
\be
h(x) \approx \frac{e^{-3\beta/2}}{2 \beta} \sqrt{r/D} \, e^{- \frac{x}{\beta}\sqrt{r/D}} \quad, \quad x \to \infty \;.
\ee

\section{Computation of the two-time correlation function}\label{appendixcorrelations}

In this appendix, we will detail the calculation of the two-time correlation function for a model described by the following dynamics:
\begin{equation} \label{eq_app_F}
\frac{dx(t)}{dt}=-\mu \, x(t) + r\, y_r(t)\, ,
\end{equation}where \begin{equation}
\displaystyle
y_r(t+dt) = \left\{
    \begin{array}{ll}
        0 \hspace*{3cm}\mbox{ with proba. } rdt \\
        y_r(t) + \xi(t)\, dt \hspace*{0.8cm} \mbox{ with proba. } (1-rdt)
    \end{array}
\right.
,
\end{equation} and $\xi(t)$ is a Gaussian white noise of zero mean $\langle \xi(t)\rangle = 0$ and delta-correlations $\langle \xi(t) \xi(t')\rangle = 2 D \delta(t-t')$, $D$ being the diffusion constant. The solution of this equation (\ref{eq_app_F}) starting from $x(0)=x_0$ reads
\be \label{sol}
x(t) = x_{0}e^{-\mu t} +r\, \int_{0}^{t} dt' \, e^{-\mu(t-t')}y_{r}(t') \;.
\ee 
Therefore the two-time correlation function is given by
\begin{equation} \label{correl}
\left<x(t_1)x(t_2)\right> = \left<\left[x_{0}e^{-\mu t_1} +r\, \int_{0}^{t_1} dt_1'\, e^{-\mu(t_1-t_1')}y_{r}(t_1') \right] \left[x_{0}e^{-\mu t_2} +r\, \int_{0}^{t_2} dt_2'\, e^{-\mu(t_2-t_2')} y_{r}(t_2') \right]   \right> \, .
\end{equation}
Since $\langle y_r(t)\rangle = 0$ for all $t$, the only nonzero terms in (\ref{correl}) are
\begin{equation}
\left<x(t_1)x(t_2)\right> = x_0^2 e^{-\mu (t_1 + t_2)} +  r^2 \, I_{C}(t_1,t_2)\, ,
\end{equation}
where
\begin{equation}\label{Ic}
 I_{C}(t_1,t_2) = \int_{0}^{t_1} dt_1'e^{-\mu(t_1-t_1')} \int_{0}^{t_2} dt_2'e^{-\mu(t_2-t_2')} \left<y_r(t_1')y_r(t_2')\right>\, .
\end{equation}When $t_2 \geq t_1$, we have \cite{correlation resetting} 
\begin{equation}\label{correly}
  \left<y_r(t_1')y_r(t_2')\right> = \left(\frac{2D}{r}\right) e^{-r(t_2'-t_1')} (1 - e^{-rt_1'}),
\end{equation} while if $t_1 \geq t_2$,
\begin{equation}
  \left<y_r(t_1')y_r(t_2')\right> = \left(\frac{2D}{r}\right) e^{-r(t_1'-t_2')} (1 - e^{-rt_2'}) \, .
\end{equation}Therefore, we divide the double integral over $t'_1$ and $t'_2$ in Eq. (\ref{Ic}) into two pieces and get, 
for $t_2 \geq t_1$,
\begin{eqnarray}
&\hspace{-1cm}I_{C}(t_1,t_2) = \int_{0}^{t_1} dt_1'e^{-\mu(t_1-t_1')} \int_{t'_1}^{t_2} dt_2'e^{-\mu(t_2-t_2')} \left[\left(\frac{2D}{r}\right) e^{-r(t_2'-t_1')} (1 - e^{-rt_1'}) \right] \\
&+    \int_{0}^{t_1} dt_1'e^{-\mu(t_1-t_1')} \int_{0}^{t'_1} dt_2'e^{-\mu(t_2-t_2')} \left[\left(\frac{2D}{r}\right) e^{-r(t_1'-t_2')} (1 - e^{-rt_2'})\right], 
\end{eqnarray}and for $t_1 \geq t_2$
\begin{eqnarray}
&\hspace{-1cm}I_{C}(t_1,t_2) = \int_{0}^{t_2} dt_2'e^{-\mu(t_2-t_2')} \int_{0}^{t'_2} dt_1'e^{-\mu(t_1-t_1')} \left[\left(\frac{2D}{r}\right) e^{-r(t_2'-t_1')} (1 - e^{-rt_1'}) \right] \\
&+ \int_{0}^{t_2} dt_2'e^{-\mu(t_2-t_2')} \int_{t'_2}^{t_1} dt_1'e^{-\mu(t_1-t_1')} \left[\left(\frac{2D}{r}\right) e^{-r(t_1'-t_2')} (1 - e^{-rt_2'}) \right] \, .
\end{eqnarray}
In the special case where $t_1 =t_2 = t$, and $x_0 = 0$, the integrals can be performed straightforwardly and one gets (for $r \neq \mu$ and $r \neq 2 \mu$)
,\begin{equation}
\text{V}(t)  =  \left<x^2(t)\right> - \left<x(t)\right>^2 = \frac{2\, D\, r}{\mu}\left[\frac{1}{\mu + r} +\frac{2\, r \, e^{-(\mu + r)t}}{\mu^2-r^2} - \frac{r\, e^{-2\mu t}}{(\mu-r)(2\mu-r)}-\frac{2\, e^{-rt}}{2\mu -r}\right]\, ,
\label{variance_time_appendix}
\end{equation}
which yields the result given in text in Eq. (\ref{variance_time}). 

\section{Moments of the time dependent distribution $p(x,t)$}\label{appendixmomentstimedep}

Using the same idea as in section \ref{statiomoments}, but now with the time dependent Fourier transform  $\hat{p}(k,t)$ of the distribution $p(x,t)$, one can derive a recursion relation for the time dependent moments of $p(x,t)$. We start from Eq. (\ref{Fourierp(x,t)}) in the text, namely
\begin{equation}\label{Fourierp(x,t)2}
\hat{p}(k,t) = e^{-rt}\, \widehat{\mathcal{P}}_0(k\, r\, e^{-\mu t},t) + r\, \int_{0}^{t}d\tau\,  e^{-r\tau}\,  \widehat{\mathcal{P}}_0(k\, r\, e^{-\mu t},\tau)\, \hat{p}(k,t-\tau)\, .
\end{equation}Let us expand $\hat{p}(k,t)$ and $\mathcal{P}_0(k\, r\, e^{-\mu t},t)$ in powers of $k$ as
\begin{equation} \label{exp_p}
\hat{p}(k,t) = \sum_{n=0}^{\infty}\frac{(ik)^{2n}}{(2n)!}\left<x^{2n}(t)\right>,
\end{equation}and 
\begin{equation} \label{exp_P}
\hat{\mathcal{P}}_0(k\, r\, e^{-\mu t},t) = \sum_{n=0}^{\infty}\frac{(ik\, r\, e^{-\mu t})^{2n}}{(2n)!}\left<A_0^{2n}(t)\right>\,,
\end{equation}
where we recall that $A_0(t) = \int_{0}^{t} dt'e^{\mu t'} B(t')$, with $B(t)$ being a standard Brownian motion starting from the origin. In particular, since $A_0(t)$ is a Gaussian random variables, one has
\begin{equation} \label{moment_A0}
\left<A_0^{2n}(t)\right> = (2n-1)!!\left<A_0^{2}(t)\right>^{n} \quad {\rm where} \quad  \left<A_0(t)^2\right> = \frac{D}{\mu^3}\, \left[e^{2\mu t}(2\mu t -3) + 4e^{\mu t} -1\right]
\end{equation}
Injecting these expansions (\ref{exp_p}) and (\ref{exp_P}) into Eq. (\ref{Fourierp(x,t)2}), we obtain,
\begin{equation}
\begin{split}
&\sum_{n=0}^{\infty}\frac{(ik)^{2n}}{(2n)!}\left<x^{2n}(t)\right> =e^{-rt}\sum_{n=0}^{\infty}\frac{(ik\, r\, e^{-\mu t})^{2n}}{(2n)!}\left<A_0^{2n}(t)\right>  \\
&+\,r\, \int_{0}^{t}d\tau\, e^{-r\tau}  \left[\sum_{p=0}^{\infty}\frac{(i k\, r\, e^{-\mu t})^{2p}}{(2p)!}\left<A_0^{2p}(\tau)\right>\right]\left[\sum_{q=0}^{\infty}\frac{(ik)^{2q}}{(2q)!}\left<x^{2q}(t-\tau)\right>\right]\, .
\end{split}
\end{equation}
Now, identifying the coefficient of $k^{2n}$ on both sides, one gets
\begin{equation}
\begin{split}
&\frac{1}{(2n)!}\left<x^{2n}(t)\right> = e^{-rt}\frac{(r\, e^{-\mu t})^{2n}}{(2n)!}\left<A_0^{2n}(t)\right> + \\
&\int_{0}^{t}d\tau\,  e^{-r\tau}  \sum_{\substack{p,q=0 \\ p+q=n}}^{\infty}\, r^{2p +1}\, \frac{1}{(2p)!(2q)!}e^{-2\mu pt}\left<A_0^{2p}(\tau)\right>\left<x^{2q}(t-\tau)\right> \;,
\end{split}
\end{equation}
which can be written as
\begin{equation}
\left<x^{2n}(t)\right> = r^{2n}\, e^{-rt}e^{-2\mu nt}\left<A_0^{2n}(t)\right> + \int_{0}^{t}d\tau\,  e^{-r\tau}  \sum_{\substack{p,q=0 \\ p+q=n}}^{\infty}\, r^{2p+1}\, \frac{(2n)!}{(2p)!(2q)!}e^{-2\mu pt}\left<A_0^{2p}(\tau)\right>\left<x^{2q}(t-\tau)\right>\, .
\end{equation}
To proceed, it is convenient to isolate the terms corresponding to $p=0$, and $q=0$ in the second term on the right hand side (r.h.s.) to obtain
\begin{equation}
\begin{split}
&\left<x^{2n}(t)\right> =  r^{2n}\, e^{-(r+2\mu n)t}\left<A_0^{2n}(t)\right> + r^{2n+1}\,  \int_{0}^{t}d\tau \, e^{-r\tau}e^{-2\mu nt}\left<A_0^{2n}(\tau)\right>  \\
&+ r\, \int_{0}^{t}d\tau\,  e^{-r\tau}\left<x^{2n}(t-\tau)\right>+\int_{0}^{t}d\tau\,  e^{-r\tau}  \sum_{\substack{p,q=1 \\ p+q=n}}^{\infty}\, r^{2p+1}\, \frac{(2n)!}{(2p)!(2q)!}e^{-2\mu pt}\left<A_0^{2p}(\tau)\right>\left<x^{2q}(t-\tau)\right>\, .
\end{split}
\end{equation}The third term on the r.h.s has a convolution structure. Using Laplace transform we can isolate $\left<x^{2n}(t)\right>$ and write:
\begin{equation}
\begin{split}
&\frac{s}{r+s}\mathcal{L}_{t\to s}\left[\left<x^{2n}(t)\right>\right] =   r^{2n}\, \mathcal{L}_{t\to s}\left[  e^{-(r+2\mu n)t}\left<A_0^{2n}(t)\right> + r\, \int_{0}^{t}d\tau \, e^{-r\tau}e^{-2\mu nt}\left<A_0^{2n}(\tau)\right>\right] \\
& +\mathcal{L}_{t \to s}\left[\int_{0}^{t}d\tau\,  e^{-r\tau}  \sum_{\substack{p,q=1 \\ p+q=n}}^{\infty}\, r^{2p+1}\, \frac{(2n)!}{(2p)!(2q)!}e^{-2\mu pt}\left<A_0^{2p}(\tau)\right>\left<x^{2q}(t-\tau)\right>\right]\, .
\end{split}
\end{equation}Finally, one can write formally
\begin{equation}\label{momentsgeneralformula}
\begin{split}
&\left<x^{2n}(t)\right>= \mathcal{L}^{-1}_{s\to t}\left[\frac{r+s}{s}\left(   r^{2n}\, \mathcal{L}_{t_1 \to s}\left[  e^{-(r+2\mu n)t_1}\left<A_0^{2n}(t_1)\right> + r\, \int_{0}^{t_1}d\tau \, e^{-r\tau}e^{-2\mu nt_1}\left<A_0^{2n}(\tau)\right>\right]\right) \right]\\
& + \mathcal{L}^{-1}_{s \to t}\left[\frac{r+s}{s}\left(\mathcal{L}_{t_1 \to s}\left[\int_{0}^{t_1}d\tau\,  e^{-r\tau}  \sum_{\substack{p,q=1 \\ p+q=n}}^{\infty}\, r^{2p+1}\, \frac{(2n)!}{(2p)!(2q)!}e^{-2\mu pt_1}\left<A_0^{2p}(\tau)\right>\left<x^{2q}(t_1-\tau)\right>\right]\right)\right]\, ,
\end{split}
\end{equation} 
Using Eq. (\ref{momentsgeneralformula}), the moments of the distribution can be computed recursively using a formal calculation software. 
The obtained expressions are rather cumbersome but we have checked that in the limit $t \to \infty$ the moments $\left<x^{2}(t)\right>$, $\left<x^{4}(t)\right>$, $\left<x^{6}(t)\right>$, $\ldots$ converge to their stationary values $\left<x^{2}\right>$, $\left<x^{4}\right>$, $\left<x^{6}\right>$, $\ldots$ computed in Eqs. (\ref{x2_app})-(\ref{x6_app}).

\end{document}